\documentclass[numberedappendix,12pt,preprint]{emulateapj}

\usepackage[]{amsmath}

\shorttitle{GLASS}
\shortauthors{Treu et al. (2015)}

\usepackage{color}			      
\definecolor{midgray}{gray}{0.4}		
\definecolor{orange}{rgb}{1,0.5,0}    

\usepackage[colorlinks=true,citecolor=midgray,linkcolor=midgray,urlcolor=midgray]{hyperref}        

\newcommand{\ep}{\epsilon_{\rm p}}

\newcommand{\simgt}{\,\rlap{\lower 3.5 pt \hbox{$\mathchar \sim$}} \raise
1pt \hbox {$>$}\,}
\newcommand{\simlt}{\,\rlap{\lower 3.5 pt \hbox{$\mathchar \sim$}} \raise
1pt \hbox {$<$}\,}

\newcommand{\lya}{Ly$\alpha$}
\newcommand{\Ha}{H$\alpha$}
\newcommand{\pasa}{PASA}

\newcommand{\Nz}{139}
\newcommand{\M}{MACSJ0717.5+3745}

\begin{document}

\title{The Grism Lens-Amplified Survey from Space (GLASS). I. Survey overview and first data release}

\author{
T.~Treu$^{1,2}$, 
K.~B.~Schmidt$^{2}$,  
G.~B.~Brammer$^3$,
B.~Vulcani$^{4}$,
X.~Wang$^{2}$,
M.~Brada\v{c}$^5$, 
M.~Dijkstra$^6$, 
A.~Dressler$^7$, 
A.~Fontana$^8$,
R.~Gavazzi$^9$, 
A.~L.~Henry$^{10}$, 
A.~Hoag$^5$, 
K.-H.~Huang$^5$,
T.~A.~Jones$^{2}$,
P.~L.~Kelly$^{11}$, 
M.~A.~Malkan$^{1}$, 
C.~Mason$^{2}$,
L.~Pentericci$^8$, 
B.~Poggianti$^{12}$, 
M.~Stiavelli$^3$,
M.~Trenti$^{13}$, 
A.~von der Linden$^{14,15}$
}
\affil{$^{1}$ Department of Physics and Astronomy, University of California, Los
Angeles, CA, USA 90095-1547}
\affil{$^{2}$ Department of Physics, University of California, Santa Barbara, CA, 93106-9530, USA}
\affil{$^{3}$ Space Telescope Science Institute, 3700 San Martin Drive, Baltimore, MD, 21218, USA}
\affil{$^{4}$ Kavli Institute for the Physics and Mathematics of the Universe (WPI), The University of Tokyo Institutes for Advanced Study (UTIAS), the University of Tokyo, Kashiwa, 277-8582,
Japan}
\affil{$^5$ Department of Physics, University of California, Davis, CA, 95616, USA}
\affil{$^6$ Institute of Theoretical Astrophysics, University of Oslo, P.O. Box 1029, N-0315 Oslo, Norway}
\affil{$^7$ The Observatories of the Carnegie Institution for Science, 813 Santa Barbara St., Pasadena, CA 91101, USA}
\affil{$^8$ INAF - Osservatorio Astronomico di Roma Via Frascati 33 - 00040 Monte Porzio Catone, I}
\affil{$^9$ Institute d'Astrophysique de Paris, F}
\affil{$^{10}$ Astrophysics Science Division, Goddard Space Flight Center,
Code 665, Greenbelt, MD 20771}
\affil{$^{11}$ Department of Astronomy, University of California, Berkeley, CA 94720-3411, USA}
\affil{$^{12}$ INAF-Astronomical Observatory of Padova, Italy}
\affil{$^{13}$ School of Physics, University of Melbourne, VIC 3010, Australia}
\affil{$^{14}$Dark Cosmology Centre, Niels Bohr Institute, University of Copenhagen Juliane Maries Vej 30, 2100 Copenhagen {\O}, DK}
\affil{$^{15}$ Kavli Institute for Particle Astrophysics and Cosmology, Stanford University, 452 Lomita Mall, Stanford, CA  94305-4085, USA}
\email{tt@astro.ucla.edu}
\begin{abstract}
We give an overview of the Grism Lens Amplified Survey from Space
(GLASS), a large Hubble Space Telescope program aimed at obtaining
grism spectroscopy of the fields of ten massive clusters of galaxies
at redshift $z=0.308-0.686$, including the Hubble Frontier Fields
(HFF).  The Wide Field Camera 3 yields near infrared spectra of the
cluster cores, covering the wavelength range $0.81-1.69\mu m$ through
grisms G102 and G141, while the Advanced Camera for Surveys in
parallel mode provides G800L spectra of the infall regions of the
clusters. The WFC3 spectra are taken at two almost orthogonal position
angles in order to minimize the effects of confusion. After
summarizing the scientific drivers of GLASS, we describe the sample
selection as well as the observing strategy and data processing
pipeline. We then utilize \M, a HFF cluster and the first one observed
by GLASS, to illustrate the data quality and the high-level data
products. Each spectrum brighter than H$_{\rm AB}=23$ is visually
inspected by at least two co-authors and a redshift is measured when
sufficient information is present in the spectra. Furthermore, we
conducted a thorough search for emission lines through all the GLASS
WFC3 spectra with the aim of measuring redshifts for sources with
continuum fainter than H$_{\rm AB}=23$. We provide a catalog of \Nz\
emission-line based spectroscopic redshifts for extragalactic sources,
including three new redshifts of multiple image systems (one probable,
two tentative). In addition to the data itself we also release
software tools that are helpful to navigate the data.
\end{abstract}


\section{Introduction}
\label{sec:intro}

Over the past 25 years, the Hubble Space Telescope (HST) has assembled
a phenomenal legacy of extragalactic surveys. Several legacy fields
have been imaged at a variety of wavelengths, becoming the focus of
ground based telescopes and communities interested in intermediate and
high redshift science
\citep{Ferguson:2000p22537,Giavalisco:2004p28701,Koekemoer:2011p9456,Scoville:9p22538,Ellis:2013p26700}. Additionally, pure-parallel fields surveys have covered large uncorrelated areas providing even larger
samples and means to control sample variance
\citep{Trenti:2011p12656,Schmidt:2014p34189,Col++13}.

A perhaps less well-known part of the HST legacy are spectroscopic surveys
carried out initially with the Near Infrared Camera and Multi-Object
Spectrograph \citep[NICMOS][]{McC++99} and the Advanced Camera for
Suverys \citep[ACS][]{Pir++13} and more recently with the Wide Field
Camera 3
\citep[WFC3][]{Atek:2010p33653,Brammer:2012p12977,Mor+15}. By virtue
of its low background and exquisite image quality compared to the
ground, HST excels in low resolution slitless spatially resolved spectroscopy,
particularly in the near infrared. Thus, it has become a workhorse for
galaxy evolution studies, including: the study of spatially resolved
star formation
\citep{Nelson:2012p12947}; the detection of rare galaxies with strong
emission lines and faint continuum emission \citep{Ate++11}; and the
measurement of absorption line redshifts for $z>1$ galaxies
\citep{vanDokkum:2011p10254,Newman:2013p33256}

Current HST spectroscopic surveys have focused on relatively shallow
spectroscopy of extragalactic fields. Although this is a very
productive way to survey large areas of the sky, the target depth has
typically been insufficient to address some of the most interesting
questions in galaxy formation and evolution. For example, HST
spectroscopic surveys have been able to detect \lya\ for the brightest
galaxies usually only during the tail-end of cosmic reionization
\citep{Rho++13}. Additionally, at lower redshifts, legacy fields and
random parallels do not contain massive clusters of galaxies and thus
cannot probe the densest environments.

The fields of massive clusters of galaxies provide a very powerful
complement to the traditional legacy fields. On the one hand they
contain massive clusters of galaxies and thus enable the study of
environmental effects. On the other hand the clusters themselves act
as gravitational telescopes by magnifying objects in the background
through the gravitational lensing effect. The lensing magnification
enables the study of intrinsically fainter and more compact galaxies
in the background than would be possible in empty fields, at the
expense of increased contamination by the foreground and a loss of
cosmic volume for a fixed solid angle. It is important to stress that
the net effect of a gravitational telescope on the number counts of a
background population (sometimes known as magnification bias) can be
both positive or negative, depending on the slope of the luminosity
function: if the differential luminosity function is steeper than
$dn/dL\propto L^{-2}$, one detects more objects behind a gravitational
telescope, if it shallower the empty field has a higher yield
\citep[e.g.,][]{SF06}.  In practice -- at the bright end of the
luminosity function where counts are rising exponentially -- the net
gain from lensing is huge and aids greatly in the detection of
galaxies at $z>6$. Even more importantly, for a fixed exposure time,
lensing allows one to probe an intrinsically fainter population, thus
providing qualitatively different insights into the source population.

Exploiting lensing magnification \citep{Tre10,CBZ15} is one of the
motivations behind many HST cluster surveys
\citep[e.g.,][]{Smi++01b,Postman:2012p27556,Bay++14} and the main
driver of the ongoing Hubble Frontier Field initiative (Lotz et
al. 2015, in preparation).

The Grism Lens Amplified Survey from Space (GLASS; PI Treu; GO 13459)
has been designed and carried out in order to address some of the most
compelling scientific questions by combining the power of deep HST
spectroscopy with the magnification effect of gravitational
lensing. The key science questions to be addressed by GLASS are

\begin{enumerate}
\item How and when did galaxies reionize the Universe (if they did)?
\item How do gas and metals cycle in and out of galaxies?
\item How does galaxy evolution depend on characteristics of the local environment?
\end{enumerate}

First results on these topics are given in the papers by
\citet{Schmidt+2015}, \citet{Jon++15}, and Vulcani et al. (2015, in
prep).
 
In addition to the key questions driving the survey design, the data
collected by GLASS enable the study of luminous and dark matter in the
cluster themselves, by providing spectroscopic redshifts of background
sources to improve the lensing models, and provide additional epochs
to search for high redshift supernovae gravitational lensed by the
clusters. Examples of these two additional science goals are given in
the early science papers by \citet[][hereafter paper
0]{Schmidt:2014p33661}, \citet{Kel++15}, and \citet{Wan+15}.

In this paper we give an overview of GLASS and we present the first
release of the data for \M, the first cluster targeted by the survey.
The paper is organized as follows. In Section~\ref{sec:sci} we review
the scientific drivers of the survey. In Section~\ref{sec:sample} we
give the rationale for selecting the sample of clusters and summarize
their main properties. In Section~\ref{sec:obs} we detail the
observational strategy as well as the data reduction. In
Section~\ref{sec:visual} we describe the postprocessing of the
spectra, including procedure and tools for visual inspection and
redshift determination. Section~\ref{sec:summary} concludes with a
summary. Present and future data products are and will be made public
through the HST archive. All magnitudes are given in the AB system and
a standard concordance cosmology with $\Omega_m=0.3$,
$\Omega_\Lambda=0.7$ and $h=0.7$ is adopted when necessary.

\section{Scientific Drivers}
\label{sec:sci}

\begin{figure}[]
\centerline{
\includegraphics[width=0.49\textwidth]{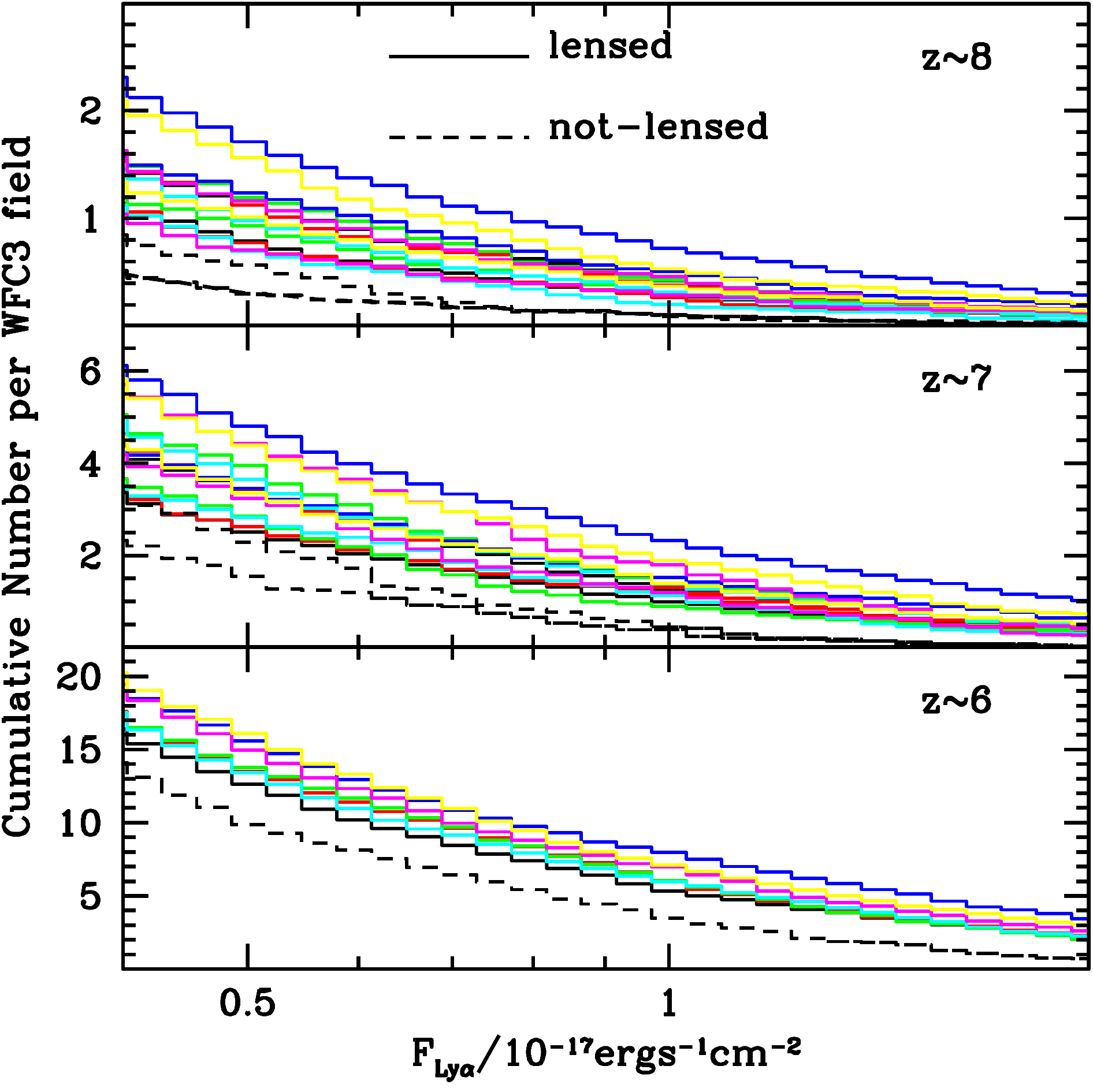}}
\caption{Predicted number of \lya\ emitters per WFC3 field in cluster and blank 
fields based on the conditional probability function of \lya\ emission
for LBGs, and the LBG luminosity function model by
\citet{MTT15}. Measuring the evolution of the luminosity function of
\lya\ emitters is one of the goals of GLASS. At the depth achievable
with the WFC3 G102 grism lensing magnification has a positive effect
at all relevant redshifts.  The dashed lines are for blank fields, the
solid lines are for specific GLASS clusters (A2744=black; A370=red;
MACS0416=green; MACS0717=blue; MACS1149=cyan; RXJ1347=yellow;
RXJ2248=magenta). At $z\sim6$ the conditional probability distribution
function is as described by \citet{Treu:2012p12658}, based on the data
by \citet{Stark:2011p27664}. At $z\sim7$ we use the two ``patchy''
measurements by \citet[][upper curve of each pair]{Treu:2012p12658}
and \citet[][lower curve]{2014ApJ...793..113P} to give a sense of the
measurement uncertainties. At $z\sim8$ the conditional probability has
been assumed to be the same as $z\sim7$. More details on the models
are given in the main text.
\label{fig:strategy1}}
\end{figure}

\subsection{Key Science Driver 1: Gas and galaxies at the epoch of reionization}

The Universe is known to undergo dramatic evolution in the redshift
range $z=11-6$ while becoming fully ionized
\citep{WMAP9,Planck2015XIII,Rob++15,MTT15}. However, how and when exactly it
gets there is the topic of much debate. The interaction between the
first galaxies and the intergalactic medium (IGM) is poorly
understood, leaving many questions unanswered. How many ionizing
photons escape from the local interstellar medium (ISM) into the IGM?
What is the topology of the ISM/IGM at these redshifts? What is the
relative contribution of galaxies vs active nuclei to ionizing budget
\citep{Mad++04,R+O04,Gia++15}?

The reduced observed \lya\ flux from galaxies at $z>6$ indicate that
the process is still ongoing at $z\approx 6-7$
\citep[e.g.,][]{McQ++07,Dij++11,Jen++13,Cho++14}. This indication
for ``late'' reionization is consistent with constraints derived from
the \lya\ forest \citep[e.g.,][]{Bec++15} and also observed
non-detection of the kSZ effect on small scales in the cosmic
microwave background \citep{Mes++12}. In detail, quantitative
constraints inferred from Lya emitting galaxies still suffer
significantly from observational uncertainties
\citep[e.g.,][]{Mes++15}. In addition, uncertainties exist on the
theoretical side which include how \lya\ photons escape from galaxies,
how this depends on galaxy properties, the role of galactic winds, and
how much Lya is reprocessed by the CGM \citep[giving rise to
`blobs/halos' see, e.g.,][for a review]{Dij14}.

Recent ground-based observations suggest dramatic and rapid evolution
of the IGM/ISM beyond $z\sim6$. An important clue is that the fraction
of dropouts that are \lya\ emitters, increases steadily out to $z\sim
6$ \citep[e.g.,][]{Stark:2011p27664}, while it declines significantly
at $z\sim7$ and above
\citep{Kashikawa:2006p28155,Fontana:2010p29506,Pentericci:2011p27723,Schenker:2012p34406,Ono:2012p27651,Cle++12,Treu:2012p12658,Treu:2013p32132}. Large
samples are now available at redshifts up to 7 where high-sensitivity
multiplexed optical spectrographs can be used, and suggest that the
IGM might indeed be partially neutral with a patchy distribution of
\lya\ optical depth \citep[][hereafter P14]{2014ApJ...793..113P}.

In contrast, spectroscopic follow-up at $z>7$ has been very difficult,
owing to the atmosphere which limits studies to a few specific
redshift windows, inspite of the recent introduction of high
sensitivity high multiplexed near infrared spectrographs
\citep[e.g., MOSFIRE][]{McLean:2012p26812}.  This difficulty is compounded by
the weak average \lya\ emission for the brighter continuum sources
typically accessible with spectroscopy in legacy fields compared to
the fainter ones \citep{Stark:2011p27664}.
 
The goal of GLASS is to target a large sample of intrinsically faint
Lyman break galaxies (LBGs) at $z>6.5$ in order to make the most
accurate measurement to date of the distribution of the \lya\ optical
depth, and thus probe the properties of the IGM/ISM.  The first result
of this investigation are discussed in detail by \citet{Schmidt+2015}.
To achieve this goal GLASS exploits lensing magnification by the
foreground clusters to reach fainter luminosities than would be
possible in blank fields and thus in principle larger number of
detections. Furthermore, theoretical \citep{Zhe+11} and observational
\citep{Ouchi:2009p27671,Gua+15} arguments indicate that high-$z$ \lya\
emitters might be significantly more spatially extended than the
associated UV continuum.  GLASS should provide the first constraints
on the spatial extent of the \lya\ emission on subarcsecond scales,
via an individual study of magnified bright sources and a stacking
analysis of the sample \citet{Schmidt+2015}

\begin{figure}[]
\centerline{
\includegraphics[width=0.49\textwidth]{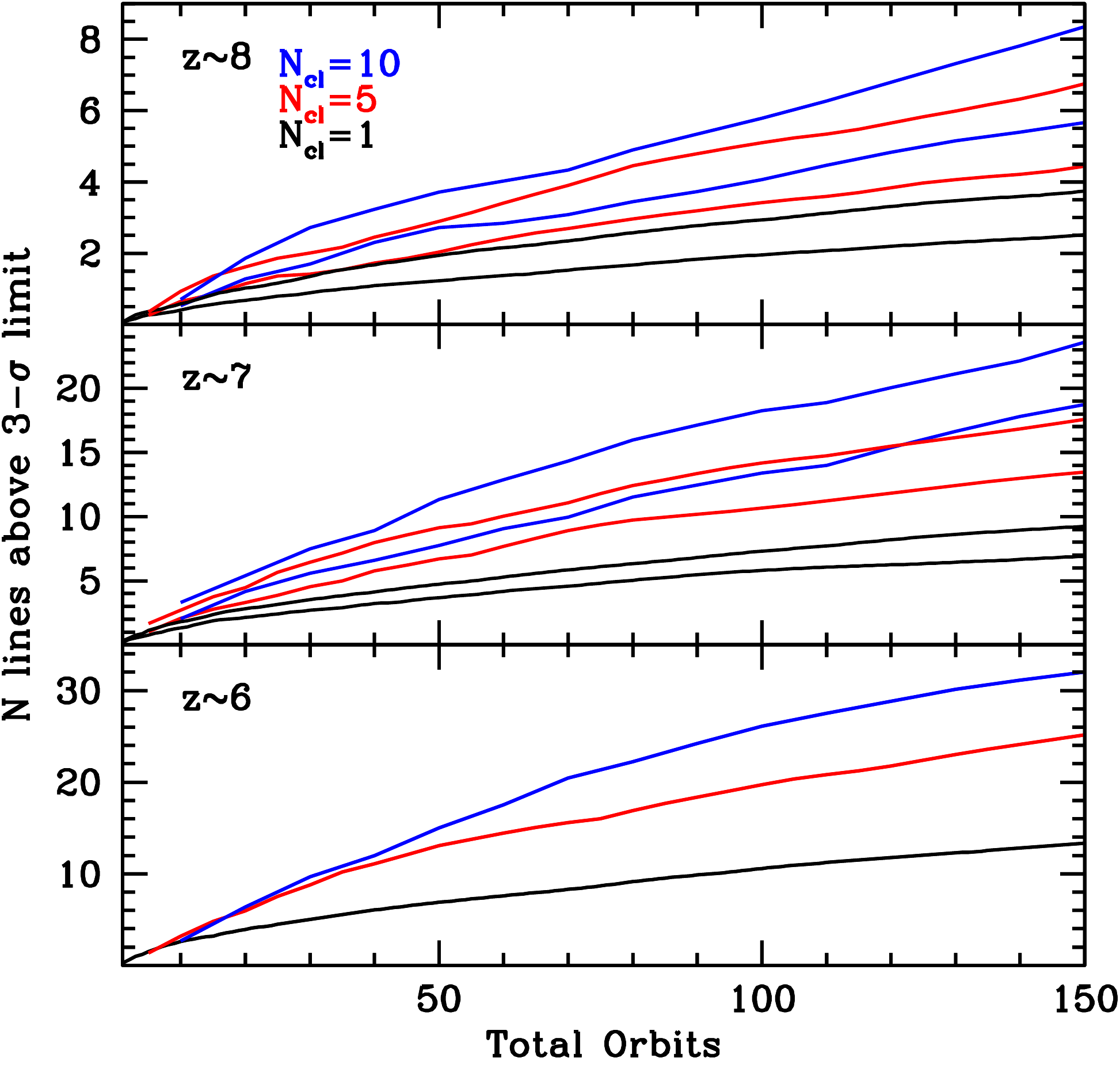}}
\caption{Number of \lya\ lines with flux above 3-$\sigma$ as a function of total G102 orbits divided over
  N$_{\rm cl}$ clusters. Once a sufficent depth is reached, the number
  of detection increases linearly with the number of clusters and less
  rapidly with exposure time owing to the loss of efficiency in
  magnification bias in addition to the normal scaling of noise and
  number counts observed in empty fields.  The predictions are based
  on the model shown in Figure~\ref{fig:strategy1} and described in
  the main text. Only the results for cluster RXJ1347 are shown for
  clarity. At $z\sim7$ and $8$ the upper and lower lines of each color
  represent $\ep=0.66$ and $\ep=0.46$ respectively, as in
  Figure~\ref{fig:strategy1}.
\label{fig:strategy2}}
\end{figure}

\subsubsection{Predicting the \lya\ number counts}
\label{sssec:model}

As a reference, Fig.~\ref{fig:strategy1} shows the predicted number of
dropouts as a function of \lya\ flux. The forecast is based on three
main ingredients: i) the luminosity function of LBGs; ii) the
conditional probability of the \lya\ equivalent width for LBGs of a
given continuum magnitude; iii) the distribution of magnifications
provided by the foreground lensing clusters. For the first ingredient
we use the model of the LBG LF recently proposed by \citet{MTT15}
building on earlier work by \citet{Trenti:2010p29335} and
\citet{Tacchella:2013p32241}. This model is in excellent agreement with 
with the actual number of LBGs detected in blank fields
\citep[e.g.,][]{Bradley:2012p23263,Oesch:2013p27877,Schmidt:2014p34189,Bouwens:2014p34683}.
For the second ingredient we use the parametrization of the
distribution of \lya\ equivalent width introduced by
\citet[][hereafter T12]{Treu:2012p12658}. Based on data from the
literature \citep{Stark:2011p27664} at $z\sim6$ the intrinsic
rest-frame distribution is:
\begin{equation}
p_6(W)=\frac{2 A}{\sqrt{2
\pi}W_{c}}e^{-\frac{1}{2}\left(\frac{W}{W_{c}}\right)^2}H(W)+(1-A)\delta(W),
\end{equation}
with W$_{c}$=47\AA, A=0.38 for sources with
$-21.75<\textrm{M}_\textrm{UV} < -20.25$ and W$_{c}$=47\AA, A=0.89 for
sources with $-20.25 < \textrm{M}_\textrm{UV}< -18.75$. A is the
fraction of emitters, $\delta$ is the delta function, and $H$ is the
step function. As an illustration and to avoid overcluttering the
figure, we consider here the ``patchy'' family of models introduce by
T12, which appears to provide the best description of the data at
$z\sim7$. For this family of models, the intrinsic distribution of
equivalenth widths is expressed in terms of $p_6$ and the parameter
$\ep$:
\begin{equation}
\begin{split}
p_{p}(W)=\ep p_6(W)+(1-\ep)\delta(W)=\\ \frac{2A\ep}{\sqrt{2\pi}W_{c}} e^{-\frac{1}{2}\left(\frac{W}{W_{c}}\right)^2}H(W)+(1-A\ep)\delta(W).
\end{split}
\end{equation}
As an illustration of observational uncertainties we consider two
values of $\ep=0.66,0.46$, taken from T12 and P14, respectively. At
$z\sim8$, only upper limits are available
\citep{Treu:2013p32132}. For simplicity we assume no evolution
between $z\sim7$ and $z\sim8$, even though the number counts might be
significantly lower at $z\sim8$.  In any case, the numbers are
sufficiently large that a smaller number of detections at $z\sim8$
would pose a stringent limit on the \lya\ optical depth.  For the
third ingredient we use magnification maps based on grid-based weak
and strong lensing models constructed by members of our team
\cite[e.g.,][and Hoag et al. 2015, in preparation]{2008ApJ...681..187B,Wan+15}, 
using the code developed by \citet{2005A&A...437...39B}, in the
adaptive version \citep{Bra++09}. Those are available for 7/10
clusters at the moment of this writing, including both HFF and non-HFF
clusters. These 7 clusters are representative of the full GLASS sample
and therefore the range of magnification maps gives a good sense of
range of expected variations from cluster to cluster. Mock catalogs of
high-redshift galaxies are ray traced through the magnification maps
to generate suitable mock catalogs in the image plane taking into
account the boost in flux as well as the change in solid angle. The
\lya\ emission is assumed to be unresolved for this excercise,
consistent with the observations
\citep{Schmidt+2015}.

The figure shows that magnification bias is positive in all cases, and
cluster fields always yield more sources than blank fields, in the
range of fluxes and redshifts considered here. Furthermore, there are
cluster to cluster variations, but they are not extreme. This is due
to the fact that the magnification effect of the most extreme lensing
clusters, such as those in the HFF sample, is somewhat mitigated by
the size of the field of view of WFC3. When the critical lines are
larger than the WFC3 field of view, the high magnification areas can
be lost to a single pointing campaign. As one can see from the range
in the predictions, cluster to cluster variations are comparable to
current uncertainties on the distribution of equivalent width of
\lya.  

\subsection{Key Science Driver 2: The Gas and Metal Cycles of Galaxies}

How gas flows in and out of galaxies is a central question in galaxy
formation and evolution as emphasized for example by the most recent
Decadal Survey \citep{NAP12951}.  The GLASS survey is particularly
well suited for addressing gas flows at redshifts $z\simeq2$,
corresponding to the peak period of cosmic star formation rate density
and hence rapid gas inflow and outflow rates
\citep[e.g.,][]{M+D14,Sha11}. We have identified a series of specific
questions that can be addressed by GLASS in order to better understand the
role of gas flows.

How do metallicity gradients evolve with cosmic time?  Metallicity
gradients are sensitive to the history of baryonic assembly: gas
accretion, mergers, star formation, and outflows.  Gradients measured
from spatially resolved emission line diagnostics have provided
significant insight into the assembly process at $z=0$
\citep[e.g.,][]{VCE92,RKB10,BKR12}. Cosmological simulations coupled with
chemical evolution models can reproduce these data, but various models
predict different behavior at earlier times
\citep[e.g.,][]{Pil++12,Gib++13}. Measurements for a large sample of
galaxies over a range of redshift thus provide important constraints
on galaxy assembly history \citep[e.g.,][]{Tro++14}.

How does the mass-metallicity-SFR relation evolve at low stellar mass?
Local galaxies lie on a tight relation where metallicity increases
with stellar mass \citep[e.g.,][]{Tremonti:2004p24667} and typically
decreases with SFR \citep[e.g.,][]{Man++10}. Both effects can be
explained by the cycling of gas and metals between galaxies and the
IGM and are more pronounced at low mass. The cosmic evolution of this
relation at low stellar mass is thus a powerful test of theories of
feedback and of metal enrichment of the IGM
\citep[e.g.,][]{F+D08,Hen++13}.

What causes the offset of star forming galaxies in the BPT diagram
\citep{BPT} at high redshifts?  Star forming galaxies and AGN at $z=0$
lie on separate well-defined loci in diagnostic line-ratio
diagrams. High redshift galaxies are often offset from the local star
forming locus toward that of AGN
\citep{Man++10,Kew++13,Sha++14,Ste++14}.  Spatially resolved data have
shown that in some galaxies this is due to a combination of star
formation and a weak active galactic nucleus
\citep{Wri++10,Trump:2011p10256}, while others show pure star formation
with offsets seen in individual giant star forming regions
\citep{Jon++13a}. Alternatively, recent work has shown that N-based
indicators could be biased tracers of Oxygen
\citep{Amo++10,A+M13,Mas++14,Sha++15}. Distinguishing these options is
critical for metallicity studies.

Progress in measuring resolved emission line diagnostics at high
redshift has been slow due to limitations in angular resolution and
sensitivity. The most promising efforts, targeting gravitationally
lensed galaxies with adaptive optics (AO), have yielded only a few
metallicity gradients to date \citep{Jon++10,Jon++13,YKR11}.
Seeing-limited data have been obtained for larger samples
\citep{Que++12,Tro++14}, and they are in tension with AO-based results,
highlighting the need for high spatial resolution. The sensitivity of
field HST surveys (e.g. WISP, 3D-HST, PEARS) is usually sufficient to
study only the very brightest systems and their angular resolution
does not benefit from lensing magnification.

The mass-metallicity-SFR relation has been measured by numerous groups
in large samples extending to $z\simeq3.5$ for the most massive
($M_*>10^{9.5}$ M$_{\odot}$) and luminous sources
\citep[e.g.,][]{Man++10}. Below $z\sim1$, faint sources can be probed with
optical spectra \citep{Hen++13}.
At higher redshifts, lensed galaxy surveys have extended this work to
lower $M_*$ and SFR \citep{Bel++13}, but are limited by small samples
(i.e., only $\sim15$ high redshift galaxies with $M_* < 10^9$
M$_{\odot}$). The vast majority of these studies rely on integrated
spectra and therefore cannot distinguish the cause of offsets in the
BPT diagram.

GLASS addresses these issues by collecting an unprecedented sample of
emission line galaxies at $0.65 < z < 3.3$, with spatially resolved
information for a significant fraction.  The questions raised above
are addressed via the integrated flux and spatial distribution of
multiple emission line diagnostics which are sensitive to the gas
metallicity and ionizing source. In particular the strong lines
[\ion{O}{2}], [\ion{O}{3}], and H$\beta$ are all observed for galaxies
at $1.3<z<2.3$ providing good diagnostics of both metallicity and
nuclear activity \citep[for example via the "blue" diagnostic
diagram;][]{Lam10}. [\ion{S}{2}] and H$\alpha$+[\ion{N}{2}] are also
valuable diagnostics available at $z<1.5$, while metallicities can
still be estimated for the strongest line emitters at $2.3<z<3.3$ via
[\ion{O}{2}] and [\ion{Ne}{3}].  As shown by \citet{Jon++15}, the
quality of the GLASS data enables accurate measurements of metallicity
gradients, and of the mass-metallicity-SFR relation down to
M$_*$=$10^7$ M$_\odot$ at $z=2$.  This progress is made possible by
the observing strategy and lensing magnification which deliver an
order of magnitude improvement in flux sensitivity and of 3-4 in
spatial resolution over previous HST grism surveys in blank fields
\citep[e.g.,][]{Hen++13}. In general, the broad wavelength coverage
of GLASS enables the comparison of metallicity estimates based on
different features, so that one can test, for example, whether the
[\ion{N}{2}] based estimates are biased with respect to [\ion{S}{2}].

\subsection{Key Science Driver 3: Environmental Dependence on Galaxy Evolution}

The epoch $0 < z < 1$ is one of rapid decline in the global star
formation rate \citep{Madau:1996p24662,Lilly:1996p24634,M+D14}, but
clusters experience an evolution in star formation activity over this
time that is even stronger than the general field. Identifying the
processes that trigger and terminate star formation in cluster
galaxies \citep{B+O84,Dre++99,Pog++99,Dre++13}, and contrasting them
to those operating in the field \citep{Coo++08,Vul+10,Muz++12,Oem++13}
is key to understanding the causes of the general decline.  For what
galaxy types and masses are environmental effects driving the
quenching of star formation, and what is instead internally driven
\citep{Pen++10}?  Is there one process that dominates over all
environments or do some play a larger role in driving galaxy evolution
in the field than they do in dense environments?  Massive clusters are
the sites where the effects of environmental processes, such as ram
pressure stripping, are expected to be strongest.  Hence, studies of
massive clusters are needed to answer these questions.

Each of the processes that have been proposed to quench star formation
in galaxies should leave a different signature on the spatial
distribution of the star formation activity within the galaxy. As an
example, the loss of the galaxy hot gas halo would lead to a
suppression of star formation over the whole disk, while ram-pressure
stripping should leave a recognizable pattern of star formation with
truncated \Ha\ disks smaller than the undisturbed stellar disk
\citep[see][for an example of ram pressure stripping]{Yag++15}. What is
needed is thus an unbiased census of the star formation activity in a
statistically significant sample of clusters, yielding both the total
star formation rates and the spatial distribution of star-forming
regions within galaxies. A most interesting epoch for such a study is
in the range $z\sim0.3-1$, where most of the evolution takes place.

\Ha\ is considered the best optical indicator of the current star 
formation rate, being much less affected by dust extinction and
metallicity effects than [\ion{O}{2}]
\citep{Kennicutt:1998p17963,Mou++06}.
For this reason, a number of \Ha\ surveys up to $z\sim1$ have been
undertaken in the field using narrow-band imaging
\citep[e.g.,][]{Sob++13} and with WFC3 grism observations \citep[e.g.,][]{Atek:2010p33653,Straughn:2011p8119}. In clusters, narrow-band \Ha\ studies are
available for just a few systems at $z=0.3-1$
\citep{Kod++04,Fin++05,Koy++11} and a few other higher-$z$ overdense
regions \citep{Koy++13}. These studies provide integrated \Ha\ fluxes,
and no spatial distribution information.  The power of spatially
resolved information is exemplified by a study of Virgo which
suggested that in the local Universe stripping of gas is the main
mechanism for quenching star formation rates in clusters \citep{K+K04}.

GLASS is designed to measure \Ha\ fluxes for all cluster star-forming
galaxies in the central 0.6-0.9 Mpc of 10 clusters at
$z=0.31-0.69$. Note that \Ha\ and [\ion{N}{2}] are blended at the
resolution of G102; this effect can be taken into account when
comparing with other samples, or corrected based on higher spectral
resolution data \citep{Tru++11,Tru++13,Fum++12}. GLASS's target
sensitivity is an order of magnitude improvement over previous studies
of emission lines in cluster galaxies \citep{Sob++13}. GLASS is
designed to reach SFR limits comparable to the deepest narrow band
studies, but for all objects and with spatially resolved information
(see Vulcani et al. 2015, in preparation, for more details). Thus,
GLASS is the first unbiased source of spatially resolved star
formation within cluster galaxies at these redshifts.  The combination
of GLASS spectroscopy and HST photometry from CLASH/HFF is used to
identify uniquely the line as
\Ha, as well as measure morphology, spatially resolved colors and
stellar masses for virtually all objects detected in emission.

\begin{figure*}[]
\centerline{
\includegraphics[width=\textwidth]{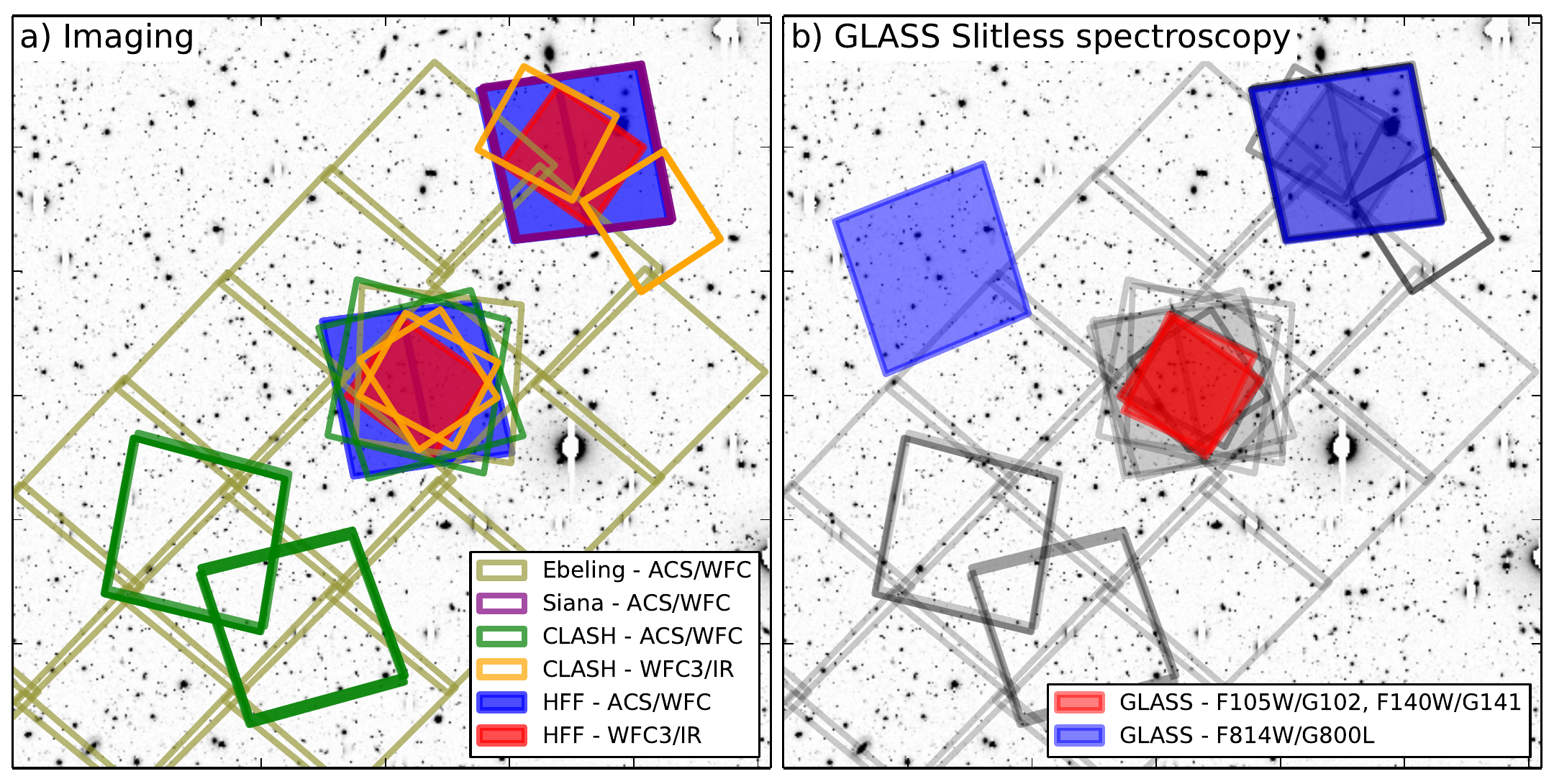}}
\caption{Observational layout of the GLASS spectroscopy in the context of existing imaging data for \M. The WFC3 grism spectroscopy is aimed at the cluster center overlapping with deep multiband imaging from CLASH and/or HFF, while the two parallel fields are approximately 90 degrees apart, with one of them coinciding with the CLASH/HFF parallels.
\label{fig:layout}}
\end{figure*}

Parallel ACS G800L Grism spectroscopy in two offset fields (at 2-3
Mpc, i.e. 1-3 virial radii; 2 position angles; see
Figure~\ref{fig:layout}) for each cluster, will cover the cluster
infall regions, a key region for some environmental processes
\citep[e.g.,][]{Tre++03}. In these regions H$\beta$ (and [\ion{O}{2}]
at $z>0.6$) cab be used as star formation indicators, while
[\ion{O}{3}] further aids in redshift identification. By combining
cluster cores, infall regions, and field, GLASS enables the study of
the spatially resolved star formation across a wide range of
environments.  Given the low number density of cluster members at such
clustercentric distances, the ACS grism is the ideal method to build
complete samples of star-forming galaxies. The parallel ACS data are
not part of this first data release and will be discussed in detail in
future publications.

\subsection{Additional Science Drivers}
\label{ssec:additional}

GLASS spectroscopy and imaging is suitable for a wide range of
applications. Although the survey design is driven by the three
science cases described above, two additional science drivers were
taken into account when possible. They are briefly summarized below.

\subsubsection{Luminous and dark matter in clusters of galaxies}

Measuring the distribution of matter in clusters of galaxies is a
powerful way to address a number of fundamental astrophysical
questions, ranging from cosmology to the interaction between
supermassive black holes and their environment. What is the
self-interaction cross section of dark matter? Is dark matter cold?
What is the net effect of energy feedback from black hole accretion on
dark matter halos?

By modeling the gravitational lensing distortion of background
galaxies one can obtain precise and accurate two-dimensional maps of
the mass distribution in the clusters themselves. Considerable effort
has gone and is going into developing modeling techniques that exploit
all the available information, and deep HST imaging data have allowed
researchers to identify more than one hundred multiple images of
background sources per clusters. However the vast majority of these
multiple images do not have spectroscopic redshift identification, and
photometric redshift must be used to transform their positions into
constraints on the gravitational potential of the
clusters. Spectroscopic redshifts help enourmously by reducing the
effective uncertainty of each potential measurement to below 1\% and
by eliminating mis-identifications and catastrophic errors on photo-z
which are common for extremely faint galaxies with unusual shapes. One
of the ancillary drivers of GLASS is to measure spectroscopic
redshifts for as many faint sources as possible in order to provide
more extended and cleaner catalogs of multiple images to be used as
input to lens models. First examples of this application of GLASS are
given by \citet{Schmidt:2014p33661} and \citet{Wan+15}.

\subsubsection{Supernovae}

Gravitational telescopes magnify all sources in the background,
including transient ones. Thus, in principle gravitational telescopes
can be used to detect and take spectra of supernovae that are fainter
and more distant than otherwise possible \citep{BKRST14}, thus
extending the look back time over which supernovae and supernova
cosmology can be studied
\citep[e.g.][]{Sul+00,Pat+14}. If the lensed supernovae are of type
Ia, knowledge of their absolute magnitude provides an opportunity to
break the mass-sheet degeneracy
\citep{ZRT14,Nor+14} (Rodney et al. 2015).

In rare circumstances, if the supernovae happens to explode within the
cluster's caustics, the supernovae can appear multiply imaged to the
observer. The difference in arrival time between the images is
measurable and can be used as a cosmic ruler to measure distances and
hence cosmology \citep{Ref64,Hol00,B+B03}, provided the gravitational
potential of the cluster can be sufficiently well constrained. This
technique has been demonstrated to be very powerful using quasars as
the variable lensed source. Typically, the deflectors are galaxies,
for which the gravitational potential can be sufficiently well
constrained by extended gravitational arcs and stellar kinematics
\citep{T+K02b,Suy++14}. Multiply imaged supernovae are significantly 
more difficult to find, and the first example has been discovered only
recently in data taken by the GLASS survey
\citep{Kel++15}.

This area of science has been done in close coordination with the
FrontierSN Program (GOs 13386 and GO 13790; PI: Rodney) for the HFF
clusters, ensuring uniform acquisition and analysis of the data.

\section{Sample selection and properties}
\label{sec:sample}

The number and choice of clusters depend on a number of factors,
starting from the steepness of the number counts of the target
population. The epoch of reionization science driver was used to
select the optimal strategy. Figure~\ref{fig:strategy2} shows the
expected number of \lya\ emitters as a function of {\it total} number
of orbits, using the model described in Section~\ref{sssec:model},
plus the actual sensitivity measured for the GLASS grism
data\footnote{As it will be shown later in the paper, the sensitivity
to \lya\ varies very rapidly between $z=5.5$ and $6.5$. We are taking
as a reference the value at $z=6$, although most objects will be
detected above $z>6$ where the sensitivity is highest.}. At
$z\sim6-8$, once a sufficient depth per cluster is reached, the number
of sources is maximized by observing more clusters rather than going
deeper.  Observing multiple clusters also minimizes cosmic variance
\citep[for $z>6$ galaxies $\lesssim 10$\% for 10 clusters, compared to
30\% for a single WFC3 field][]{Trenti:2008p32309}. Observing only 4
clusters (e.g. the initial HFF targets) for a longer individual
exposure reduces the number of detections because magnification
becomes less effective. Of course, even though we chose the number of
sources as metric, this is not the only reasonable choice. By going
deeper on a smaller number of clusters one would have probed an
intrinisically different population of galaxies.

In order to find the 10 best clusters to target with GLASS we turned
to the CLASH and HFF initiatives for the unique data that are being
accumulated by a variety of groups with HST and other facilities.
This vast array of data makes the identification of emission lines
much easier than for clusters with just a few HST filters,
e.g. allowing for the use of the dropout technique to confirm that a
single emission line is indeed \lya, and of photo-$z$s to resolve
cases with multiple line identifications. We further pruned the list
of GLASS and CLASH clusters, using cluster redshift as an additional
criterion. In fact, clusters at $z\approx0.3-0.8$ are the best
gravitational telescopes for two reasons: i) the size of the critical
lines is well matched to the WFC3 field of view; ii) cluster galaxies
appear smaller and fainter than in lower redshift clusters, thus
minimizing the foreground contamination for background sources. The
final requirement is given by the ability to observe H$\alpha$ in
cluster galaxies ($z=0.22-1.59$). Based on these criteria we selected
the sample of ten clusters at $z=0.31-0.69$ listed in
Table~\ref{tab:sample}. The table lists the main (but not only)
sources of HST imaging and Spitzer Imaging, as well as basic features
like redshift, X-ray temperature and weak lensing virial mass. Much
ancillary data is available for these clusters at virtually all
wavelengths.

\begin{deluxetable*}{lcccccccc} \tablecolumns{9}
\tablecaption{Sample properties}
\tablehead{
\colhead{cluster} & \colhead{RA} & \colhead{DEC} & \colhead{z} & \colhead{HST imaging} & \colhead{{\it Spitzer} imaging} & \colhead{L$_{\rm X}$} & \colhead{M$_{\rm 500}$} & \colhead{First Release}  \\
\colhead{} & \colhead{(J2000)} & \colhead{(J2000)} & \colhead{} & \colhead{} & \colhead{} & \colhead{(10$^{44}$erg s$^{-1}$)} & \colhead{(10$^{14}$M$_\odot$)} & \colhead{}}
\startdata
A2744          & 00:14:23.4  & -30:23:26 & 0.308 &    HFF1	  & SFF     & 15.28$\pm$0.39& 17.6$\pm$2.3 & F15   \\  
A370           & 02:39:52.8  & -01:34:36 & 0.375 &    HFF3	  & SFF	    &  8.56$\pm$0.37& 11.7$\pm$2.1 & W16   \\
M0416.1-2403   & 04:16:09.4  & -24:04:04 & 0.420 &    CLASH/HFF1  & SURFSUP &  8.11$\pm$0.50&  9.1$\pm$2.0 & W16   \\
M0717.5+3745   & 07:17:31.6  & +37:45:18 & 0.548 &    CLASH/HFF2  & SFF	    & 24.99$\pm$0.92& 24.9$\pm$2.7 & This paper \\
M0744.9+3927   & 07:44:52.8  & +39:27:24 & 0.686 &    CLASH	  & SURFSUP & 18.94$\pm$0.61& 12.5$\pm$1.6 & W16   \\
M1149.6+2223   & 11:49:35.9  & +22:23:55 & 0.544 &    CLASH/HFF2  & SURFSUP & 17.25$\pm$0.68& 18.7$\pm$3.0 & F15   \\
RXJ1347.5-1145 & 13:47:30.6  & -11:45:10 & 0.451 &    CLASH       & SURFSUP & 47.33$\pm$1.2 & 21.7$\pm$3.0 & S15   \\
M1423.8+2404   & 14:23:47.8  & +24:04:40 & 0.545 &    CLASH       & SURFSUP & 13.96$\pm$0.52& 6.64$\pm$0.88& S15   \\
M2129.4-0741   & 21:29:26.1  & -07:41:29 & 0.570 &    CLASH       & SURFSUP & 13.69$\pm$0.57& 10.6$\pm$1.4 & S15   \\
RXJ2248        & 22:48:44.3  & -44:31:36 & 0.348 &    CLASH/HFF3  & SFF	    & 30.81$\pm$1.57& 22.5$\pm$3.3 & S15   \\
\enddata
\tablecomments{\label{tab:sample} For each cluster we list J2000 coordinates, redshift, the main sources of HST and SST imaging, as well as M$_{500}$ from M$_{\rm gas}$ and X-ray luminosity \citep[from][]{Mant++10}.
The last column lists the target date for the first public data
release (F=Fall, S=Summer, W=Winter).  }
\end{deluxetable*}

\section{Observations and data reduction}
\label{sec:obs}

\begin{figure}[]
\centerline{
\includegraphics[width=0.49\textwidth]{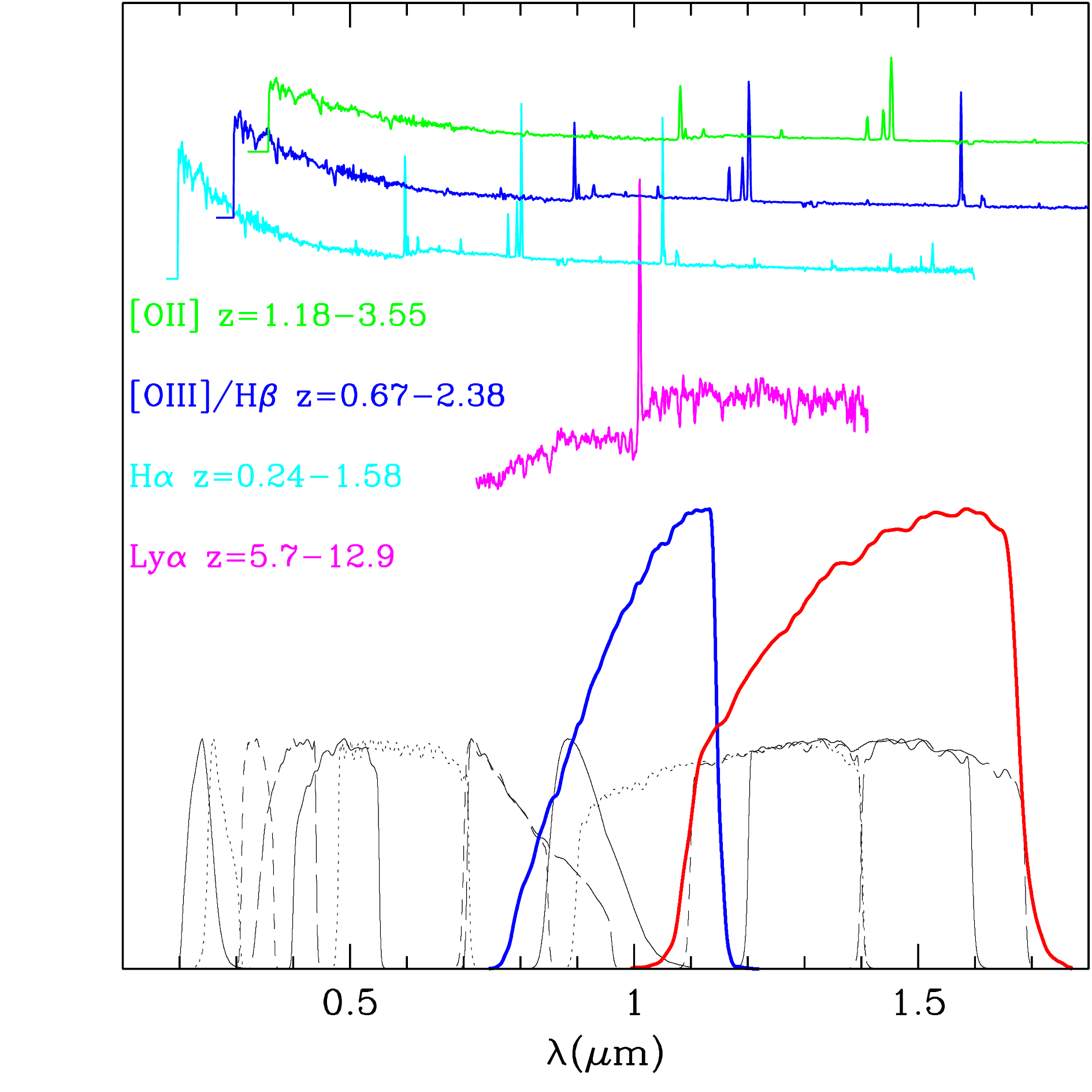}}
\caption{Illustration of the wavelength coverage provided by the two WFC3 grism (in red and blue, respectively G141 and G102). The continuous wavelength coverage between $\sim 0.81-1.69 \mu$m enables all three main science drivers. The transmission curves of the filters used by the CLASH imaging program are shown as light black curves for comparison. Four example spectra, one for each of the diagnostic features shown in the caption, are overplotted for illustration.
\label{fig:wave}}
\end{figure}

The two grisms G102 and G141 provide continuous wavelength coverage in
the range $0.81-1.69 \mu$m (defined as the range where the sensitivity
is $>20$\% of the peak), thus enabling the detection of \lya\ above
$z=5.7$, metallicity gradients over a range of redshifts, and \Ha\ for
the cluster members (see Figure~\ref{fig:wave}).

The depth of 10 orbits in G102 was driven by the need to reach a
sensitivity sufficient to probe the \lya\ luminosity function at
$z\gtrsim6$. The depth of four orbits in G141 was chosen to reach
approximately uniform sensitivity across the two grisms, which is
appropriate to measure typical line ratios required for metallicity
determinations \citep{Atek:2010p33653,Hen++13}.  In this section we
detail the observational strategy (\ref{ssec:phaseII}) and the data
reduction process (\ref{ssec:reduction}).

\subsection{Phase II design}
\label{ssec:phaseII}

The main pointing was chosen to overlap with existing or planned HST
imaging observations. The observations of each cluster were then
divided into two sets of visits at different roll angles, to help
dealing with the effects of contamination in the crowded cluster
fields. One position angle was chosen so that its ACS parallel field
would land in a HFF or CLASH parallel field, while the second roll
angle was chosen to be at $90\pm10$ degrees from the first, so as to
resolve contamination from overlapping spectra, while maximizing the
overlap between the two WFC3 grism pointings.

A dither strategy analogous to that followed by the 3D-HST survey
shown in Figure~3 of \cite{Brammer:2012p12977} has been adopted, in
order to maximize spatial and spectral resolution and defect
removal. For each visit at least 4 subexposures were taken for each
grism (typically half orbit each) with semi-integer pixel offsets, so
that they could be combined by interlacing as described in the next
section.

A paired direct image exposure was taken with each grism exposure
(F105W or F140W for G102 or G141, respectively) without offsetting the
telescope for image alignment and spectral calibration. For visits
that contained only G102 spectroscopy, some of the pre-imaging was
obtained in F140W in order to ensure uniform filter coverage of
transient events like Supernovae. This choice had virtually no impact
of the depth of the F105W exposures or accuracy of the calibration of
the G102 spectroscopy while at the same time assisting for the
ancillary science case. Furthermore, after aligning the grism visits
to existing observations from the CLASH and HFF programs, the deeper
images from those programs can be used as the reference for the
spectral extractions and modeling.

The effective exposure times for \M{} in G102, G141, F105W and F140W
at the two different position angles are listed in
Table~\ref{tab:exptime} and are typical of all GLASS clusters. The
notional depth of the G102 and G141 grism exposures are 10 and 4
orbits respectively, including alignment images and overheads. 

\begin{figure}
\begin{center}
\includegraphics[width=0.49\textwidth]{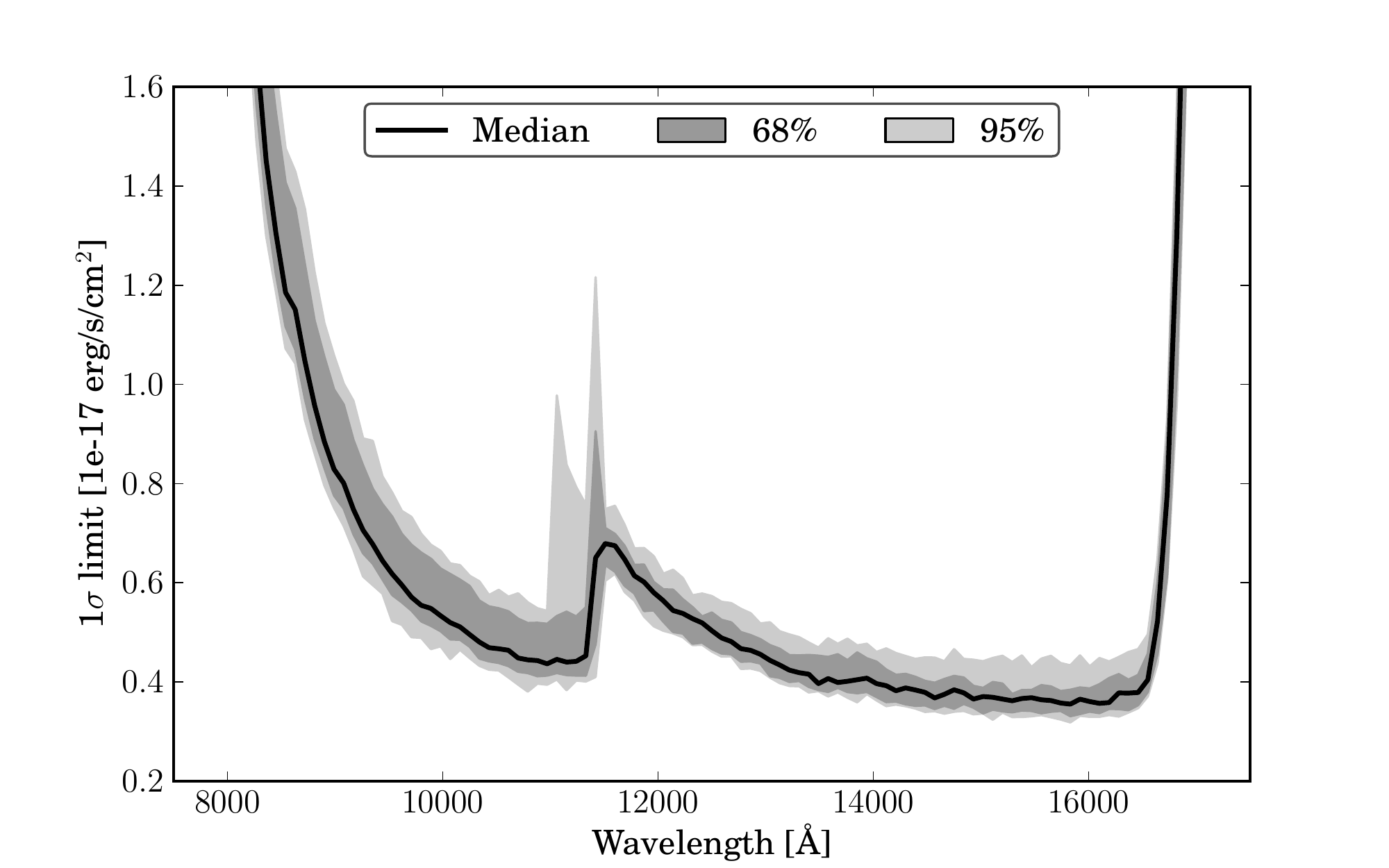}\\
\caption{The observed 1$\sigma$ noise level (not accounted for lensing magnification) for a sample of non-detections high-$z$ dropouts from the first 6 GLASS clusters (see \cite{Schmidt+2015} for details). 
An extraction parameter of approximately 0.6\arcsec$\times$100\AA{}
running over the full G102+G141 wavelength range was used.  Due to
significantly varying contamination for individual objects between the
two observed GLASS position angles, the noise level is shown for
spectra extracted from a single P.A.  Hence, the displayed limits
correspond to the depth obtained from half of the GLASS data on each
object.  Combining spectra at the two P.A. for contamination free
wavelength ranges will lower the noise level by a factor of
$\sqrt{2}$. For more extended galaxies, more representative noise
levels can be obtained using a larger extraction aperture, e.g., of
1.2\arcsec$\times$200\AA{}, which will decrease the sensitivity by a
factor 2.  }
\label{fig:sensitivity}
\end{center}
\end{figure}

%
The depth of the spectroscopic data is illustrated in
Figure~\ref{fig:sensitivity} which gives the sensitivity for an
unresolved emission line in a clean part of the field of view.  The
sensitivity estimates were estimated for a sample of non-detections
$z\gtrsim7$ compact dropouts from the first 6 GLASS clusters as
presented by \cite{Schmidt+2015}.  We used extraction apertures of
5~(spatial) by 3~(spectral) native pixels, which corresponds to
apertures of $\sim$0.6\arcsec$\times$100\AA{} similar to what was used
by \cite{Schmidt:2014p33661}.  For more extended emission and larger
objects, a larger aperture of, e.g., 10~(spatial) by 6~(spectral)
native pixels, might be more representative of the sensitivity
estimates. Such an aperture will increase the noise level by a factor
of 2.  These noise levels are in fair agreement with previously
published sensitivities of the \emph{HST} NIR grisms
\citep{Brammer:2012p12977,Trump:2014p41000}.  We note that the depth
of the spectroscopic data varies significantly from cluster to cluster
owing to the variable background level, due to the combined zodial
light and atmospheric emission.
The depth of the imaging data is comparable to that of CLASH. All
imaging data obtained as part of the GLASS program are combined with
imaging data taken as part of the CLASH and HFF programs.

\begin{deluxetable}{lccc} \tablecolumns{4}
\tablecaption{\M{} Exposure Times}
\tablehead{\colhead{Filter} & \colhead{P.A.=20} & \colhead{P.A.=280} & \colhead{Total} \\
~ &$t_\textrm{exp}/[\textrm{s}]$ & $t_\textrm{exp}/[\textrm{s}]$ & $t_\textrm{exp}/[\textrm{s}]$  }
\startdata
G102\footnote{Exposure times after correction for He Earth-glow (see Section~\ref{ssec:reduction})}          
                     & 9629  & 10829   & 20459 \\
G141$^a$     & 3812  &  4312  & 8123  \\
F105W          & 1979  & 1979   & 3959 \\
F140W          & 712  &  712  &  1423 \\
\enddata
\tablecomments{\label{tab:exptime} The table summarizes the exposure times of the GLASS data for \M. These are typical of the GLASS dataset for all 10 clusters. The position angles listed correspond to the HST keyword PA\_V3. The P.A. of the y-axis corresponds to PA\_V3+44.69. Additional imaging data are available from the HST archive.}
\end{deluxetable}

\subsection{Data reduction}
\label{ssec:reduction}

\begin{figure*}
\begin{center}
\includegraphics[width=0.7\textwidth]{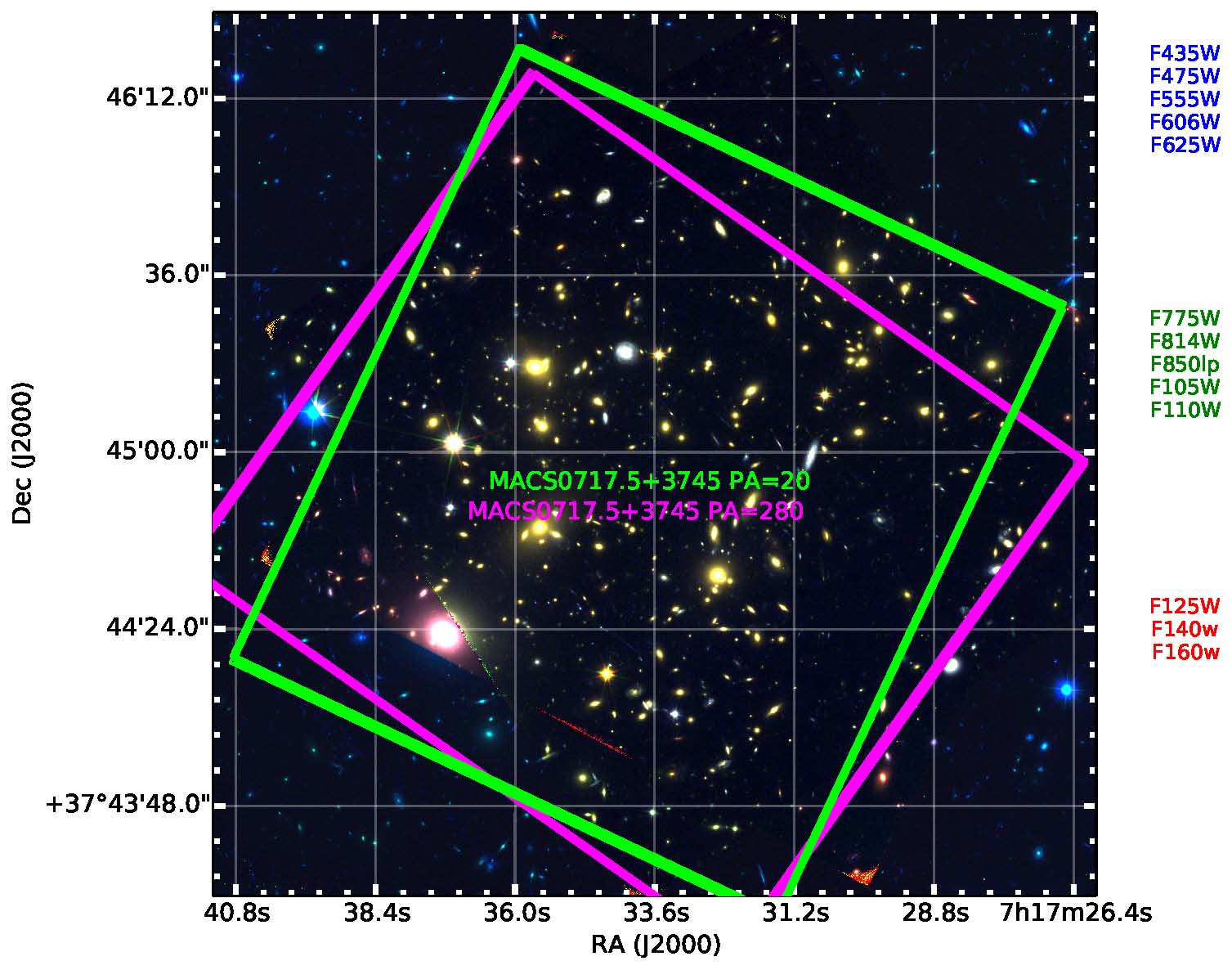}\\
\includegraphics[width=0.45\textwidth]{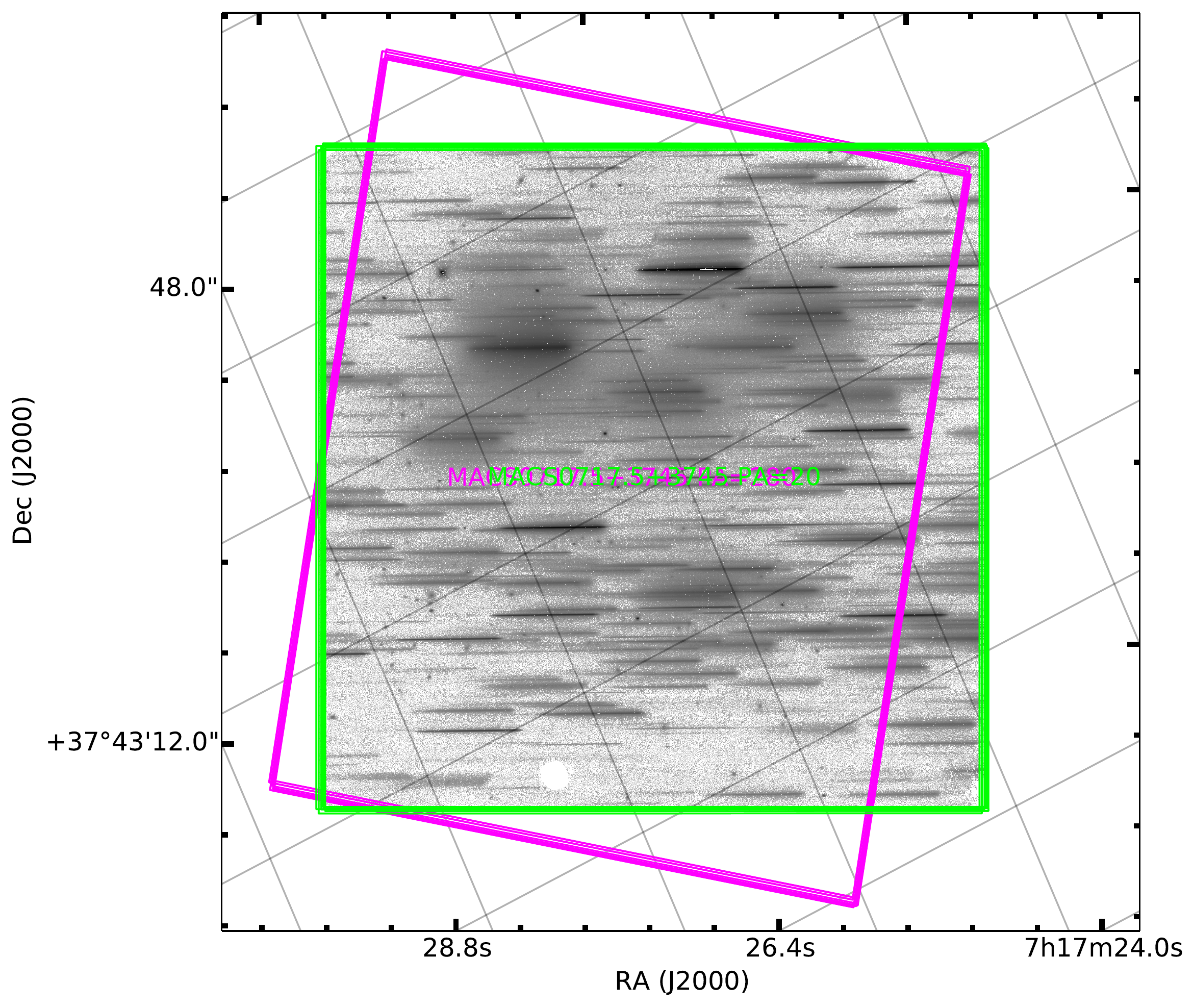}
\includegraphics[width=0.45\textwidth]{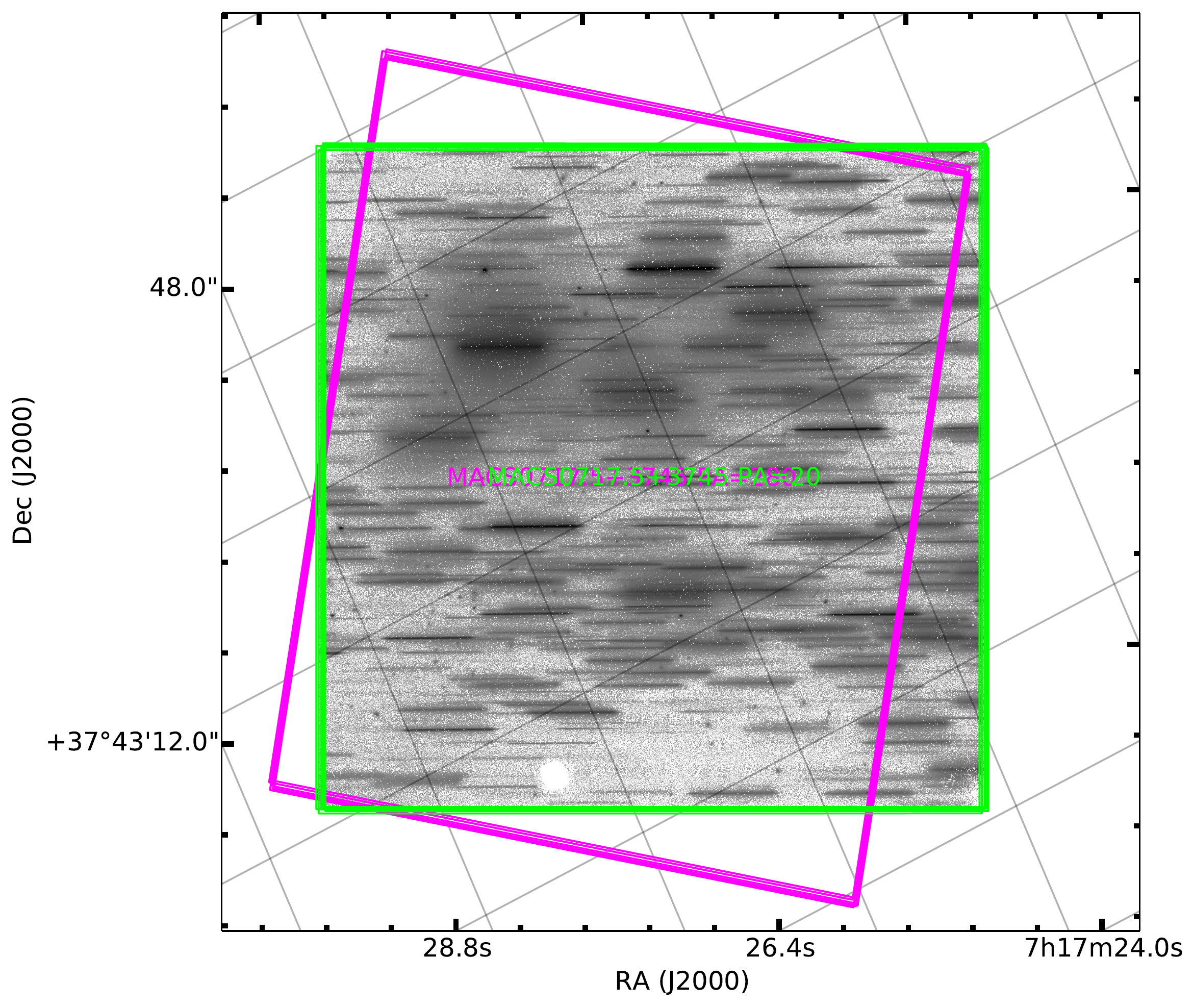}\\
\includegraphics[width=0.45\textwidth]{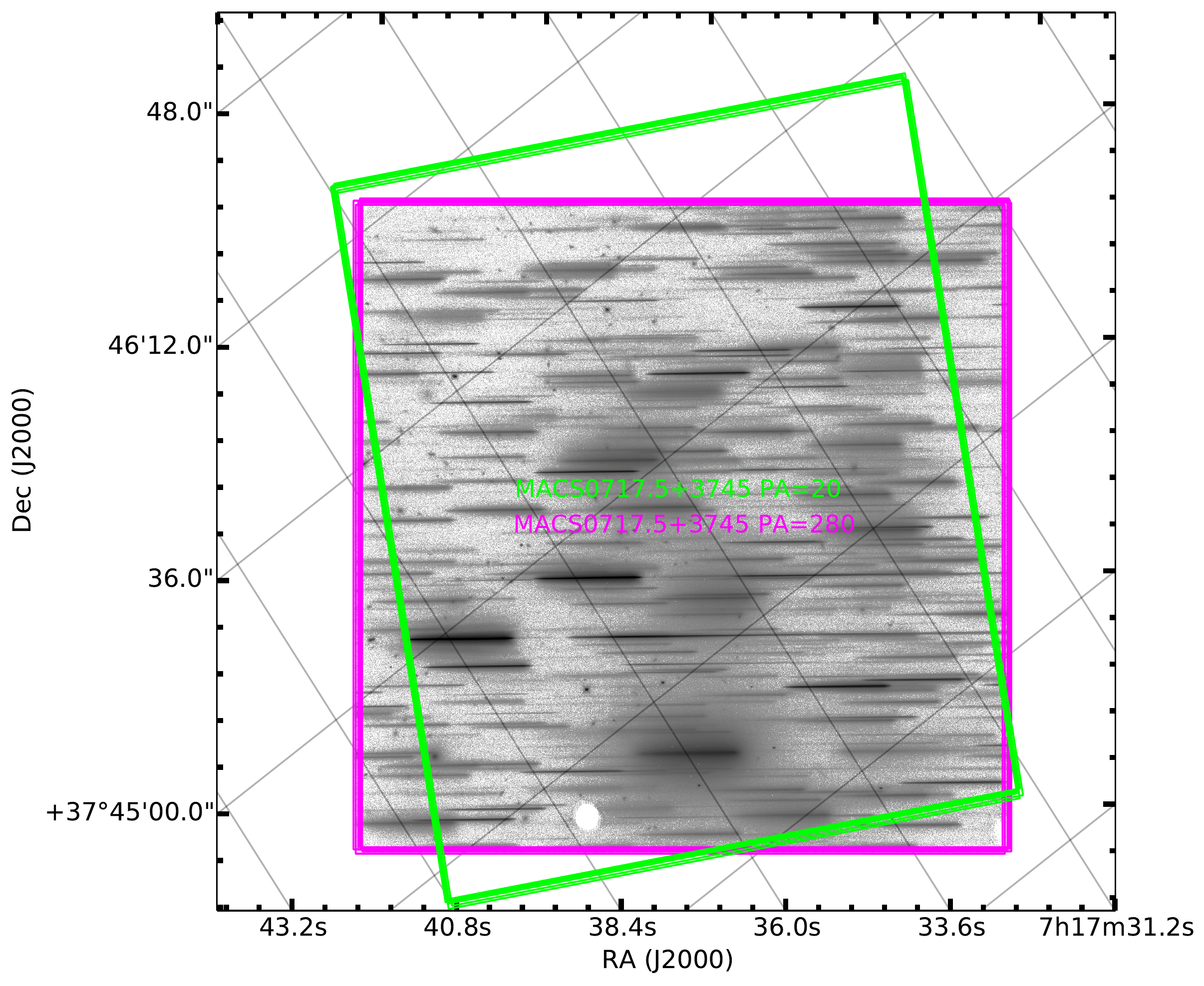}
\includegraphics[width=0.45\textwidth]{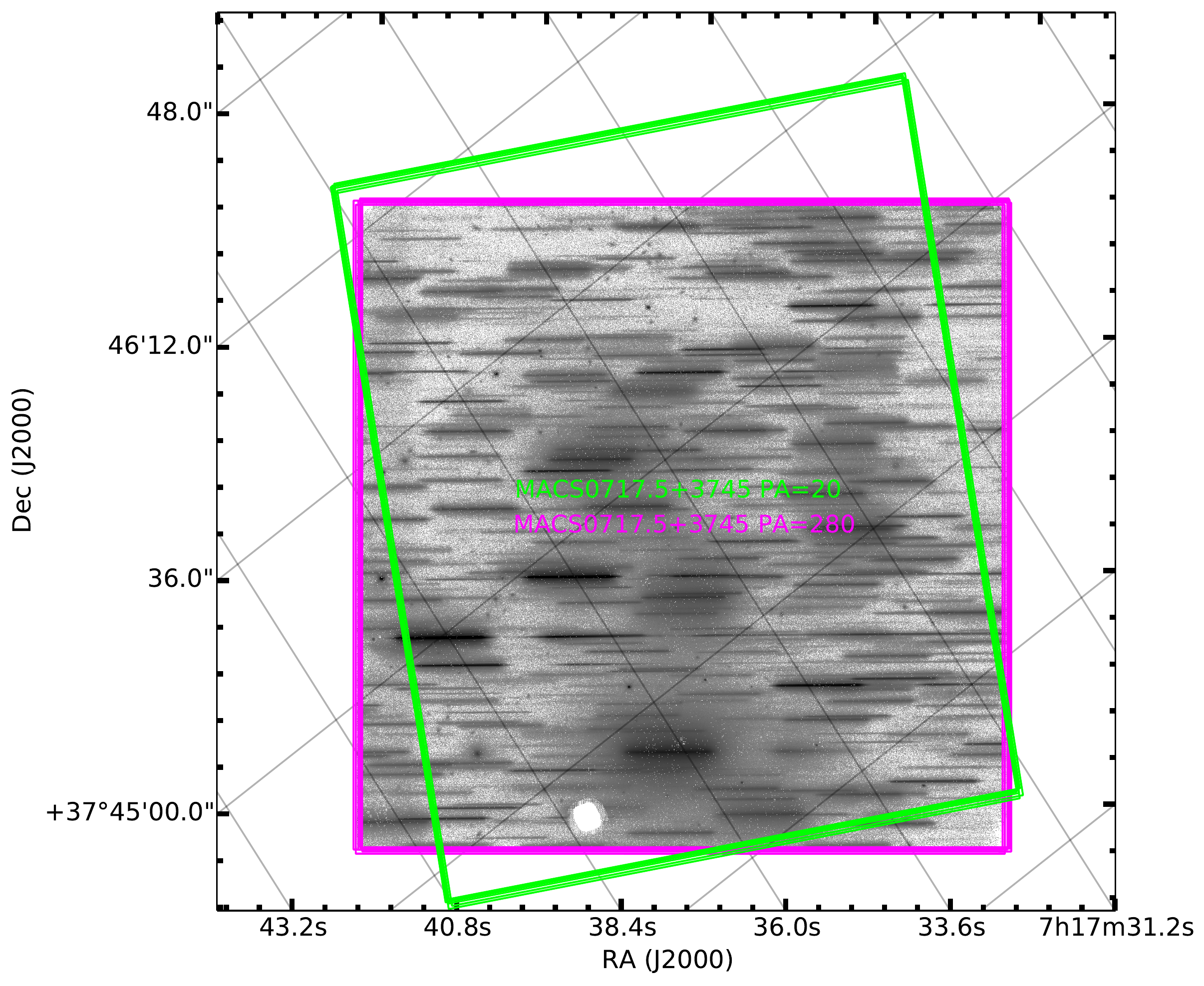}
\caption{The \M{} GLASS fields-of-view.
The color composite image (top) is based on the CLASH imaging with blue, green and red channels as noted on the right.
The four bottom panels show the final interlaced sky-subtracted G102 (left) and G141(right) grism mosaics at the position angle 20 (top; green polygons) and 280 (bottom; magenta polygons) degrees.
The individual spectra are extracted from these mosaics based on the extent of the corresponding object in the direct image mosaics (not shown here).}
\label{fig:image}
\end{center}
\end{figure*}

As noted in Section~\ref{ssec:phaseII}, the GLASS observations are
designed to comply with the 3D-HST observing strategy and were
processed with an updated version of the 3D-HST reduction
pipeline\footnote{http://code.google.com/p/threedhst/} described by
\citet{Brammer:2012p12977}.  The updated pipeline combines the
individual exposures into mosaics using AstroDrizzle
\citep{Gonzaga:2012p26307}, replacing the MultiDrizzle package
\citep{Koekemoer:2003p31861} used in earlier versions of the pipeline.
The individual exposures and visits are aligned using
\verb+tweakreg+ and grism sky backgrounds are subtracted using master
sky images as described by Brammer (2015, in preparation),
\cite{Kummel:2011p33451}, and
\cite{Brammer:2012p12977}.  The direct images were
sky subtracted by fitting a 2nd order polynomial to each of the
source-subtracted exposures.  Each exposure is then interlaced to a
final image with a pixel size of $\approx
0\farcs06\times\sim$12(22)\AA{} for the G102(G141) grisms.  Before
sky-subtraction and interlacing each individual exposure was checked
and corrected for elevated backgrounds due to the He Earth-glow
described by \citet{Brammer:2014p34990}. The pipeline is re-run in a
way that does not flag bad reads as cosmic rays, as discussed by
Brammer (2015, in preparation), using the script available here
\url{https://github.com/gbrammer/wfc3/blob/master/reprocess_wfc3.py}. The final interlaced and sky-subtracted mosaics of
the G102 and G141 grism for the two GLASS position angles are shown in
the 4 bottom panels of Figure~\ref{fig:image}.

From these final mosaics, the spectra of each individual object are
extracted by predicting the position and extent of each
two-dimensional spectrum based on the \verb+SExtractor+
\citep{Bertin:1996p12964} segmentation of the corresponding direct
images.  As this is done for every single object, the contamination,
i.e., the dispersed light from neighboring objects in the direct image
field-of-view, can be estimated and accounted for. We note that a
complete description of the 3D-HST image preparation pipeline,
spectral extractions, and spectral fitting, will be provided by
Momcheva et al. (2015, in prep)

\section{Data quality, visual inspection, and construction of the redshift catalog}
\label{sec:visual}

Given the richness and complexity of the data set, we find that visual
inspection of a magnitude limited subset of the data is a very useful
addition to the automatic processing.  It was decided to carry out
visual inspection of the redshift determinations as well, in order to
flag catastrophic failures in the automated redshift
determination. The automated redshift determination and subsequent
visual inspection are described in Section~\ref{ssec:redshifts}. In
addition to the complete magnitude limited catalogs we also carried
out a visual search for isolated emission lines through the entire
dataset. This search yielded additional redshifts and is described in
Section~\ref{ssec:blind}.

\subsection{Visual inspection strategy}
\label{ssec:visual}

Visual inspection carried out with the publicly available Graphic User
Interface GIG (GLASS Inspection GUI) described in
Appendix~\ref{sec:GiG}.  GIG provides a convenient and efficient way
to browse the dataset. The inspection is aimed at assessing the
quality of the data and it is therefore independent of any additional
analysis like photometric redshifts.

\begin{figure}[]
\centerline{
\includegraphics[width=0.49\textwidth]{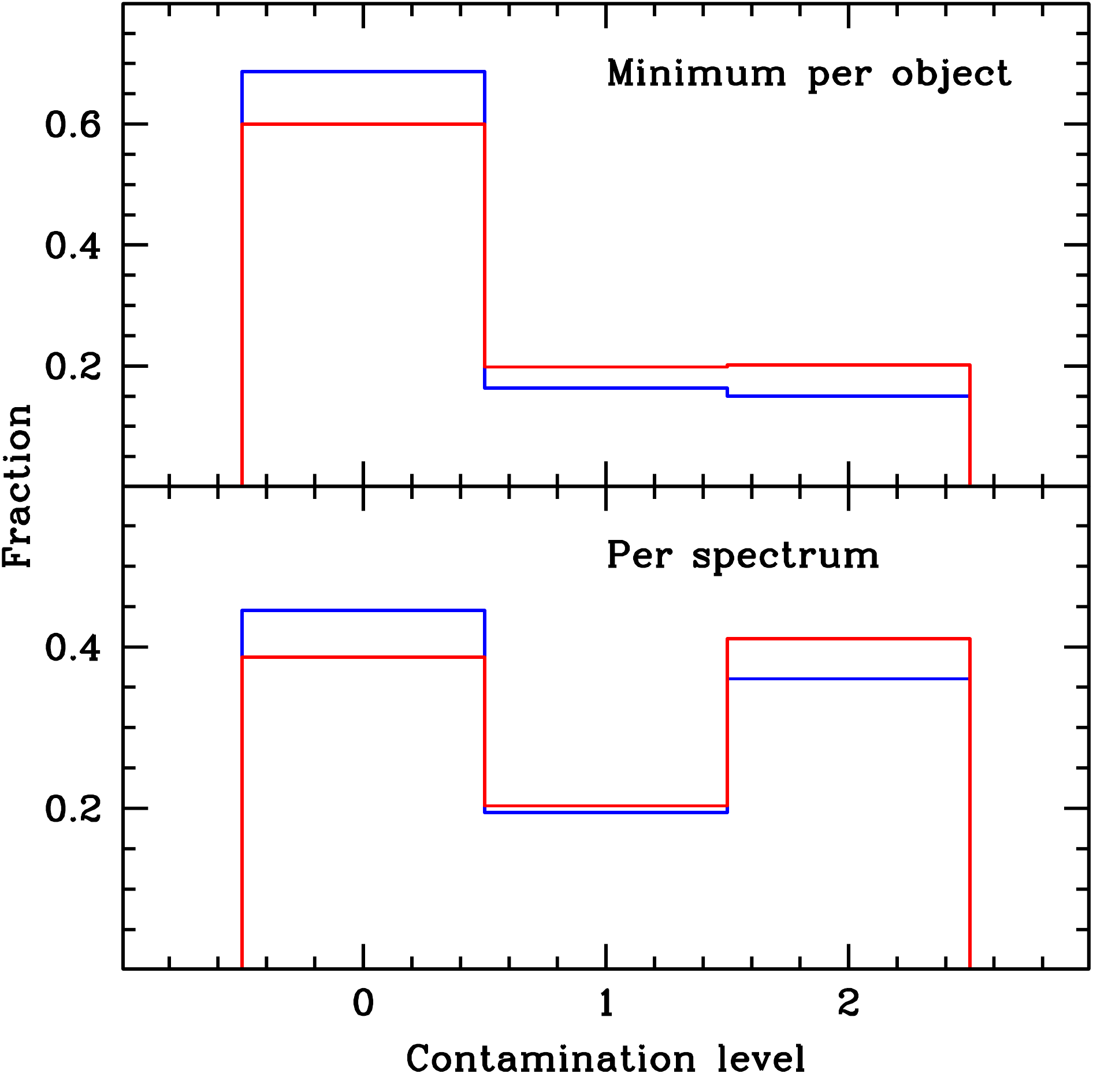}}
\caption{Contamination statistics for the \M\ catalog. The distribution of contamination levels (0=mild, 1=moderate, 2=severe) assigned by one classifier is shown by object and by spectrum, for the G102 (blue) and G141 (red) grisms. Approximately 40\% of the spectra suffer from mild contamination and are therefore ``clean''. However, thanks to the two position angle observing strategy more than 60\% of the objects have at least one clean spectrum.
\label{fig:histoCont}}
\end{figure}

The main goals of the GLASS team visual inspection were the following:

\begin{enumerate}
\item Identify and flag catastrophic failures of the pipeline. Even though these are rare in numbers, they can create substantial problems in subsequent analysis if they are not properly flagged. Examples include systematic errors in the reconstruction of the contamination spectra and artifacts due to edge effects.
\item Assess the degree of contamination in the spectra. The degree of contamination of a spectrum depends on a combination of factors, for example the relative surface brightness of the main object to that of the contaminants, their distance on the sky and on the presence of spectral features. Although some of these quantities can be calculated by the pipeline it is useful to provide an additional comprehensive human assessment of the degree of contamination. After significant experiments we decided to divide the spectra in three classes of contamination mild, moderate and severe. Roughly speaking, the three categories refer to spectra where the area of the detector with contamination sufficient to affect the target spectrum is considered to be respectively $<10$\%, $10-40$\% $>40$\% although the assessment takes into account all the factors listed above. As shown in Figure~\ref{fig:histoCont}, approximately 40\% of all spectra have mild contamination. However, thanks to two-PA strategy, there is at least one clean spectrum for more than 60\% of all the objects in the catalog.
\item Flag and identify strong emission lines and the presence of a continuum. During the inspection of the spectra a box is ticked to denote the detection of a continuum, and the wavelengths of identified lines are marked. This information can be used in the next step of determining redshifts, allowing one, for example, to visually inspect only the spectra where continuum or emission lines are present.
\item Note additional features. The user can also provide additional comments, or flag the object as being a star or defect etc.
\end{enumerate}

\begin{figure}[]
\centerline{
\includegraphics[width=0.49\textwidth]{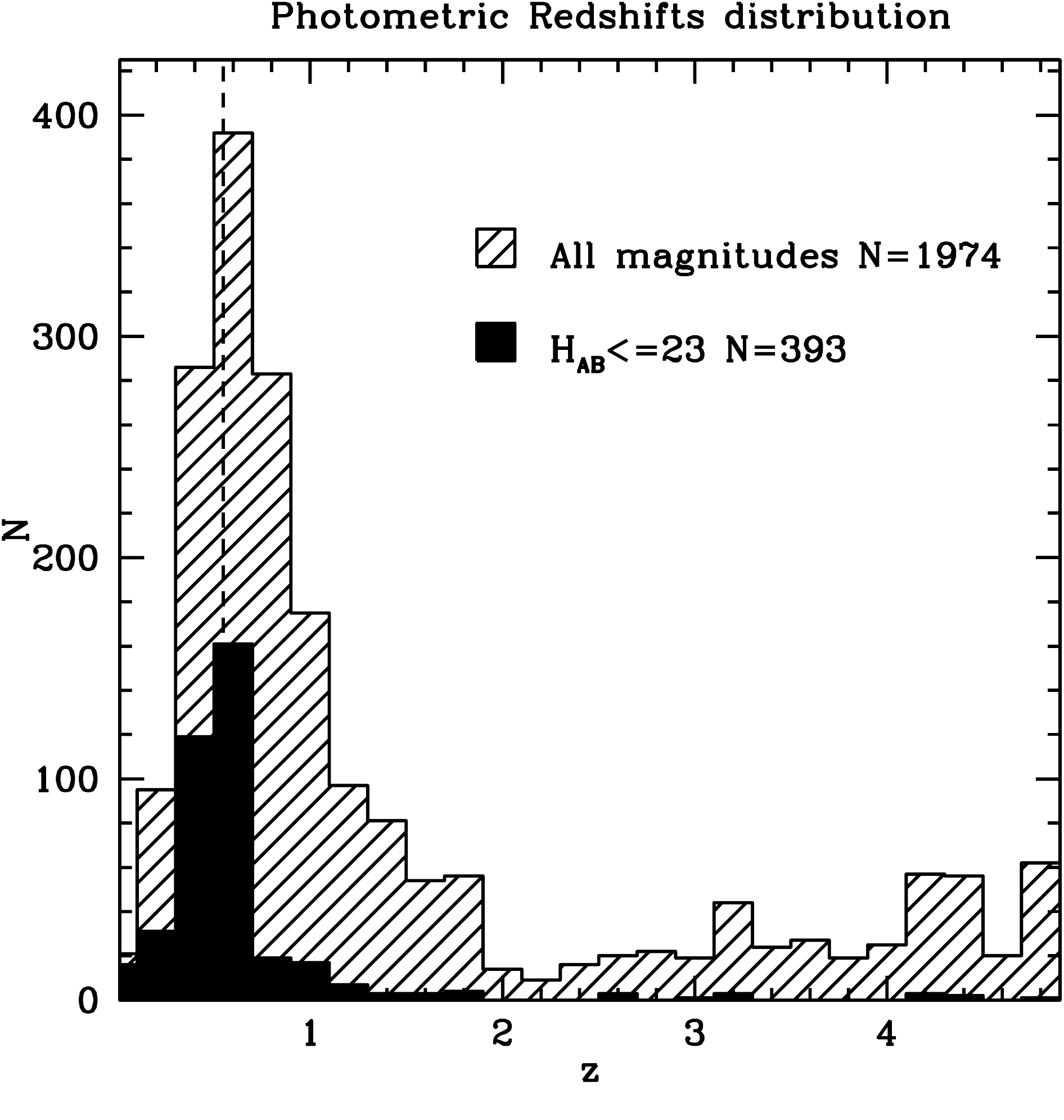}}
\caption{
\label{fig:photoz} Distribution of photometric redshifts for the all the sources in the CLASH catalog of \M. The cluster redshift ($z=0.548$) is indicated by a vertical dashed line.} 
\end{figure}

In practice, in order to provide some safeguard against the inevitable
subjectivity associated with visual inspection, each cluster catalog
is inspected by at least two team members. For \M\, co-authors
T.T. and B.V.  visually classified the entire dataset of 1151 objects
down to magnitude $H_{\rm AB}<24$ (F140W). This deep classification
was conducted on the first cluster for exploration purposes. For the
remaining clusters full visual inspection will be limited to H$_{\rm
AB}=23$, plus a search for emission lines in the fainter objects).
For reference the distribution of photometric redshifts, taken from
the CLASH catalog, of the parent sample, and the sample limited to
H$_{\rm}<23$ is shown in Figure~\ref{fig:photoz}.  The two inspection
catalogs were merged by averaging numeric flags, i.e. if one inspector
assigned mild (0) contamination and the other assigned severe (1)
contamination, the entry in the combined catalog is 0.5.

\medskip

We caution that the inspection performed by the team is of general
purpose, and even though it should provide a useful guidance for
anyone interested in the GLASS data, it may not suffice for very
specific applications. For example, a dedicated re-inspection of
photometrically selected high redshift galaxies was carried out by the
team while looking for faint \lya\ emission \citep{Schmidt+2015}.

\begin{figure}[]
\centerline{
\includegraphics[width=0.49\textwidth]{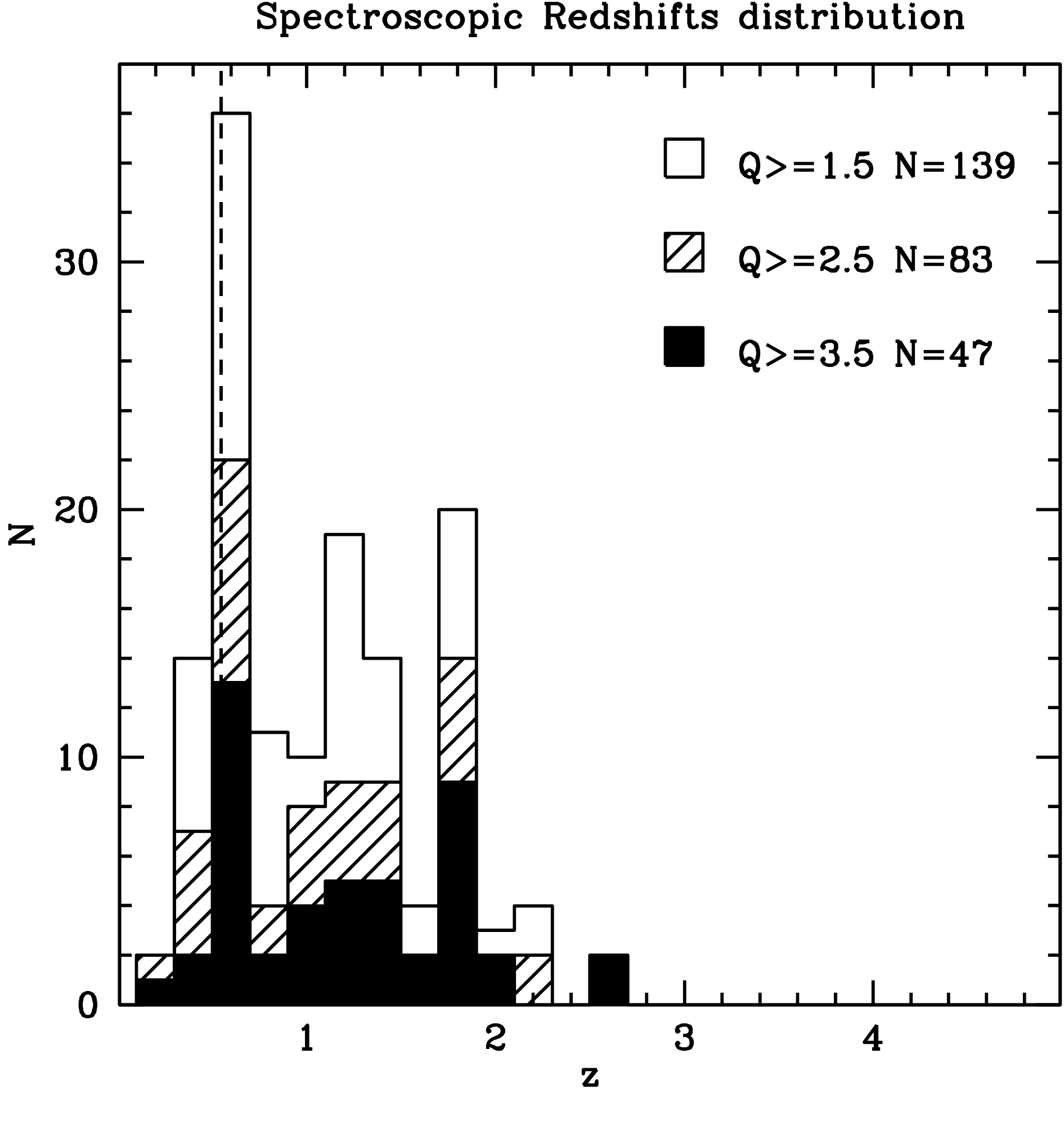}}
\caption{
\label{fig:blindz} Redshift distribution of emission line sources identified from the WFC3 GLASS spectra in the field of \M. Note the peak at the cluster redshift $z=0.548$ (vertical dashed line). Stars are not included in the histogram and tally.}
\end{figure}
\medskip

\begin{deluxetable*}{lcccccccc} \tablecolumns{7}
\tablewidth{0pt} 
\tablecaption{Redshift catalog}
\tablehead{
\colhead{ID$^\star$} & \colhead{RA}     & \colhead{DEC}     & \colhead{$z$} & \colhead{quality} & \colhead{mult} & \colhead{Notes}\\
\colhead{}  & \colhead{(J2000)} & \colhead{(J2000)} & \colhead{}  & \colhead{}        & \colhead{} & \colhead{}}
\startdata
036      &    109.39293627      &     37.77049774   &  0.0000   &  4.0   &  0.0     &    star   \\
055      &    109.39810224      &    37.770416861  &   0.8785  &   3.5  &   0.0    &    \\
057      &    109.40038631      &    37.770271353  &   0.0000  &   3.0  &   0.0    &    star   \\
145      &    109.40230016      &    37.766627582  &   1.3575  &   2.0  &   0.0    &    \\
173      &    109.39848457      &    37.766442326  &   0.5560  &   3.0  &   0.5    &    \\
234      &    109.39639585      &    37.764684116  &   0.5490  &   4.0  &   0.0    &    \\
236      &    109.39366519      &    37.764470231  &   0.3900  &   4.0  &   0.0    &    \\
272      &    109.39400607      &    37.764400472  &   0.3895  &   4.0  &   0.0    &    \\
273      &    109.40514662      &    37.764210184  &   0.0000  &   4.0  &   0.0    &    star  \\
299      &    109.39108321      &    37.763095453  &   1.9000  &   1.5  &   0.0    &    \\
307      &    109.39619713      &     37.76335404   &  2.5400   &  4.0   &  1.0     &    image 13.3  \\
\enddata
\tablecomments{\label{tab:catalog} First entries of the redshift catalog. 
The full catalog is given in its entirety in the electronic edition. 
The column ``quality'' contains the quality flag as described in the text. 
The column ``mult'' is set to one when a single line is detected and there are multiple possible redshift interpretations. 
The column ``note'' lists special comments about the object, e.g. if the object is part of a known multiply image system.
$^\star$ In the final catalog ids are formatted as either 0xxxx or 9xxxx. 
The IDs with a leading 9 refers to a reduction, which used more aggressive SExtractor de-blending and detection parameters. 
This was done to accommodate the detection of objects near bright objects, which are not assigned individual IDs when using the CLASH SExtractor parameter file (IDs with a leading 0). }
\end{deluxetable*}
\vspace{1cm}

\subsection{Redshift determination}
\label{ssec:redshifts}

Redshift determination is performed in two steps. In the first step,
templates are fit to each of the four available grism spectra
independently (G102 and G141 at two PAs each) to determine a posterior
distribution function for the redshift. If available, photometric
redshift distributions can be used as input priors to the grism fits
in order to reduce computational time. Quite frequently, the
information content of the four exposures is rather uneven, for
example because some of the spectra are strongly contaminated, or
maybe because emission lines fall in only one of the two grisms (G102
or G141) and no continuum is present in the other. Rather than
combining blindly the four pdfs the team decided to undergo a step of
visual inspection using the dedicated publicly available GLASS
inspection GUI for redshifts (GIGz; described in
Appendix~\ref{sec:GiG}).  With the help of GIGz the user can flag
which grism fits are reliable or alternatively enter a redshift by
hand if the redshift is misidentified by the automatic procedure. As
described in the appendix GIGz is fully interactive. This is carried
out for all galaxies down to a magnitude limit of H$<23$, as well for
additional samples as described in the next two sections.

Using GIGz we assigned a quality flag to the redshift according to the
following scheme (4=secure; 3=probable; 2=possible; 1=tentative, but
likely an artifact; 0=no-z).  These quality criteria take into account
the signal to noise ratio of the detection, the possibility that the
line is a contaminant, and the identification of the feature with a
specific emission line. For example, Q=4 indicates multiple emission
have been detected with high signal-to-noise ratio and thus there is
no doubt of the redshift identification; Q=3 indicates that either a
single strong emission line is robustly detected and the redshift
identification is supported by the photometric redshift, or that more
than one feature is marginally detected; Q=2 indicates a single line
detection of marginal quality; Q=1 indicates that there is someting
but it is most likely an artifact. We also ticked boxes to indicate
secure identification of some of the more common and stronger
lines. In some cases there is ambiguity about the identification of a
single emission lines. Those instances are marked during the
inspection process. This procedure is carried out independently by
each inspector and then their outputs are combined.

\subsection{Search for emission lines in faint sources}
\label{ssec:blind}

In addition to the systematic visual inspection of the magnitude
limited catalog, a search for emission lines was carried out by three
of the authors (T.T., K.B.S., B.V.), based on visual inspection of all
the jpegs of the two dimensional spectra of objects with continuum
fainter than the $H<23$ limit. For each object, the authors inspected
the spectra, the contamination model, and the contamination subtracted
model. The two PAs and stacked spectra were simultaneously inspected
together with postage stamp images of the system to allow for quick
identification of spurious contaminants, like zero order spectra from
other sources. Candidate emission lines were flagged for further
inspection using GiGz, as described above. With the aim of providing a
robust redshift catalog, as opposed to a complete redshift catalog, we
took the conservative approach of considering as real lines only those
identified by both human classifiers. This subjective and quick
procedure is sufficiently fast that allows a single investigator to
look at all the over 20,000 spectra obtained by the survey and at the
same time provides a first catalog of faint emission line objects. A
more systematic search for emission lines using machine based methods
(see, e.g., Maseda et al. 2015, in preparation) is left for future
work.

\subsection{Search for multiple image redshifts}
\label{ssec:mulz}

Given the importance of measuring spectroscopic redshifts for multiply
imaged systems, all multiple image candidates compiled in the recent
paper by \citep{Die++14} were subject to an additional round of visual
inspection by two of the authors (T.T. and X.W.). We confirm the
redshifts of several previously known systems, including from our own
GLASS survey (paper 0) and those measured by other teams
\citep{Ma++08,Limousin:2012p33689,EMB14}, and measure three new
redshifts. One of them (image 5.1, GLASS ID=1108, $z=0.928$) is
classified as probable and differs substantially from the value
$z=4.3$ assumed by \citet{Die++14}. Two of the redshifts are tentative
and should not be used until further confirmation (29.2, GLASS ID=378
$z=1.73$; 55.2, GLASS ID=82, $z=0.47$, which would be in the
foreground if confirmed).

\subsection{Properties of the spectroscopic sample and example spectra}
\label{sec:spec}

\begin{figure*}
  \centering
  \includegraphics[width=\textwidth]{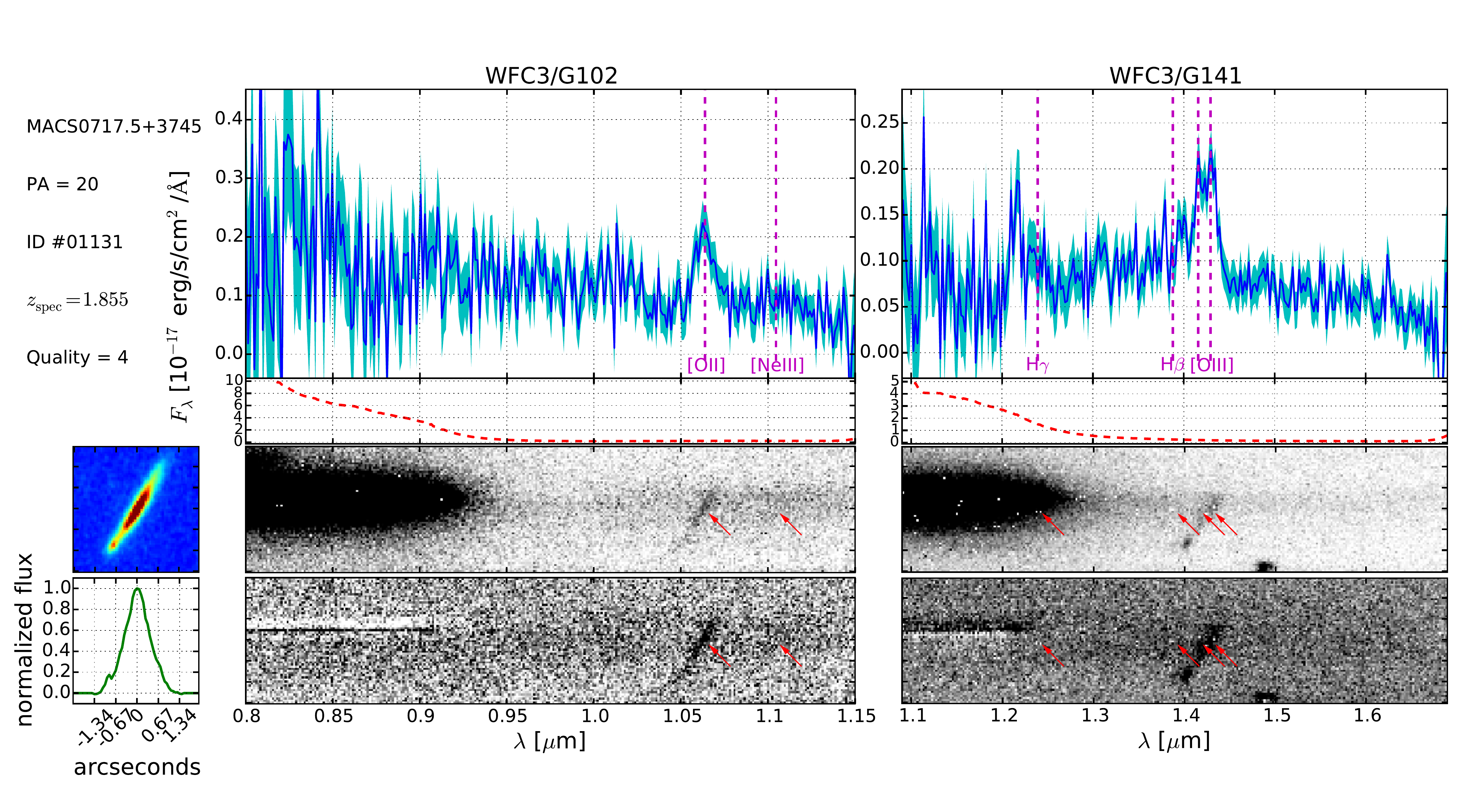}
  \includegraphics[width=\textwidth]{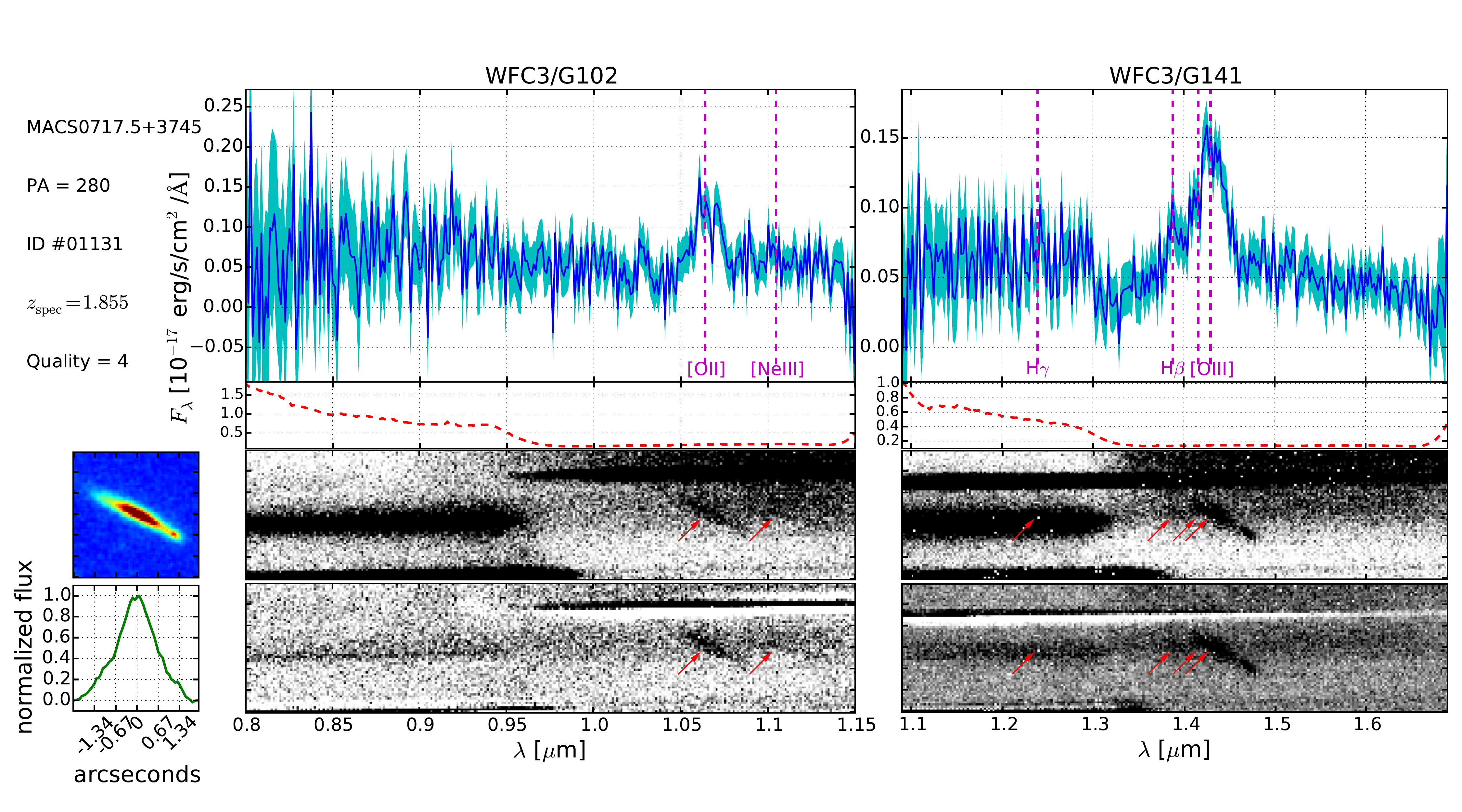}
  \caption{Spectra for GLASS ID \#1131 (Also known as arc 4.1). In
  each sub-figure, the two panels on top show the 1-dimensional
  spectra, where the observed flux and contamination model are denoted
  by blue solid and red dashed lines respectively. The cyan shaded
  region represents the noise level. For the four panels directly
  underneath, the middle two display the interlaced 2-dimensional
  spectra whereas the bottom two have contamination subtracted. In the
  1- and 2-dimensional spectra, the identified emission lines are
  denoted by vertical dashed lines in magenta and arrows in red
  respectively. The two panels on the left show the 2-dimensional
  postage stamp created from the HFF co-adds through drizzling (top)
  and the 1-dimensional collapsed image (bottom). Note that these two
  panels share the same x-axis along the grism dispersion
  direction. Some ancillary information is also shown in the upper
  left corner in each sub-figure, including the redshift quality flag
  Q.}  \label{fig:ex1}
\end{figure*}

\begin{figure*}
  \centering
  \includegraphics[width=\textwidth]{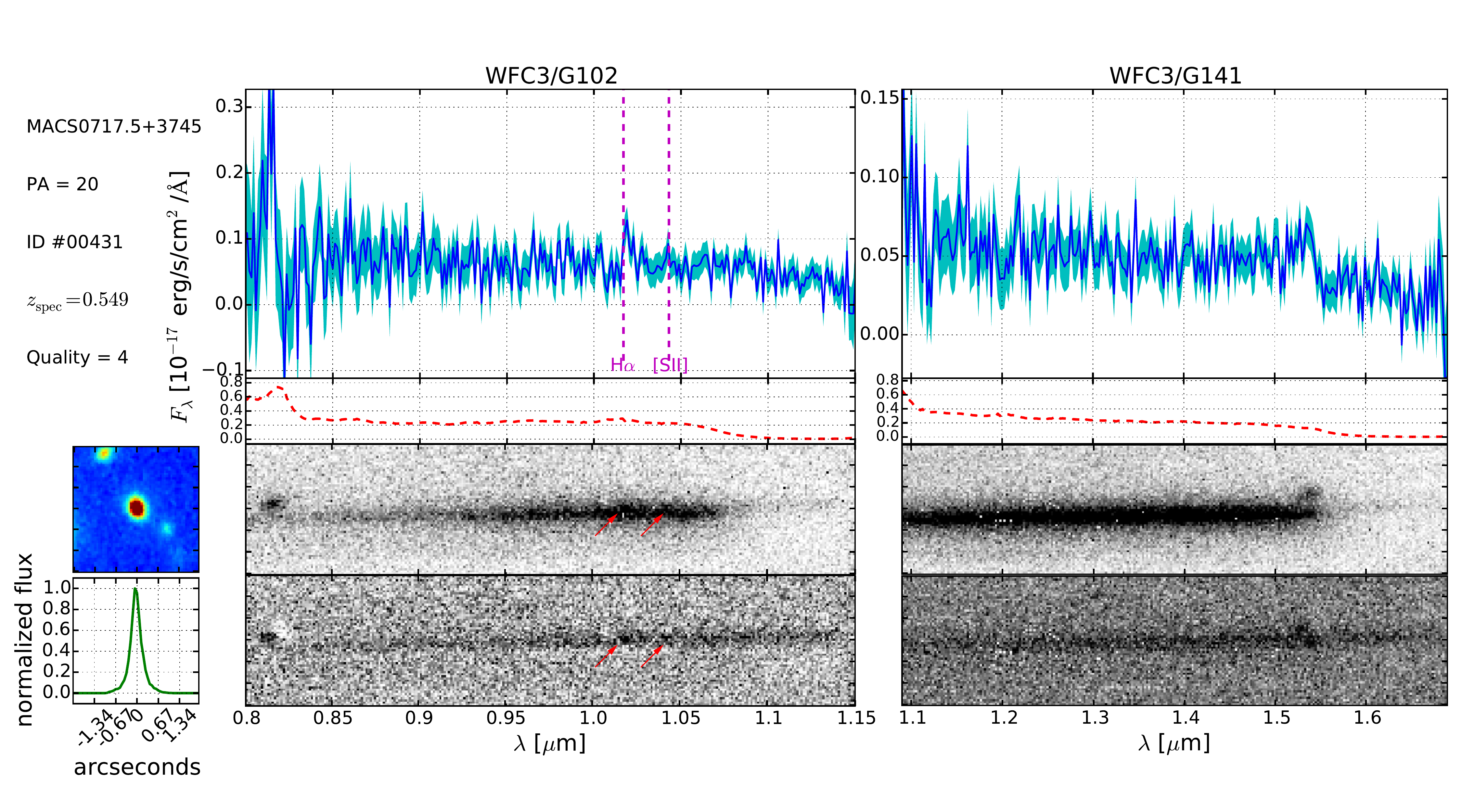}
  \includegraphics[width=\textwidth]{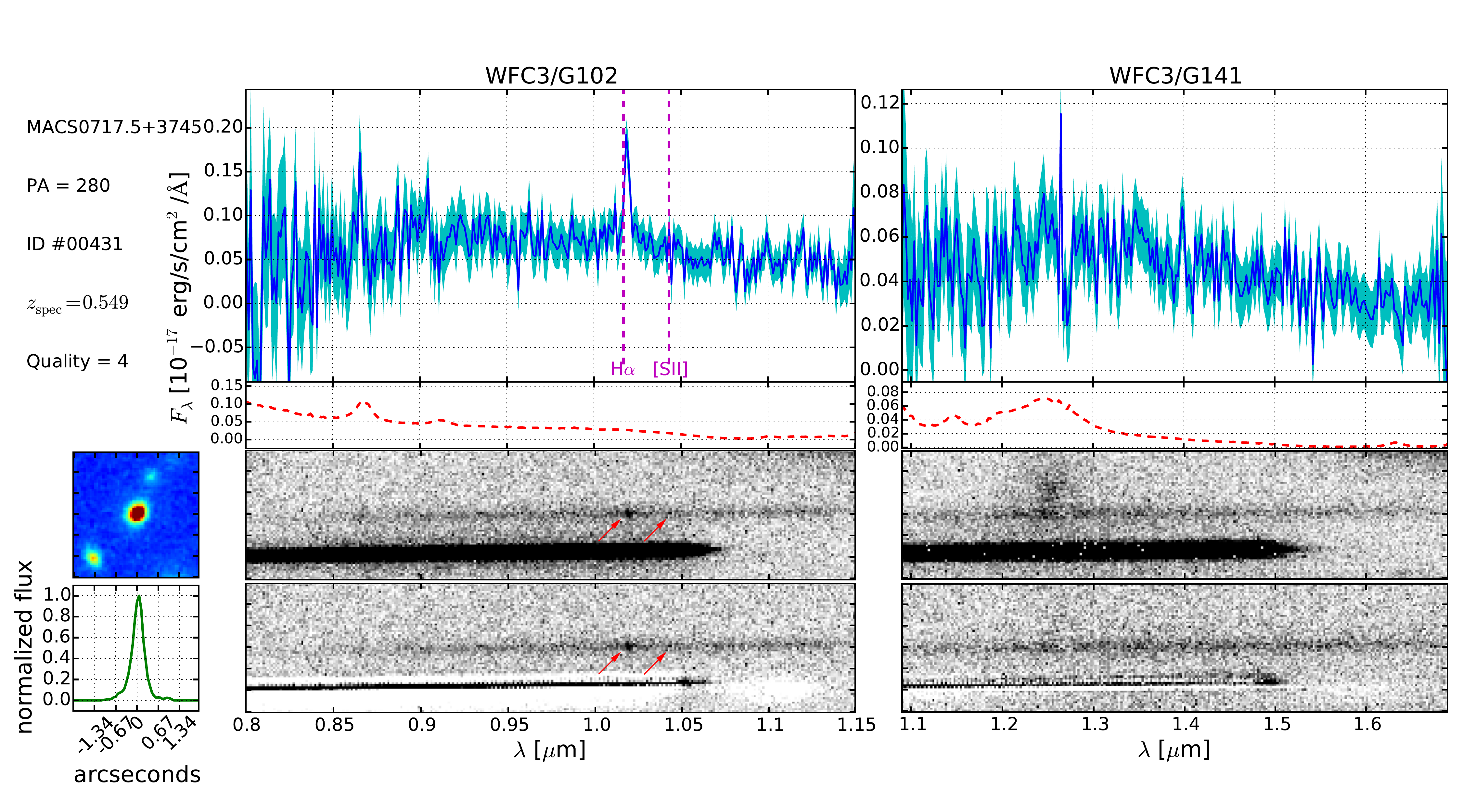}
  \caption{As Figure~\ref{fig:ex1} for cluster member GLASS ID 431.}
	\label{fig:ex2}
\end{figure*}

\begin{figure*}
  \centering
  \includegraphics[width=\textwidth]{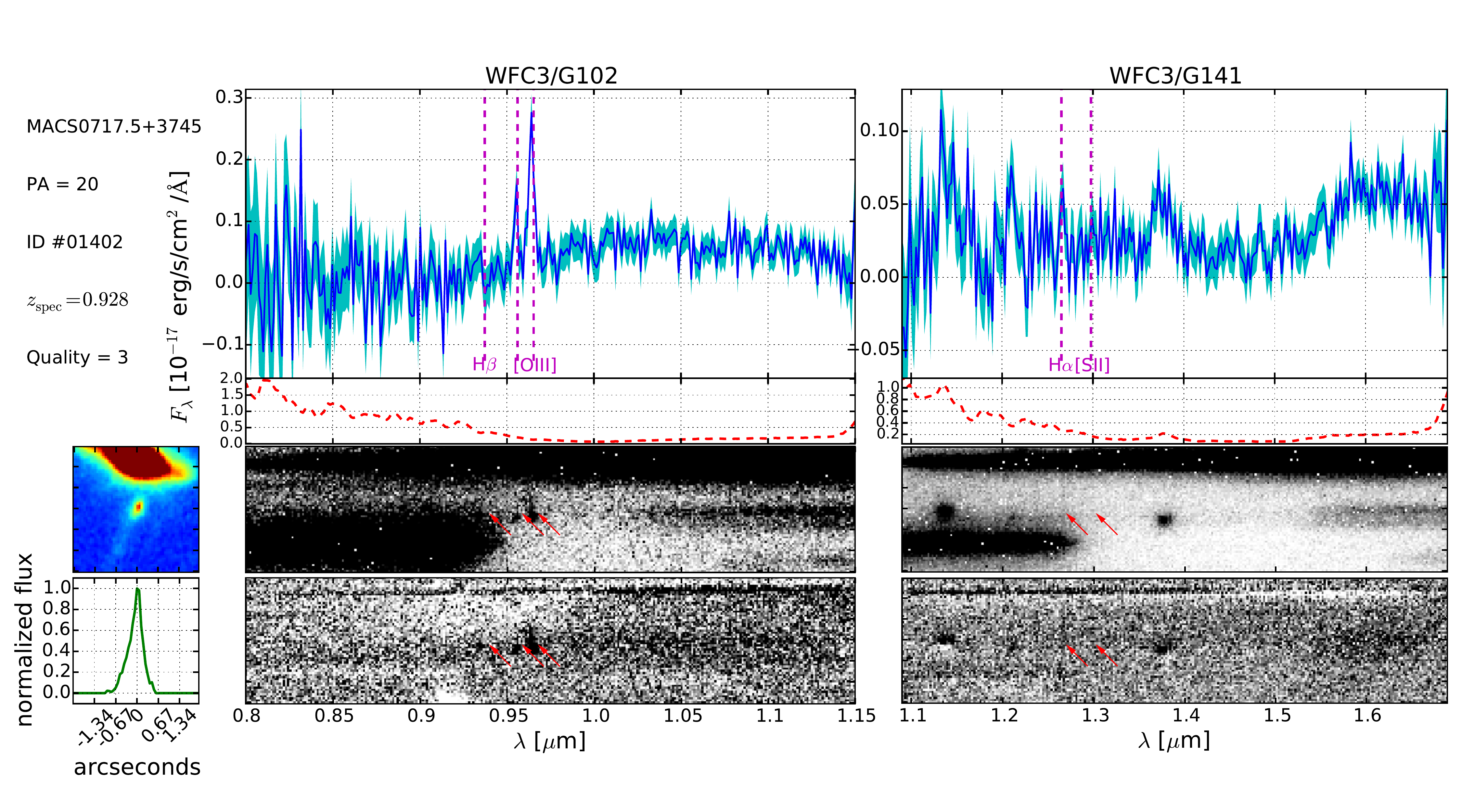}
  \includegraphics[width=\textwidth]{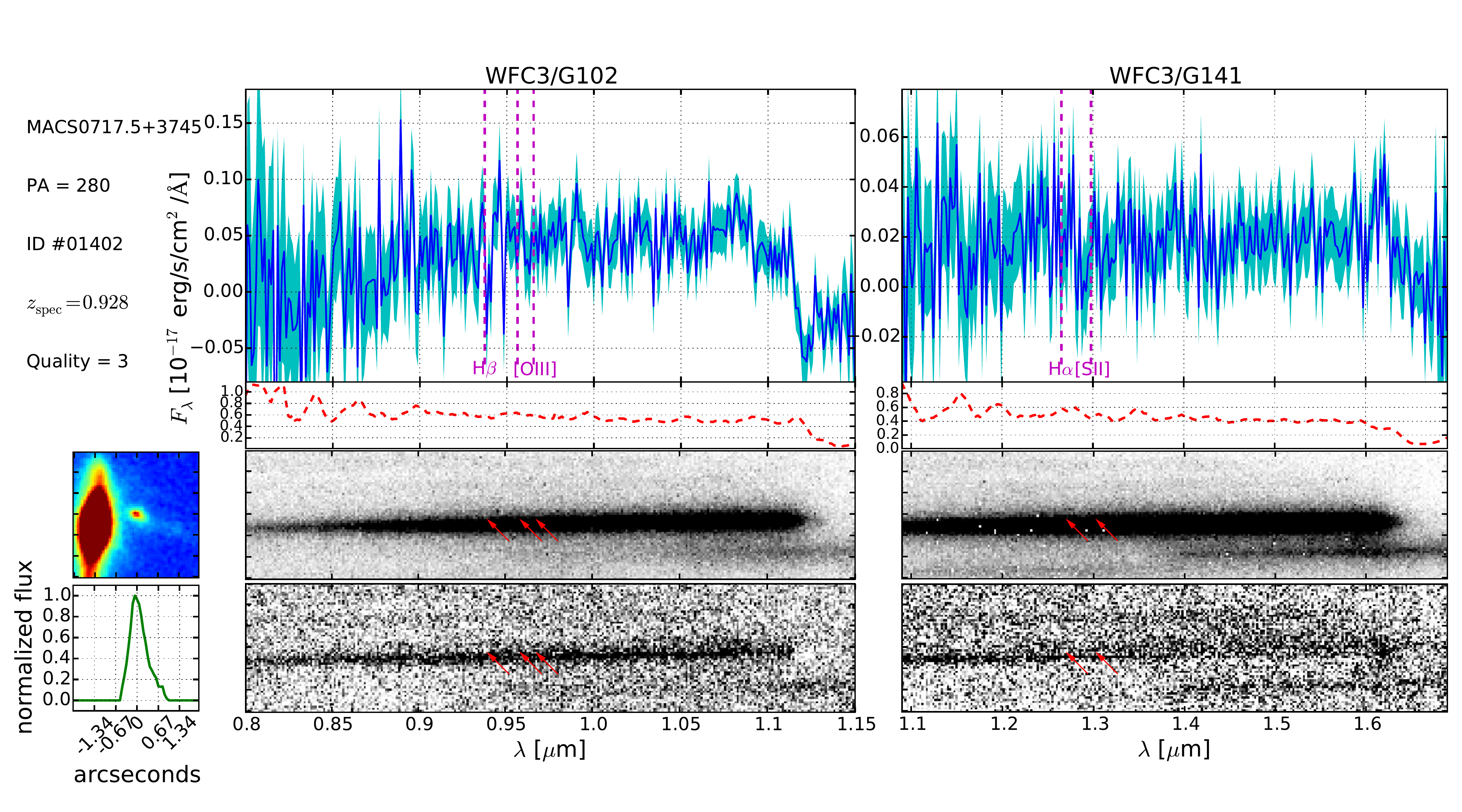}
  \caption{As Figure~\ref{fig:ex1} for GLASS ID 1402 (arc 12.1). The
  data at PA=280 suffer from strong contamination from the nearby
  galaxy.} \label{fig:ex3}
\end{figure*}

\begin{figure*}
  \centering
  \includegraphics[width=\textwidth]{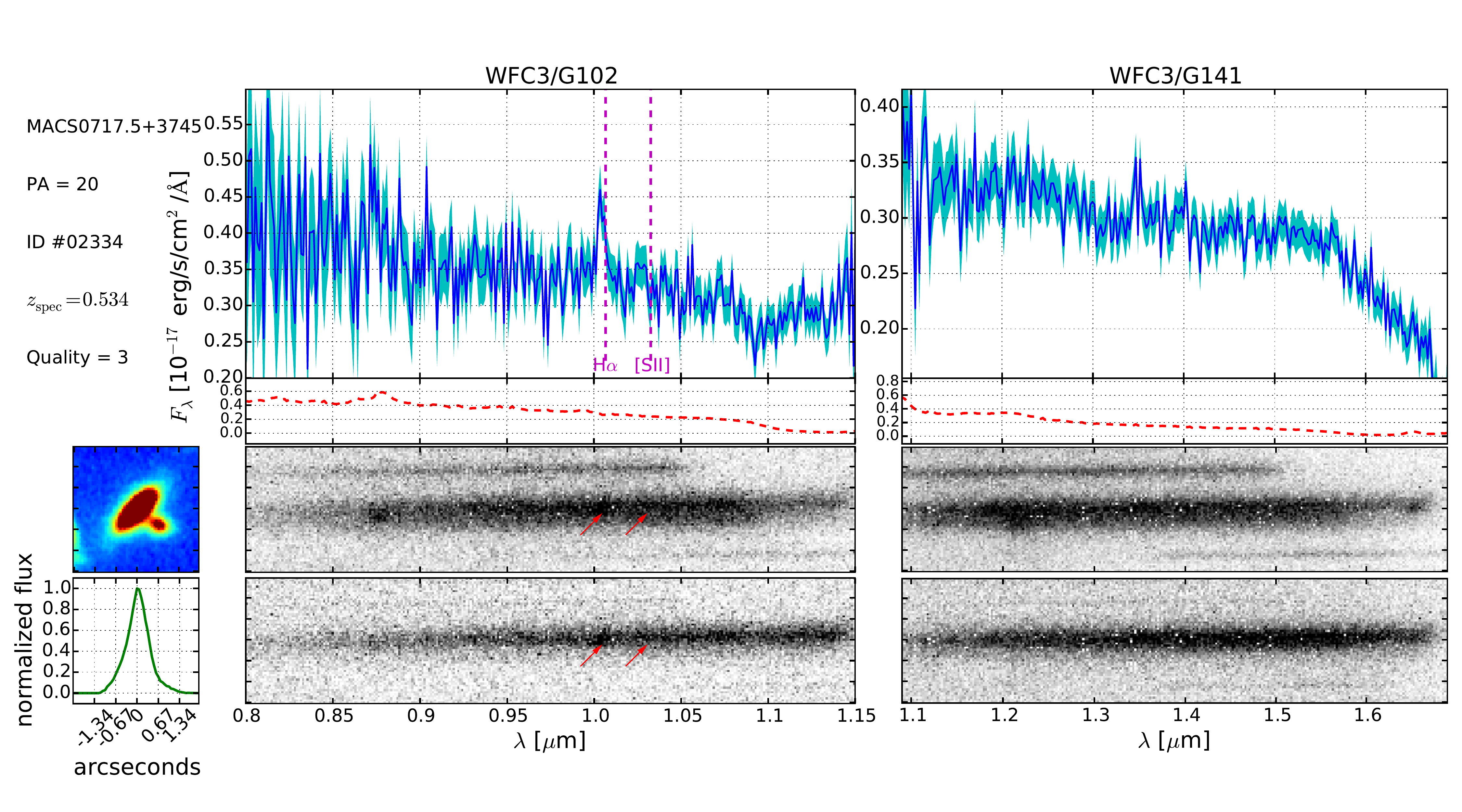}
  \includegraphics[width=\textwidth]{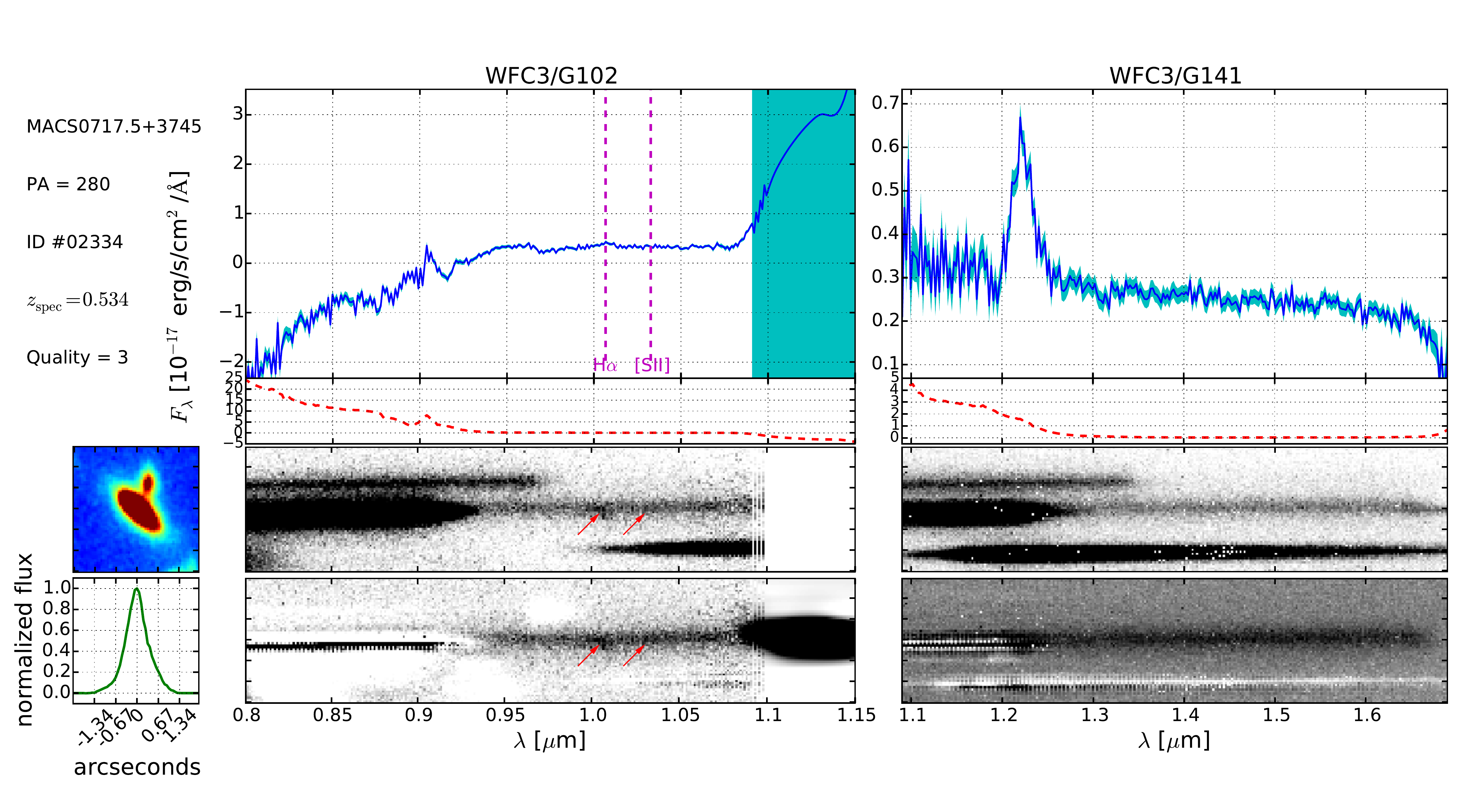}
  \caption{As Figure~\ref{fig:ex1} for cluster member GLASS ID 2334. }
  \label{fig:ex4}
\end{figure*}

\begin{figure*}
  \centering
  \includegraphics[width=\textwidth]{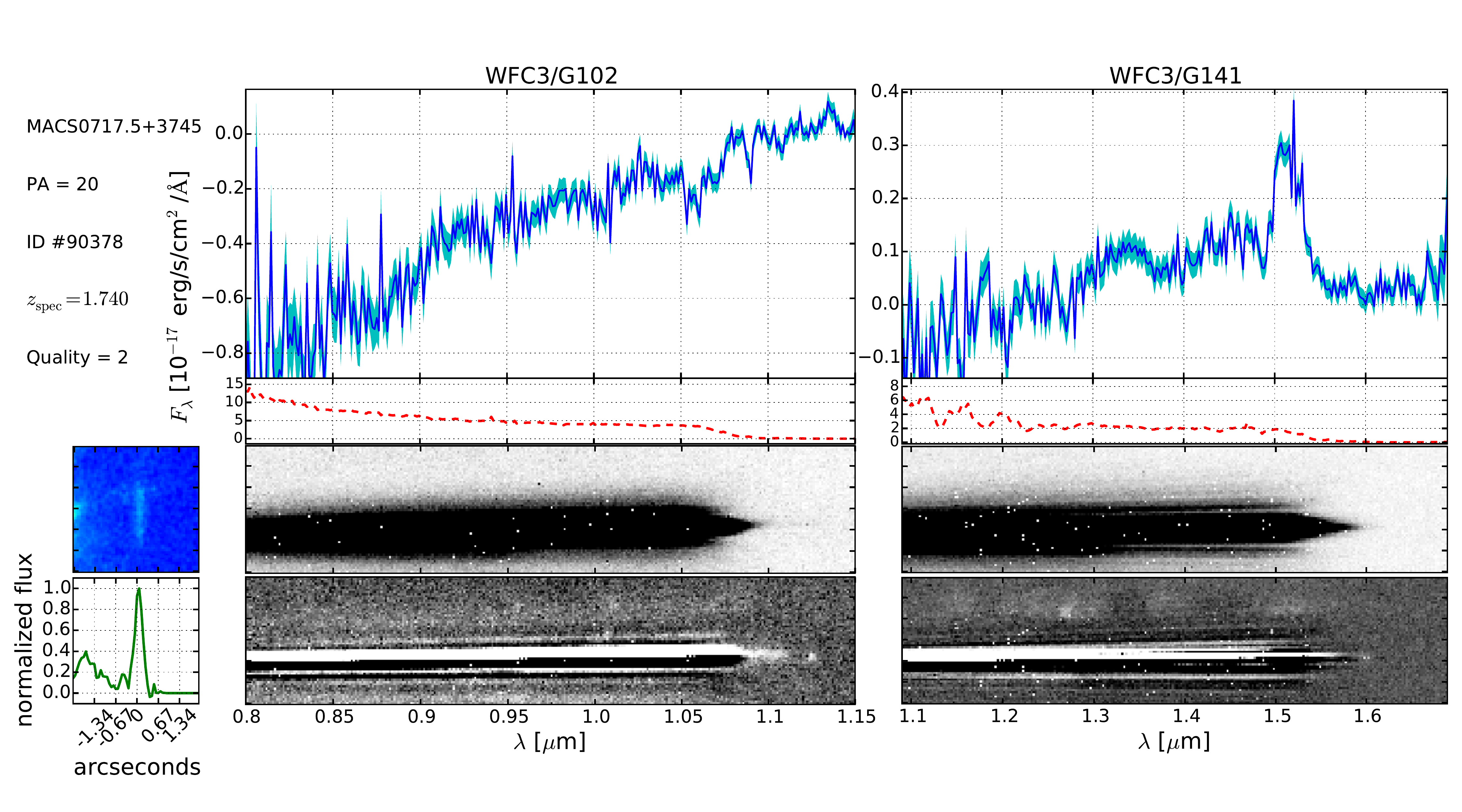}
  \includegraphics[width=\textwidth]{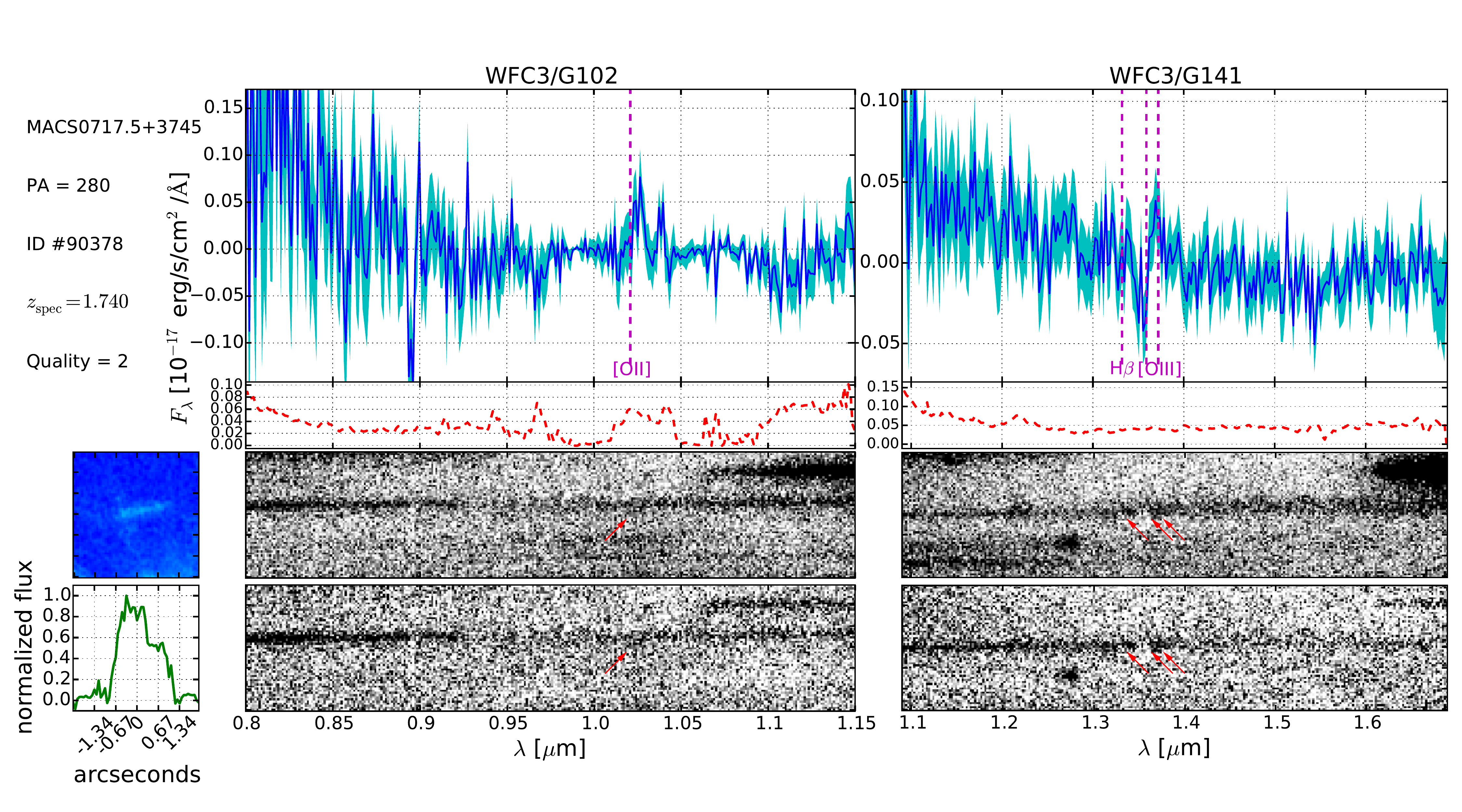}
  \caption{As Figure~\ref{fig:ex1} for GLASS ID 90378 (arc 29.2). Data
  at PA=20 suffer from strong contamination from a nearby star. The
  data at PA=280 show low signal-to-noise ratio emission lines, thus
  justifying quality flag Q=2.}  \label{fig:ex5}
\end{figure*}

\begin{figure*}
  \centering
  \includegraphics[width=\textwidth]{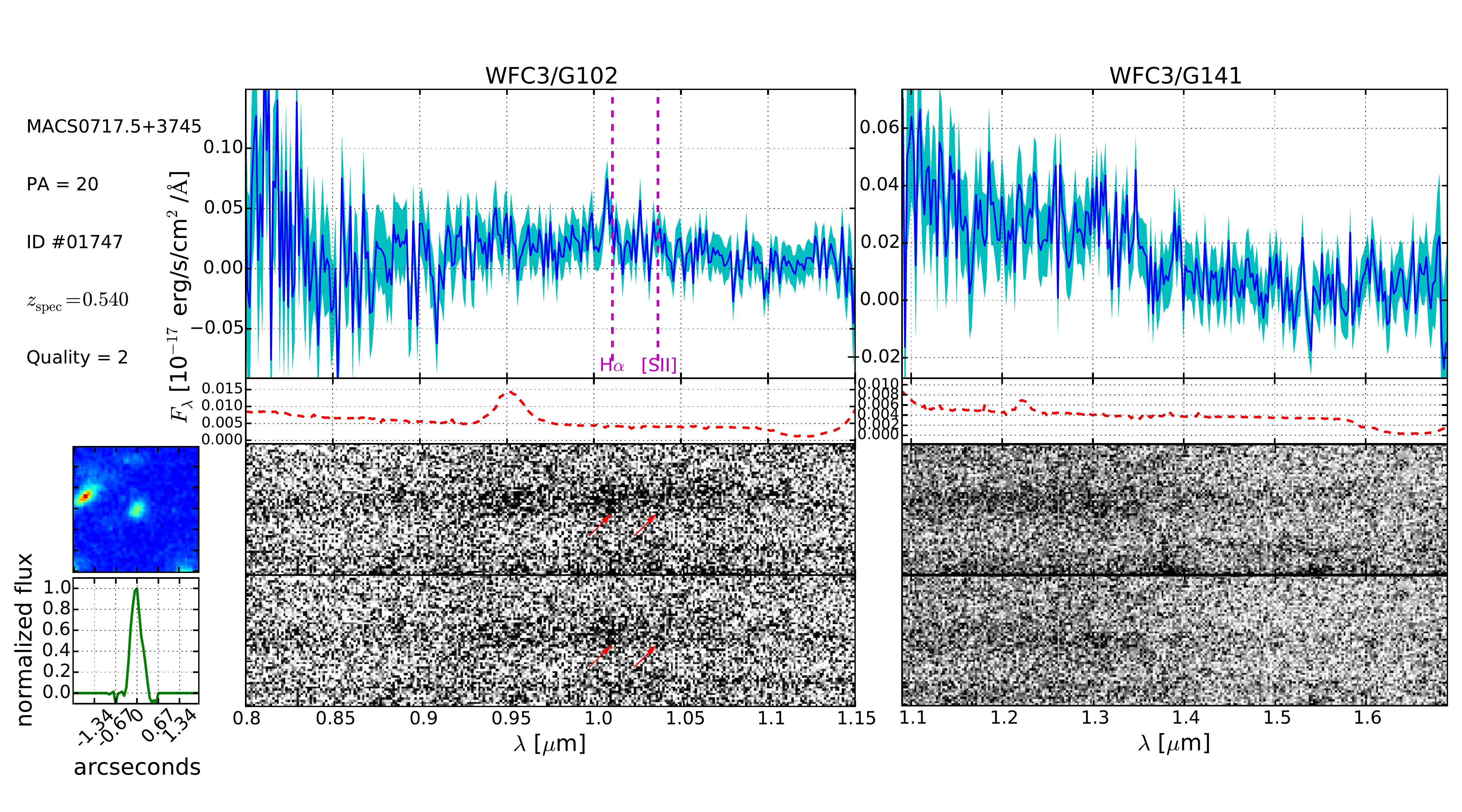}
  \includegraphics[width=\textwidth]{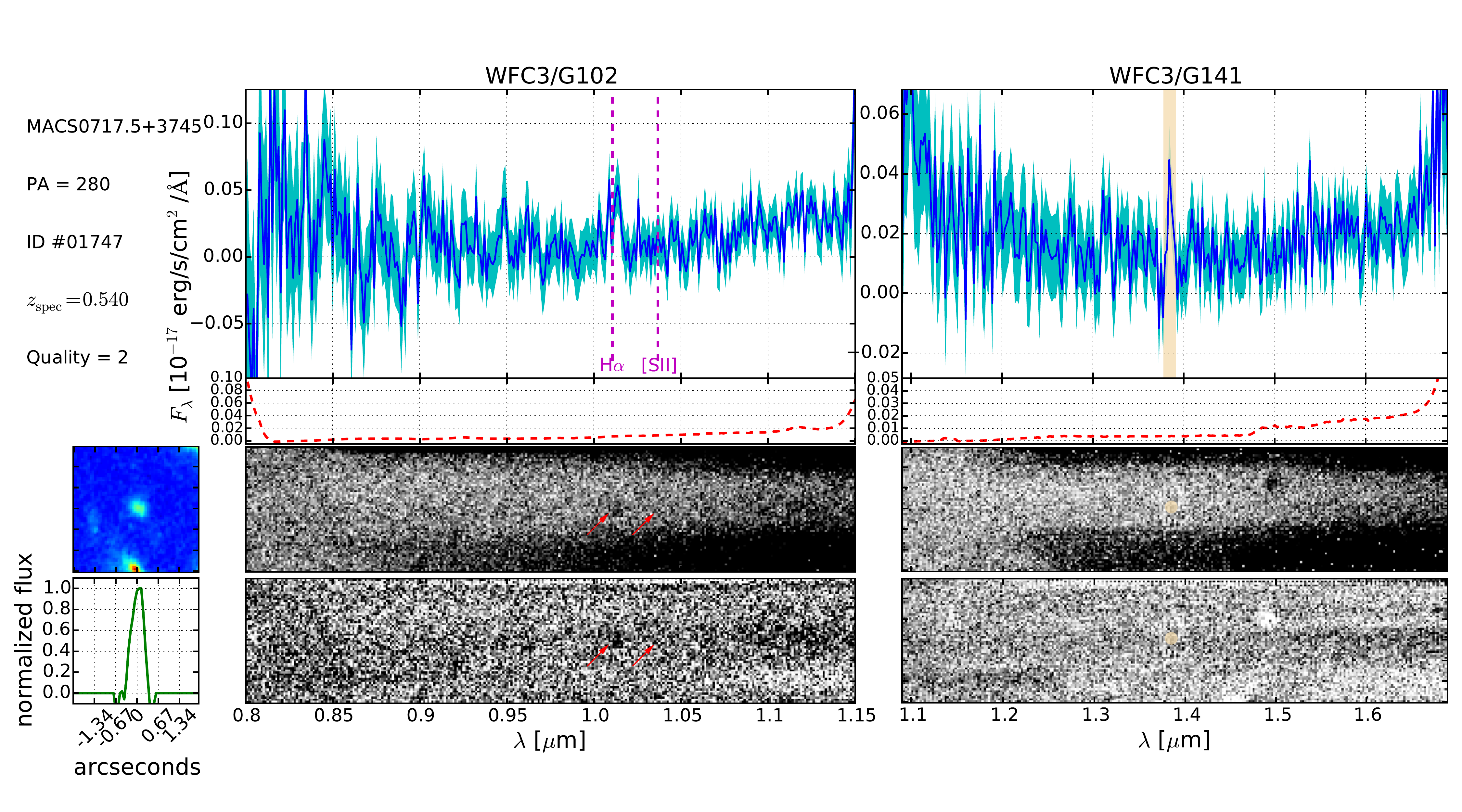}
  \caption{As Figure~\ref{fig:ex1} for cluster member GLASS ID 1747. A
  low signal to noise ratio emission line is detected at both PA, and
  tentatively identified as Balmer H$\alpha$ at the cluster reddshift.}
  \label{fig:ex6}
\end{figure*}

A histogram of redshifts based on emission lines detected in the WFC3
grism spectra is shown in Figure~\ref{fig:blindz}. Besides a clear
peak associated with the cluster redshift, the distribution covers a
broad range in redshifts. Some examples of spectra are shown in
Figures~\ref{fig:ex1} to~Fig.\ref{fig:ex6}. We show one example each
of quality 4,3,2 redshifts, including multiply image candidates when
possible. The examples are meant to illustrate the diversity of the
data, including incomplete datasets, arising from edge effects or
strong contamination.

\section{Publicly available data and tools}\label{sec:datatools}

\subsection{Release plan}

For each cluster we plan three releases:

\begin{enumerate}
\item Release 1. WFC3 images and spectra. In order to 
provide the most useful match to the photometry provided by the CLASH
collaboration for the first GLASS, the spectra are extracted based on
photometric catalogs generated by running SEXtractor \citep{B+A96} on
the GLASS IR detection image using the same parameters as those
adopted by CLASH (Coe, D. 2015, private communication; available at
the GLASS website).  All the spectra are visually inspected to
magnitude F140W$<$23AB. The structure of the data and naming of the
files is identical to that produced for 3D-HST, with the addition of
stack files that combine the two PAs into a single spectrum. More
details are available from the 3D-HST paper and at the HST archive. In
addition, a catalog is provided in electronic form. The first few
entries are shown in Table~\ref{tab:catalog} for guidance.

\item Release 2. ACS images and spectra in the parallel fields. 
The spectra are extracted based on catalogs generated from the GLASS
ACS images themselves. Visual inspection will be performed to a
magnitude limit to be defined. The data and catalog structure will be
as similar as possible to that of the WFC3 data and will be described
in a future publication.

\item Release 3. WFC3 and ACS spectra. The spectra are extracted based 
on catalogs generated from the final images of the clusters, after the
completion of the HFF campaigns. Spectra of newly identified objects
are visually inspected and redshifts are recalculated based on updated
photometric redshifts.
\end{enumerate}

Barring the unexpected, the first release is scheduled to happen
within approximately one year from the acquisition of the data (see
Table~\ref{tab:sample}; release will start after this paper is
accepted). The second release is tentatively scheduled to follow by
approximately 6-12 months. The third release will depend on
availability of other datasets and availability of funds. The publicly
released data will be available through the Mikulski Archive for Space
Telescopes (MAST) at URL
\url{https://archive.stsci.edu/prepds/glass/}.






\section{Summary}
\label{sec:summary}

This paper provides an overview of the GLASS survey. After reviewing
the scientific motivations that drove the survey design, the paper
describes the data and data processing steps.  We use \M, the first
cluster observed by GLASS (and part of the HFF campaign), to
illustrate the data set.  The main results of this paper can be
summarized as follows:

\begin{enumerate}
\item The line flux sensitivity of GLASS reaches $\approx4-5\cdot 10^{-18}$ erg s$^{-1}$ cm$^{-2}$ (1-$\sigma$) for clean
 spectra of compact sources in the most sensitive wavelength range,
 for each of the two position angles. The observing strategy,
 consisting of two orthogonal position angles, is effective in
 reducing the effects of foreground contamination. Whereas $\sim$40\%
 of the individual spectra are classified as suffering from mild
 contamination (i.e. being virtually clean), more than 60\% of the
 objects are observed as clean at at least one position angle. For
 those objects for which clean spectra are available at both position
 angles, the sensitivity reaches $\approx3\cdot 10^{-18}$ erg s$^{-1}$
 cm$^{-2}$ (1-$\sigma$).  

\item The first data release for cluster \M\ is complete, 
and high level data products made publicly available. Spectra for 1151
galaxies down to magnitude $H_{\rm AB}<24$ (F140W) have been visually
inspected by members of our team to ensure quality control.

\item A visual search for emission lines has been carried out 
through the entire dataset, including galaxies fainter than the
inspection limit. In total, we measure emission line redshifts for
\Nz\ extragalactic sources. The redshift catalog is made public in
electronic format as part of the first data release.

\item A dedicated search for redshift of candidate multiple imaged sources reveals 
three new redshifts, as well as confirming several previously known
systems.

\item In addition to the high-level data products we make available two Graphic User Interfaces (GiG and GiGz) which allow for efficient browsing of the dataset, and interactive redshift determination. The GUIs are described in the appendix to this paper.
\end{enumerate}

\acknowledgements

Support for GLASS (HST-GO-13459) was provided by NASA through a grant
from the Space Telescope Science Institute, which is operated by the
Association of Universities for Research in Astronomy, Inc., under
NASA contract NAS 5-26555. We are very grateful to the staff of the
Space Telescope for their assistance in planning, scheduling and
executing the observations, and in setting up the GLASS public release
website. We thank the referee for helpful suggestions that improved
the paper. T.T. gratefully acknowledges the hospitality of the
American Academy in Rome and of the Observatorio di Monteporzio
Catone, where parts of this manuscript were written. B.V.
acknowledges the support from the World Premier International Research
Center Initiative (WPI), MEXT, Japan and the Kakenhi Grant-in-Aid for
Young Scientists (B) (26870140) from the Japan Society for the
Promotion of Science (JSPS).


\begin{appendix}

\section{The GLASS inspection GUI}
\label{sec:GiG}

In this appendix we describe the GLASS inspection GUI (GiG) and the GLASS inspection GUI for redshifts (GiGz) v1.0, released as part of the first public data release from GLASS described in this paper. This \verb+Python+-based software provides a convenient and efficient way of inspecting and browsing the GLASS data products and is made publicly available for download at \url{https://github.com/kasperschmidt/GLASSinspectionGUIs}. A more detailed and continuously updated description of the software package can be found in the \href{https://github.com/kasperschmidt/GLASSinspectionGUIs/blob/master/README.pdf}{GiG README} which is also available at \url{https://github.com/kasperschmidt/GLASSinspectionGUIs}.

\subsection{The GLASS inspection GUI (GiG)}

\begin{figure*}
\begin{center}
\includegraphics[width=0.8\textwidth,angle=270]{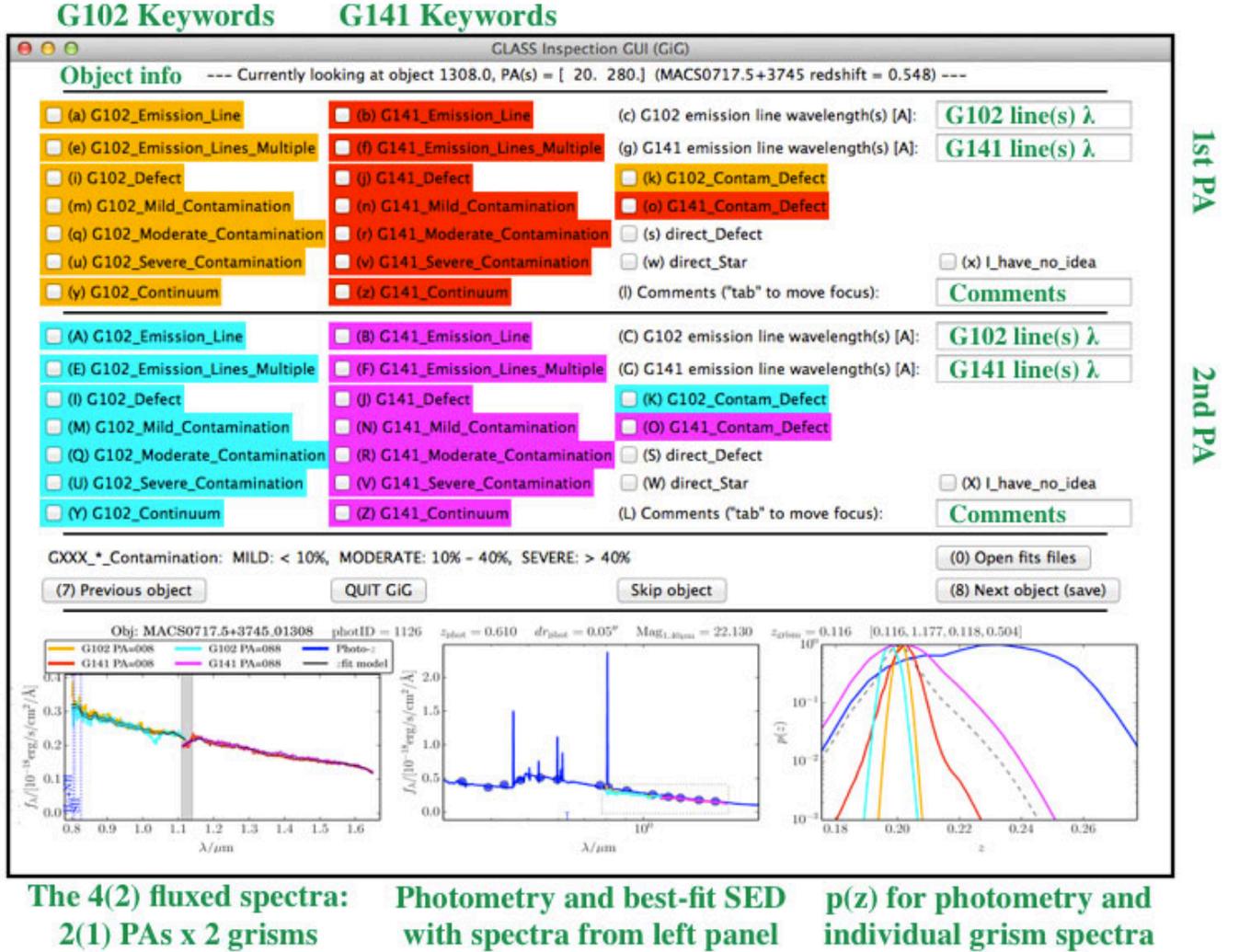}\\
\caption{Overview of the GiG interface. The green text comments on the individual parts of the GUI.
The software is publicly available for download at \url{https://github.com/kasperschmidt/GLASSinspectionGUIs}
}
\label{fig:GiG}
\end{center}
\end{figure*}

GiG is designed to inspect and browse the main products from the data
reduction pipeline described in Section~\ref{ssec:reduction}. Overview
and descriptions of the products are available from the 3D-HST papers
and the HST MAST archive.  In brief, they include the extracted
two-dimensional spectra, stacks of the spectra at the two separate
position angles in 2D, extracted one-dimensional spectra, and various
diagnostic plots.
GiG (and GiGz described below) are self-contained in the Python script \verb+visualinspection.py+ and only depend on standard publicly available python packages (see the \href{https://github.com/kasperschmidt/GLASSinspectionGUIs/blob/master/README.pdf}{GiG README} for details).
A general overview of the interface of GiG v2.2 is shown in Figure~\ref{fig:GiG}.
GiG allows the user to quickly visualize all data products, including
direct inspection of twodimensional fits files via interfacing with
ds9. In addition to visualization, GiG allows the user to rate the
contamination of the individual spectra at the two GLASS PAs, indicate
the presence of a continuum, mark defects in the spectra or
contamination models and mark emission lines including noting the
redshift at which the lines were identified.

The contamination levels are rated as mild, moderate, or severe corresponding to roughly $<$10\%, 10-40\% and $>$40 flux contamination of the central region where the spectral trace of the object displayed is expected to be.
As this rating is somewhat subjective an automatically generated estimated of the level of flux in the pixels 'belong' to the spectrum is stored in the GiG output file.
This contamination level is defined as the fraction of pixels in the fits-extension of the extracted 2D spectra containing the contamination model (masked by the object model extension) with $>10^{-3}$e/s/pixel.

Apart from the manually set keywords and the automatic contamination estimate, 
the spectral coverage of each spectrum is also estimated and stored automatically in the output file.

\subsection{The GLASS inspection GUI for redshifts (GiGz)}

\begin{figure*}
\begin{center}
\includegraphics[width=0.8\textwidth,angle=270]{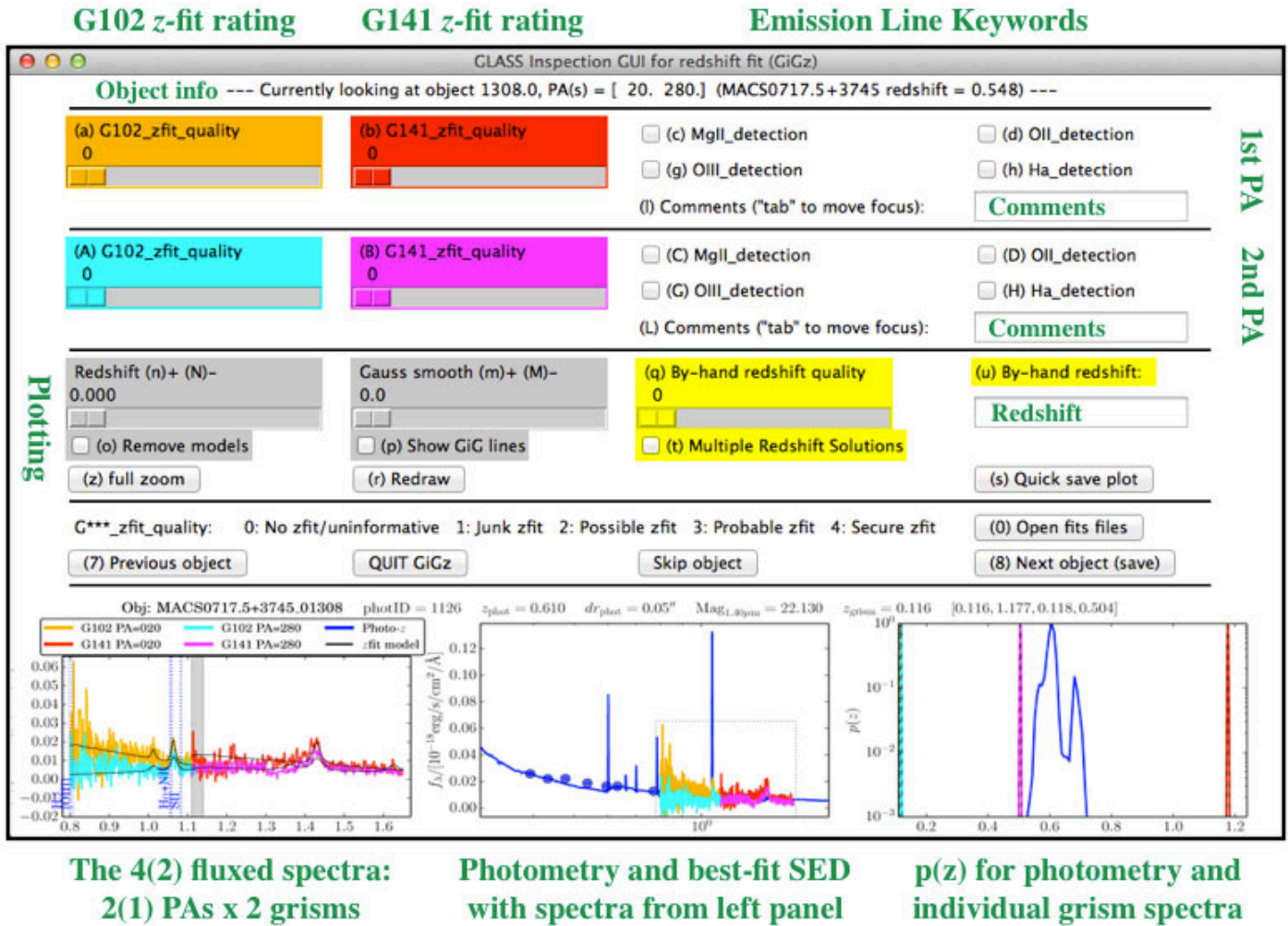}\\
\caption{Overview of the GiGz interface. The green text comments on the individual parts of the GUI.
The software is publicly available for download at \url{https://github.com/kasperschmidt/GLASSinspectionGUIs}
}
\label{fig:GiGz}
\end{center}
\end{figure*}

GiGz is developed for general inspection of the extracted 1D spectra,
the inspection of the redshift fits generated as described by
\citep{Brammer:2012p12977}, and for manual redshift fitting of any
detected emission lines.
The default GUI window shown for GiGz v1.0 in Figure~\ref{fig:GiGz}
enables quality assessment of the redshift fits and flagging of
particular lines identified for easy identification of, e.g.,
[\ion{O}{3}] emitters in the data set.  GiGz also includes an
interactive plotting interface which can be controlled from the main
window.  This enables a convenient way of plotting and inspecting the
extracted 1D spectra, the redshift models, and any emission lines
identified in the GiG inspection. Twodimensional fits files of the
spectra can also be inspected through GiG via a ds9 interface.

\end{appendix}


\begin{thebibliography}{}
\expandafter\ifx\csname natexlab\endcsname\relax\def\natexlab#1{#1}\fi

\bibitem[{{Amor{\'{\i}}n} {et~al.}(2010){Amor{\'{\i}}n}, {P{\'e}rez-Montero},
  \& {V{\'{\i}}lchez}}]{Amo++10}
{Amor{\'{\i}}n}, R.~O., {P{\'e}rez-Montero}, E., \& {V{\'{\i}}lchez}, J.~M.
  2010, \apjl, 715, L128

\bibitem[{{Andrews} \& {Martini}(2013)}]{A+M13}
{Andrews}, B.~H., \& {Martini}, P. 2013, \apj, 765, 140

\bibitem[{Atek {et~al.}(2010)Atek, Malkan, McCarthy, Teplitz, Scarlata, Siana,
  Henry, Colbert, Ross, Bridge, Bunker, Dressler, Fosbury, Martin, \&
  Shim}]{Atek:2010p33653}
Atek, H., Malkan, M., McCarthy, P., {et~al.} 2010, The Astrophysical Journal,
  723, 104

\bibitem[{{Atek} {et~al.}(2011){Atek}, {Siana}, {Scarlata}, {Malkan},
  {McCarthy}, {Teplitz}, {Henry}, {Colbert}, {Bridge}, {Bunker}, {Dressler},
  {Fosbury}, {Hathi}, {Martin}, {Ross}, \& {Shim}}]{Ate++11}
{Atek}, H., {Siana}, B., {Scarlata}, C., {et~al.} 2011, \apj, 743, 121

\bibitem[{{Baldwin} {et~al.}(1981){Baldwin}, {Phillips}, \& {Terlevich}}]{BPT}
{Baldwin}, J.~A., {Phillips}, M.~M., \& {Terlevich}, R. 1981, \pasp, 93, 5

\bibitem[{{Bayliss} {et~al.}(2014){Bayliss}, {Rigby}, {Sharon}, {Wuyts},
  {Florian}, {Gladders}, {Johnson}, \& {Oguri}}]{Bay++14}
{Bayliss}, M.~B., {Rigby}, J.~R., {Sharon}, K., {et~al.} 2014, \apj, 790, 144

\bibitem[{{Becker} {et~al.}(2015){Becker}, {Bolton}, {Madau}, {Pettini},
  {Ryan-Weber}, \& {Venemans}}]{Bec++15}
{Becker}, G.~D., {Bolton}, J.~S., {Madau}, P., {et~al.} 2015, \mnras, 447, 3402

\bibitem[{{Belli} {et~al.}(2013){Belli}, {Jones}, {Ellis}, \&
  {Richard}}]{Bel++13}
{Belli}, S., {Jones}, T., {Ellis}, R.~S., \& {Richard}, J. 2013, \apj, 772, 141

\bibitem[{{Bennett} {et~al.}(2013){Bennett}, {Larson}, {Weiland}, {Jarosik},
  {Hinshaw}, {Odegard}, {Smith}, {Hill}, {Gold}, {Halpern}, {Komatsu}, {Nolta},
  {Page}, {Spergel}, {Wollack}, {Dunkley}, {Kogut}, {Limon}, {Meyer}, {Tucker},
  \& {Wright}}]{WMAP9}
{Bennett}, C.~L., {Larson}, D., {Weiland}, J.~L., {et~al.} 2013, \apjs, 208, 20

\bibitem[{Bertin \& Arnouts(1996)}]{Bertin:1996p12964}
Bertin, E., \& Arnouts, S. 1996, Astronomy and Astrophysics Supplement, 117,
  393

\bibitem[{{Bertin} \& {Arnouts}(1996)}]{B+A96}
{Bertin}, E., \& {Arnouts}, S. 1996, \aaps, 117, 393

\bibitem[{{Bolton} \& {Burles}(2003)}]{B+B03}
{Bolton}, A.~S., \& {Burles}, S. 2003, \apj, 592, 17

\bibitem[{Bouwens {et~al.}(2014)Bouwens, Illingworth, Oesch, Trenti, Labbe',
  Bradley, Carollo, van Dokkum, Gonzalez, Holwerda, Franx, Spitler, Smit, \&
  Magee}]{Bouwens:2014p34683}
Bouwens, R.~J., Illingworth, G.~D., Oesch, P.~A., {et~al.} 2014, eprint arXiv,
  1403, 4295

\bibitem[Brada{\v c} et al.(2005)]{2005A&A...437...39B} Brada{\v c}, M., Schneider, P., Lombardi, M., \& Erben, T.\ 2005, \aap, 437, 39 

\bibitem[Brada{\v c} et al.(2008)]{2008ApJ...681..187B} Brada{\v c}, M., 
Schrabback, T., Erben, T., et al.\ 2008, \apj, 681, 187 

\bibitem[{{Brada{\v c}} {et~al.}(2009){Brada{\v c}}, {Treu}, {Applegate},
  {Gonzalez}, {Clowe}, {Forman}, {Jones}, {Marshall}, {Schneider}, \&
  {Zaritsky}}]{Bra++09}
{Brada{\v c}}, M., {Treu}, T., {Applegate}, D., {et~al.} 2009, \apj, 706, 1201

\bibitem[{Brada{\v c} {et~al.}(2012)Brada{\v c}, Vanzella, Hall, Treu, Fontana,
  Gonzalez, Clowe, Zaritsky, Stiavelli, \& Cl{\'e}ment}]{Bradac:2012p28826}
Brada{\v c}, M., Vanzella, E., Hall, N., {et~al.} 2012, The Astrophysical
  Journal Letters, 755, L7

\bibitem[{Bradley {et~al.}(2012)Bradley, Trenti, Oesch, Stiavelli, Treu,
  Bouwens, Shull, Holwerda, \& Pirzkal}]{Bradley:2012p23263}
Bradley, L.~D., Trenti, M., Oesch, P.~A., {et~al.} 2012, The Astrophysical
  Journal, 760, 108

\bibitem[{Bradley {et~al.}(2013)Bradley, Zitrin, Coe, Bouwens, Postman,
  Balestra, Grillo, Monna, Rosati, Seitz, Host, Lemze, Moustakas, Moustakas,
  Shu, Zheng, Broadhurst, Carrasco, Jouvel, Koekemoer, Medezinski, Meneghetti,
  Nonino, Smit, Umetsu, Bartelmann, Benitez, Donahue, Ford, Infante,
  Jimenez-Teja, Kelson, Lahav, Maoz, Melchior, Merten, \&
  Molino}]{Bradley:2013p32053}
Bradley, L.~D., Zitrin, A., Coe, D., {et~al.} 2013, eprint arXiv, 1308, 1692

\bibitem[{{Brammer} {et~al.}(2014){Brammer}, {Kelly}, {Rodney}, {Schmidt}, \&
  {Treu}}]{BKRST14}
{Brammer}, G., {Kelly}, P., {Rodney}, S., {Schmidt}, K.~B., \& {Treu}, T. 2014,
  The Astronomer's Telegram, 5728, 1

\bibitem[{Brammer {et~al.}(2014)Brammer, Pirzkal, McCullough, \&
  MacKenty}]{Brammer:2014p34990}
Brammer, G.~B., Pirzkal, N., McCullough, P.~R., \& MacKenty, J.~W. 2014, STScI
  IRS

\bibitem[{Brammer {et~al.}(2012)Brammer, van Dokkum, Franx, Fumagalli, Patel,
  Rix, Skelton, Kriek, Nelson, Schmidt, Bezanson, Cunha, Erb, Fan, Schreiber,
  Illingworth, Labb{\'e}, Leja, Lundgren, Magee, Marchesini, McCarthy,
  Momcheva, Muzzin, Quadri, Steidel, Tal, Wake, Whitaker, \&
  Williams}]{Brammer:2012p12977}
Brammer, G.~B., van Dokkum, P.~G., Franx, M., {et~al.} 2012, The Astrophysical
  Journal Supplement, 200, 13

\bibitem[{{Bresolin} {et~al.}(2012){Bresolin}, {Kennicutt}, \&
  {Ryan-Weber}}]{BKR12}
{Bresolin}, F., {Kennicutt}, R.~C., \& {Ryan-Weber}, E. 2012, \apj, 750, 122

\bibitem[{{Butcher} \& {Oemler}(1984)}]{B+O84}
{Butcher}, H., \& {Oemler}, Jr., A. 1984, \apj, 285, 426

\bibitem[{{Choudhury} {et~al.}(2014){Choudhury}, {Puchwein}, {Haehnelt}, \&
  {Bolton}}]{Cho++14}
{Choudhury}, T.~R., {Puchwein}, E., {Haehnelt}, M.~G., \& {Bolton}, J.~S. 2014,
  ArXiv e-prints, arXiv:1412.4790

\bibitem[{{Cl{\'e}ment} {et~al.}(2012){Cl{\'e}ment}, {Cuby}, {Courbin},
  {Fontana}, {Freudling}, {Fynbo}, {Gallego}, {Hibon}, {Kneib}, {Le F{\`e}vre},
  {Lidman}, {McMahon}, {Milvang-Jensen}, {Moller}, {Moorwood}, {Nilsson},
  {Pentericci}, {Venemans}, {Villar}, \& {Willis}}]{Cle++12}
{Cl{\'e}ment}, B., {Cuby}, J.-G., {Courbin}, F., {et~al.} 2012, \aap, 538, A66

\bibitem[{{Coe} {et~al.}(2014){Coe}, {Bradley}, \& {Zitrin}}]{CBZ15}
{Coe}, D., {Bradley}, L., \& {Zitrin}, A. 2014, ArXiv 1405.0011,
  arXiv:1405.0011

\bibitem[{{Colbert} {et~al.}(2013){Colbert}, {Teplitz}, {Atek}, {Bunker},
  {Rafelski}, {Ross}, {Scarlata}, {Bedregal}, {Dominguez}, {Dressler}, {Henry},
  {Malkan}, {Martin}, {Masters}, {McCarthy}, \& {Siana}}]{Col++13}
{Colbert}, J.~W., {Teplitz}, H., {Atek}, H., {et~al.} 2013, \apj, 779, 34

\bibitem[{{Cooper} {et~al.}(2008){Cooper}, {Newman}, {Weiner}, {Yan},
  {Willmer}, {Bundy}, {Coil}, {Conselice}, {Davis}, {Faber}, {Gerke},
  {Guhathakurta}, {Koo}, \& {Noeske}}]{Coo++08}
{Cooper}, M.~C., {Newman}, J.~A., {Weiner}, B.~J., {et~al.} 2008, \mnras, 383,
  1058

\bibitem[{Council(2010)}]{NAP12951}
Council, N.~R. 2010, New Worlds, New Horizons in Astronomy and Astrophysics
  (Washington, DC: The National Academies Press)

\bibitem[{{Diego} {et~al.}(2014){Diego}, {Broadhurst}, {Zitrin}, {Lam}, {Lim},
  {Ford}, \& {Zheng}}]{Die++14}
{Diego}, J.~M., {Broadhurst}, T., {Zitrin}, A., {et~al.} 2014, ArXiv e-prints,
  arXiv:1410.7019

\bibitem[{{Dijkstra}(2014)}]{Dij14}
{Dijkstra}, M. 2014, \pasa, 31, 40

\bibitem[{{Dijkstra} {et~al.}(2011){Dijkstra}, {Mesinger}, \&
  {Wyithe}}]{Dij++11}
{Dijkstra}, M., {Mesinger}, A., \& {Wyithe}, J.~S.~B. 2011, \mnras, 414, 2139

\bibitem[{{Dressler} {et~al.}(2013){Dressler}, {Oemler}, {Poggianti},
  {Gladders}, {Abramson}, \& {Vulcani}}]{Dre++13}
{Dressler}, A., {Oemler}, Jr., A., {Poggianti}, B.~M., {et~al.} 2013, \apj,
  770, 62

\bibitem[{{Dressler} {et~al.}(1999){Dressler}, {Smail}, {Poggianti}, {Butcher},
  {Couch}, {Ellis}, \& {Oemler}}]{Dre++99}
{Dressler}, A., {Smail}, I., {Poggianti}, B.~M., {et~al.} 1999, \apjs, 122, 51

\bibitem[{{Ebeling} {et~al.}(2014){Ebeling}, {Ma}, \& {Barrett}}]{EMB14}
{Ebeling}, H., {Ma}, C.-J., \& {Barrett}, E. 2014, \apjs, 211, 21

\bibitem[{Ellis {et~al.}(2013)Ellis, McLure, Dunlop, Robertson, Ono, Schenker,
  Koekemoer, Bowler, Ouchi, Rogers, Curtis-Lake, Schneider, Charlot, Stark,
  Furlanetto, \& Cirasuolo}]{Ellis:2013p26700}
Ellis, R.~S., McLure, R.~J., Dunlop, J.~S., {et~al.} 2013, The Astrophysical
  Journal Letters, 763, L7

\bibitem[{Ferguson {et~al.}(2000)Ferguson, Dickinson, \&
  Williams}]{Ferguson:2000p22537}
Ferguson, H.~C., Dickinson, M., \& Williams, R. 2000, Annu. Rev. Astro.
  Astrophys., 38, 667

\bibitem[{{Finlator} \& {Dav{\'e}}(2008)}]{F+D08}
{Finlator}, K., \& {Dav{\'e}}, R. 2008, \mnras, 385, 2181

\bibitem[{{Finn} {et~al.}(2005){Finn}, {Zaritsky}, {McCarthy}, {Poggianti},
  {Rudnick}, {Halliday}, {Milvang-Jensen}, {Pell{\'o}}, \& {Simard}}]{Fin++05}
{Finn}, R.~A., {Zaritsky}, D., {McCarthy}, Jr., D.~W., {et~al.} 2005, \apj,
  630, 206

\bibitem[{Fontana {et~al.}(2010)Fontana, Vanzella, Pentericci, Castellano,
  Giavalisco, Grazian, Boutsia, Cristiani, Dickinson, Giallongo, Maiolino,
  Moorwood, \& Santini}]{Fontana:2010p29506}
Fontana, A., Vanzella, E., Pentericci, L., {et~al.} 2010, The Astrophysical
  Journal Letters, 725, L205

\bibitem[{{Fumagalli} {et~al.}(2012){Fumagalli}, {Patel}, {Franx}, {Brammer},
  {van Dokkum}, {da Cunha}, {Kriek}, {Lundgren}, {Momcheva}, {Rix}, {Schmidt},
  {Skelton}, {Whitaker}, {Labbe}, \& {Nelson}}]{Fum++12}
{Fumagalli}, M., {Patel}, S.~G., {Franx}, M., {et~al.} 2012, \apjl, 757, L22

\bibitem[{{Giallongo} {et~al.}(2015){Giallongo}, {Grazian}, {Fiore}, {Fontana},
  {Pentericci}, {Vanzella}, {Dickinson}, {Kocevski}, {Castellano}, {Cristiani},
  {Ferguson}, {Finkelstein}, {Grogin}, {Hathi}, {Koekemoer}, {Newman}, \&
  {Salvato}}]{Gia++15}
{Giallongo}, E., {Grazian}, A., {Fiore}, F., {et~al.} 2015, ArXiv e-prints,
  arXiv:1502.02562

\bibitem[{Giavalisco {et~al.}(2004)Giavalisco, Ferguson, Koekemoer, Dickinson,
  Alexander, Bauer, Bergeron, Biagetti, Brandt, Casertano, Cesarsky,
  Chatzichristou, Conselice, Cristiani, Costa, Dahlen, de~Mello, Eisenhardt,
  Erben, Fall, Fassnacht, Fosbury, Fruchter, Gardner, Grogin, Hook,
  Hornschemeier, Idzi, Jogee, Kretchmer, Laidler, Lee, Livio, Lucas, Madau,
  Mobasher, Moustakas, Nonino, Padovani, Papovich, Park, Ravindranath, Renzini,
  Richardson, Riess, Rosati, Schirmer, Schreier, Somerville, Spinrad, Stern,
  Stiavelli, Strolger, Urry, Vandame, Williams, \&
  Wolf}]{Giavalisco:2004p28701}
Giavalisco, M., Ferguson, H.~C., Koekemoer, A.~M., {et~al.} 2004, The
  Astrophysical Journal, 600, L93

\bibitem[{{Gibson} {et~al.}(2013){Gibson}, {Pilkington}, {Brook}, {Stinson}, \&
  {Bailin}}]{Gib++13}
{Gibson}, B.~K., {Pilkington}, K., {Brook}, C.~B., {Stinson}, G.~S., \&
  {Bailin}, J. 2013, \aap, 554, A47

\bibitem[{Gonzaga {et~al.}(2012)Gonzaga, Hack, Fruchter, \&
  Mack}]{Gonzaga:2012p26307}
Gonzaga, S., Hack, W., Fruchter, A., \& Mack, J. 2012

\bibitem[{{Guaita} {et~al.}(2015){Guaita}, {Melinder}, {Hayes}, {{\"O}stlin},
  {Gonzalez}, {Micheva}, {Adamo}, {Mas-Hesse}, {Sandberg},
  {Ot{\'{\i}}-Floranes}, {Schaerer}, {Verhamme}, {Freeland}, {Orlitov{\'a}},
  {Laursen}, {Cannon}, {Duval}, {Rivera-Thorsen}, {Herenz}, {Kunth}, {Atek},
  {Puschnig}, {Gruyters}, \& {Pardy}}]{Gua+15}
{Guaita}, L., {Melinder}, J., {Hayes}, M., {et~al.} 2015, \aap, 576, A51

\bibitem[{{Henry} {et~al.}(2013){Henry}, {Scarlata}, {Dom{\'{\i}}nguez},
  {Malkan}, {Martin}, {Siana}, {Atek}, {Bedregal}, {Colbert}, {Rafelski},
  {Ross}, {Teplitz}, {Bunker}, {Dressler}, {Hathi}, {Masters}, {McCarthy}, \&
  {Straughn}}]{Hen++13}
{Henry}, A., {Scarlata}, C., {Dom{\'{\i}}nguez}, A., {et~al.} 2013, \apjl, 776,
  L27

\bibitem[{{Holz}(2001)}]{Hol00}
{Holz}, D.~E. 2001, \apjl, 556, L71

\bibitem[{{Jensen} {et~al.}(2013){Jensen}, {Laursen}, {Mellema}, {Iliev},
  {Sommer-Larsen}, \& {Shapiro}}]{Jen++13}
{Jensen}, H., {Laursen}, P., {Mellema}, G., {et~al.} 2013, \mnras, 428, 1366

\bibitem[{{Jones} {et~al.}(2010){Jones}, {Ellis}, {Jullo}, \&
  {Richard}}]{Jon++10}
{Jones}, T., {Ellis}, R., {Jullo}, E., \& {Richard}, J. 2010, \apjl, 725, L176

\bibitem[{{Jones} {et~al.}(2013{\natexlab{a}}){Jones}, {Ellis}, {Richard}, \&
  {Jullo}}]{Jon++13}
{Jones}, T., {Ellis}, R.~S., {Richard}, J., \& {Jullo}, E. 2013{\natexlab{a}},
  \apj, 765, 48

\bibitem[{{Jones} {et~al.}(2015){Jones}, {Wang}, {Schmidt}, {Treu}, {Brammer},
  {Brada{\v c}}, {Dressler}, {Henry}, {Malkan}, {Pentericci}, \&
  {Trenti}}]{Jon++15}
{Jones}, T., {Wang}, X., {Schmidt}, K.~B., {et~al.} 2015, \aj, 149, 107

\bibitem[{{Jones} {et~al.}(2013{\natexlab{b}}){Jones}, {Ellis}, {Schenker}, \&
  {Stark}}]{Jon++13a}
{Jones}, T.~A., {Ellis}, R.~S., {Schenker}, M.~A., \& {Stark}, D.~P.
  2013{\natexlab{b}}, \apj, 779, 52

\bibitem[{Kashikawa {et~al.}(2006)Kashikawa, Shimasaku, Malkan, Doi, Matsuda,
  Ouchi, Taniguchi, Ly, Nagao, Iye, Motohara, Murayama, Murozono, Nariai, Ohta,
  Okamura, Sasaki, Shioya, \& Umemura}]{Kashikawa:2006p28155}
Kashikawa, N., Shimasaku, K., Malkan, M.~A., {et~al.} 2006, The Astrophysical
  Journal, 648, 7

\bibitem[{{Kelly} {et~al.}(2015){Kelly}, {Rodney}, {Treu}, {Foley}, {Brammer},
  {Schmidt}, {Zitrin}, {Sonnenfeld}, {Strolger}, {Graur}, {Filippenko}, {Jha},
  {Riess}, {Bradac}, {Weiner}, {Scolnic}, {Malkan}, {von der Linden}, {Trenti},
  {Hjorth}, {Gavazzi}, {Fontana}, {Merten}, {McCully}, {Jones}, {Postman},
  {Dressler}, {Patel}, {Cenko}, {Graham}, \& {Tucker}}]{Kel++15}
{Kelly}, P.~L., {Rodney}, S.~A., {Treu}, T., {et~al.} 2015, Science, 347, 1123

\bibitem[{Kennicutt(1998)}]{Kennicutt:1998p17963}
Kennicutt, R.~C. 1998, Annu. Rev. Astro. Astrophys., 36, 189

\bibitem[{{Kewley} {et~al.}(2013){Kewley}, {Maier}, {Yabe}, {Ohta}, {Akiyama},
  {Dopita}, \& {Yuan}}]{Kew++13}
{Kewley}, L.~J., {Maier}, C., {Yabe}, K., {et~al.} 2013, \apjl, 774, L10

\bibitem[{{Kodama} {et~al.}(2004){Kodama}, {Balogh}, {Smail}, {Bower}, \&
  {Nakata}}]{Kod++04}
{Kodama}, T., {Balogh}, M.~L., {Smail}, I., {Bower}, R.~G., \& {Nakata}, F.
  2004, \mnras, 354, 1103

\bibitem[{Koekemoer {et~al.}(2003)Koekemoer, Fruchter, Hook, \&
  Hack}]{Koekemoer:2003p31861}
Koekemoer, A.~M., Fruchter, A.~S., Hook, R.~N., \& Hack, W. 2003, The 2002 HST
  Calibration Workshop : Hubble after the Installation of the ACS and the
  NICMOS Cooling System, 337

\bibitem[{Koekemoer {et~al.}(2011)Koekemoer, Faber, Ferguson, Grogin, Kocevski,
  Koo, Lai, Lotz, Lucas, McGrath, Ogaz, Rajan, Riess, Rodney, Strolger,
  Casertano, Castellano, Dahlen, Dickinson, Dolch, Fontana, Giavalisco,
  Grazian, Guo, Hathi, Huang, Wel, Yan, Acquaviva, Alexander, Almaini, Ashby,
  Barden, Bell, Bournaud, Brown, Caputi, Cassata, Challis, Chary, Cheung,
  Cirasuolo, Conselice, Cooray, Croton, Daddi, Dav{\'e}, de~Mello, Ravel,
  Dekel, Donley, Dunlop, Dutton, Elbaz, Fazio, Filippenko, Finkelstein, Frazer,
  Gardner, Garnavich, Gawiser, Gruetzbauch, Hartley, H{\"a}ussler, Herrington,
  Hopkins, Huang, Jha, Johnson, Kartaltepe, Khostovan, Kirshner, Lani, Lee, Li,
  Madau, McCarthy, McIntosh, McLure, Mcpartland, Mobasher, Moreira, Mortlock,
  Moustakas, Mozena, Nandra, Newman, Nielsen, Niemi, Noeske, Papovich,
  Pentericci, Pope, Primack, Ravindranath, Reddy, Renzini, Rix, Robaina,
  Rosario, Rosati, Salimbeni, Scarlata, Siana, Simard, Smidt, Snyder,
  Somerville, Spinrad, Straughn, Telford, Teplitz, Trump, Vargas, Villforth,
  Wagner, Wandro, Wechsler, Weiner, Wiklind, Wild, Wilson, Wuyts, \&
  Yun}]{Koekemoer:2011p9456}
Koekemoer, A.~M., Faber, S.~M., Ferguson, H.~C., {et~al.} 2011, The
  Astrophysical Journal Supplement, 197, 36

\bibitem[{{Koopmann} \& {Kenney}(2004)}]{K+K04}
{Koopmann}, R.~A., \& {Kenney}, J.~D.~P. 2004, \apj, 613, 866

\bibitem[{{Koyama} {et~al.}(2011){Koyama}, {Kodama}, {Nakata}, {Shimasaku}, \&
  {Okamura}}]{Koy++11}
{Koyama}, Y., {Kodama}, T., {Nakata}, F., {Shimasaku}, K., \& {Okamura}, S.
  2011, \apj, 734, 66

\bibitem[{{Koyama} {et~al.}(2013){Koyama}, {Kodama}, {Tadaki}, {Hayashi},
  {Tanaka}, {Smail}, {Tanaka}, \& {Kurk}}]{Koy++13}
{Koyama}, Y., {Kodama}, T., {Tadaki}, K.-i., {et~al.} 2013, \mnras, 428, 1551

\bibitem[{K{\"u}mmel {et~al.}(2011)K{\"u}mmel, Kuntschner, Walsh, \&
  Bushouse}]{Kummel:2011p33451}
K{\"u}mmel, M., Kuntschner, H., Walsh, J.~R., \& Bushouse, H. 2011, ST-ECF
  Instrument Science Report WFC3-2011-01, 1

\bibitem[Lamareille(2010)]{Lam10} Lamareille, F.\ 2010, \aap, 509, A53

\bibitem[{Lilly {et~al.}(1996)Lilly, Fevre, Hammer, \&
  Crampton}]{Lilly:1996p24634}
Lilly, S.~J., Fevre, O.~L., Hammer, F., \& Crampton, D. 1996, Astrophysical
  Journal Letters v.460, 460, L1

\bibitem[{Limousin {et~al.}(2012)Limousin, Ebeling, Richard, Swinbank, Smith,
  Jauzac, Rodionov, Ma, Smail, Edge, Jullo, \& Kneib}]{Limousin:2012p33689}
Limousin, M., Ebeling, H., Richard, J., {et~al.} 2012, Astronomy {\&}
  Astrophysics, 544, 71

\bibitem[{{Ma} {et~al.}(2008){Ma}, {Ebeling}, {Donovan}, \& {Barrett}}]{Ma++08}
{Ma}, C.-J., {Ebeling}, H., {Donovan}, D., \& {Barrett}, E. 2008, \apj, 684,
  160

\bibitem[{{Madau} \& {Dickinson}(2014)}]{M+D14}
{Madau}, P., \& {Dickinson}, M. 2014, \araa, 52, 415

\bibitem[{Madau {et~al.}(1996)Madau, Ferguson, Dickinson, Giavalisco, Steidel,
  \& Fruchter}]{Madau:1996p24662}
Madau, P., Ferguson, H.~C., Dickinson, M.~E., {et~al.} 1996, Monthly Notices
  RAS, 283, 1388

\bibitem[{{Madau} {et~al.}(2004){Madau}, {Rees}, {Volonteri}, {Haardt}, \&
  {Oh}}]{Mad++04}
{Madau}, P., {Rees}, M.~J., {Volonteri}, M., {Haardt}, F., \& {Oh}, S.~P. 2004,
  \apj, 604, 484

\bibitem[{{Mannucci} {et~al.}(2010){Mannucci}, {Cresci}, {Maiolino}, {Marconi},
  \& {Gnerucci}}]{Man++10}
{Mannucci}, F., {Cresci}, G., {Maiolino}, R., {Marconi}, A., \& {Gnerucci}, A.
  2010, \mnras, 408, 2115

\bibitem[{{Mantz} {et~al.}(2010){Mantz}, {Allen}, {Ebeling}, {Rapetti}, \&
  {Drlica-Wagner}}]{Mant++10}
{Mantz}, A., {Allen}, S.~W., {Ebeling}, H., {Rapetti}, D., \& {Drlica-Wagner},
  A. 2010, \mnras, 406, 1773

\bibitem[Mason et al.(2015)]{MTT15} Mason, C., Trenti, M., 
\& Treu, T.\ 2015, ApJ, submitted, arXiv:1508.01204 

\bibitem[{{Masters} {et~al.}(2014){Masters}, {McCarthy}, {Siana}, {Malkan},
  {Mobasher}, {Atek}, {Henry}, {Martin}, {Rafelski}, {Hathi}, {Scarlata},
  {Ross}, {Bunker}, {Blanc}, {Bedregal}, {Dom{\'{\i}}nguez}, {Colbert},
  {Teplitz}, \& {Dressler}}]{Mas++14}
{Masters}, D., {McCarthy}, P., {Siana}, B., {et~al.} 2014, \apj, 785, 153

\bibitem[{{McCarthy} {et~al.}(1999){McCarthy}, {Yan}, {Freudling}, {Teplitz},
  {Malumuth}, {Weymann}, {Malkan}, {Fosbury}, {Gardner}, {Storrie-Lombardi},
  {Thompson}, {Williams}, \& {Heap}}]{McC++99}
{McCarthy}, P.~J., {Yan}, L., {Freudling}, W., {et~al.} 1999, \apj, 520, 548

\bibitem[{McLean {et~al.}(2012)McLean, Steidel, Epps, Konidaris, Matthews,
  Adkins, Aliado, Brims, Canfield, Cromer, Fucik, Kulas, Mace, Magnone,
  Rodriguez, Rudie, Trainor, Wang, Weber, \& Weiss}]{McLean:2012p26812}
McLean, I.~S., Steidel, C.~C., Epps, H.~W., {et~al.} 2012, Ground-based and
  Airborne Instrumentation for Astronomy IV. Proceedings of the SPIE, 8446,
  doi:10.1117/12.924794

\bibitem[{{McQuinn} {et~al.}(2007){McQuinn}, {Hernquist}, {Zaldarriaga}, \&
  {Dutta}}]{McQ++07}
{McQuinn}, M., {Hernquist}, L., {Zaldarriaga}, M., \& {Dutta}, S. 2007, \mnras,
  381, 75

\bibitem[{{Mesinger} {et~al.}(2015){Mesinger}, {Aykutalp}, {Vanzella},
  {Pentericci}, {Ferrara}, \& {Dijkstra}}]{Mes++15}
{Mesinger}, A., {Aykutalp}, A., {Vanzella}, E., {et~al.} 2015, \mnras, 446, 566

\bibitem[{{Mesinger} {et~al.}(2012){Mesinger}, {McQuinn}, \&
  {Spergel}}]{Mes++12}
{Mesinger}, A., {McQuinn}, M., \& {Spergel}, D.~N. 2012, \mnras, 422, 1403

\bibitem[Meylan et al.(2006)]{SF06} Meylan, G., Jetzer, P., 
North, P., et al.\ 2006, Saas-Fee Advanced Course 33: Gravitational 
Lensing: Strong, Weak and Micro,  


\bibitem[Morris et al.(2015)]{Mor+15} Morris, A.~M., Kocevski, 
D.~D., Trump, J.~R., et al.\ 2015, \aj, 149, 178

\bibitem[{{Moustakas} {et~al.}(2006){Moustakas}, {Kennicutt}, \&
  {Tremonti}}]{Mou++06}
{Moustakas}, J., {Kennicutt}, Jr., R.~C., \& {Tremonti}, C.~A. 2006, \apj, 642,
  775

\bibitem[{{Muzzin} {et~al.}(2012){Muzzin}, {Wilson}, {Yee}, {Gilbank},
  {Hoekstra}, {Demarco}, {Balogh}, {van Dokkum}, {Franx}, {Ellingson}, {Hicks},
  {Nantais}, {Noble}, {Lacy}, {Lidman}, {Rettura}, {Surace}, \&
  {Webb}}]{Muz++12}
{Muzzin}, A., {Wilson}, G., {Yee}, H.~K.~C., {et~al.} 2012, \apj, 746, 188

\bibitem[{Nelson {et~al.}(2012)Nelson, van Dokkum, Brammer, Schreiber, Franx,
  Fumagalli, Patel, Rix, Skelton, Bezanson, Cunha, Kriek, Labbe, Lundgren,
  Quadri, \& Schmidt}]{Nelson:2012p12947}
Nelson, E.~J., van Dokkum, P.~G., Brammer, G., {et~al.} 2012, The Astrophysical
  Journal Letters, 747, L28

\bibitem[{Newman {et~al.}(2013)Newman, Ellis, Andreon, Treu, Raichoor, \&
  Trinchieri}]{Newman:2013p33256}
Newman, A.~B., Ellis, R.~S., Andreon, S., {et~al.} 2013, eprint arXiv, 1310,
  6754, submitted to ApJ

\bibitem[{{Nordin} {et~al.}(2014){Nordin}, {Rubin}, {Richard}, {Rykoff},
  {Aldering}, {Amanullah}, {Atek}, {Barbary}, {Deustua}, {Fakhouri},
  {Fruchter}, {Goobar}, {Hook}, {Hsiao}, {Huang}, {Kneib}, {Lidman}, {Meyers},
  {Perlmutter}, {Saunders}, {Spadafora}, {Suzuki}, \& {Supernova Cosmology
  Project}}]{Nor+14}
{Nordin}, J., {Rubin}, D., {Richard}, J., {et~al.} 2014, \mnras, 440, 2742

\bibitem[{{Oemler} {et~al.}(2013){Oemler}, {Dressler}, {Gladders}, {Fritz},
  {Poggianti}, {Vulcani}, \& {Abramson}}]{Oem++13}
{Oemler}, Jr., A., {Dressler}, A., {Gladders}, M.~G., {et~al.} 2013, \apj, 770,
  63

\bibitem[{Oesch {et~al.}(2013)Oesch, Bouwens, Illingworth, Labb{\'e}, Franx,
  van Dokkum, Trenti, Stiavelli, Gonzalez, \& Magee}]{Oesch:2013p27877}
Oesch, P.~A., Bouwens, R.~J., Illingworth, G.~D., {et~al.} 2013, The
  Astrophysical Journal, 773, 75

\bibitem[{Ono {et~al.}(2012)Ono, Ouchi, Mobasher, Dickinson, Penner, Shimasaku,
  Weiner, Kartaltepe, Nakajima, Nayyeri, Stern, Kashikawa, \&
  Spinrad}]{Ono:2012p27651}
Ono, Y., Ouchi, M., Mobasher, B., {et~al.} 2012, The Astrophysical Journal,
  744, 83

\bibitem[{Ouchi {et~al.}(2009)Ouchi, Mobasher, Shimasaku, Ferguson, Fall, Ono,
  Kashikawa, Morokuma, Nakajima, Okamura, Dickinson, Giavalisco, \&
  Ohta}]{Ouchi:2009p27671}
Ouchi, M., Mobasher, B., Shimasaku, K., {et~al.} 2009, The Astrophysical
  Journal, 706, 1136

\bibitem[{{Patel} {et~al.}(2014){Patel}, {McCully}, {Jha}, {Rodney}, {Jones},
  {Graur}, {Merten}, {Zitrin}, {Riess}, {Matheson}, {Sako}, {Holoien},
  {Postman}, {Coe}, {Bartelmann}, {Balestra}, {Ben{\'{\i}}tez}, {Bouwens},
  {Bradley}, {Broadhurst}, {Cenko}, {Donahue}, {Filippenko}, {Ford},
  {Garnavich}, {Grillo}, {Infante}, {Jouvel}, {Kelson}, {Koekemoer}, {Lahav},
  {Lemze}, {Maoz}, {Medezinski}, {Melchior}, {Meneghetti}, {Molino},
  {Moustakas}, {Moustakas}, {Nonino}, {Rosati}, {Seitz}, {Strolger}, {Umetsu},
  \& {Zheng}}]{Pat+14}
{Patel}, B., {McCully}, C., {Jha}, S.~W., {et~al.} 2014, \apj, 786, 9

\bibitem[{{Peng} {et~al.}(2010){Peng}, {Lilly}, {Kova{\v c}}, {Bolzonella},
  {Pozzetti}, {Renzini}, {Zamorani}, {Ilbert}, {Knobel}, {Iovino}, {Maier},
  {Cucciati}, {Tasca}, {Carollo}, {Silverman}, {Kampczyk}, {de Ravel},
  {Sanders}, {Scoville}, {Contini}, {Mainieri}, {Scodeggio}, {Kneib}, {Le
  F{\`e}vre}, {Bardelli}, {Bongiorno}, {Caputi}, {Coppa}, {de la Torre},
  {Franzetti}, {Garilli}, {Lamareille}, {Le Borgne}, {Le Brun}, {Mignoli},
  {Perez Montero}, {Pello}, {Ricciardelli}, {Tanaka}, {Tresse}, {Vergani},
  {Welikala}, {Zucca}, {Oesch}, {Abbas}, {Barnes}, {Bordoloi}, {Bottini},
  {Cappi}, {Cassata}, {Cimatti}, {Fumana}, {Hasinger}, {Koekemoer},
  {Leauthaud}, {Maccagni}, {Marinoni}, {McCracken}, {Memeo}, {Meneux}, {Nair},
  {Porciani}, {Presotto}, \& {Scaramella}}]{Pen++10}
{Peng}, Y.-j., {Lilly}, S.~J., {Kova{\v c}}, K., {et~al.} 2010, \apj, 721, 193

\bibitem[{Pentericci {et~al.}(2011)Pentericci, Fontana, Vanzella, Castellano,
  Grazian, Dijkstra, Boutsia, Cristiani, Dickinson, Giallongo, Giavalisco,
  Maiolino, Moorwood, Paris, \& Santini}]{Pentericci:2011p27723}
Pentericci, L., Fontana, A., Vanzella, E., {et~al.} 2011, The Astrophysical
  Journal, 743, 132

\bibitem[Pentericci et al.(2014)]{2014ApJ...793..113P} Pentericci, L., 
Vanzella, E., Fontana, A., et al.\ 2014, \apj, 793, 113 

\bibitem[{{Pilkington} {et~al.}(2012){Pilkington}, {Few}, {Gibson}, {Calura},
  {Michel-Dansac}, {Thacker}, {Moll{\'a}}, {Matteucci}, {Rahimi}, {Kawata},
  {Kobayashi}, {Brook}, {Stinson}, {Couchman}, {Bailin}, \&
  {Wadsley}}]{Pil++12}
{Pilkington}, K., {Few}, C.~G., {Gibson}, B.~K., {et~al.} 2012, \aap, 540, A56

\bibitem[{{Pirzkal} {et~al.}(2013){Pirzkal}, {Rothberg}, {Ly}, {Malhotra},
  {Rhoads}, {Grogin}, {Dahlen}, {Noeske}, {Meurer}, {Walsh}, {Hathi}, {Cohen},
  {Bellini}, {Holwerda}, {Straughn}, {Mechtley}, \& {Windhorst}}]{Pir++13}
{Pirzkal}, N., {Rothberg}, B., {Ly}, C., {et~al.} 2013, \apj, 772, 48

\bibitem[{{Planck Collaboration}(2015)}]{Planck2015XIII}
{Planck Collaboration}. 2015, ArXiv e-prints, arXiv:1502.01589

\bibitem[{{Poggianti} {et~al.}(1999){Poggianti}, {Smail}, {Dressler}, {Couch},
  {Barger}, {Butcher}, {Ellis}, \& {Oemler}}]{Pog++99}
{Poggianti}, B.~M., {Smail}, I., {Dressler}, A., {et~al.} 1999, \apj, 518, 576

\bibitem[{Postman {et~al.}(2012)Postman, Coe, Ben{\'\i}tez, Bradley,
  Broadhurst, Donahue, Ford, Graur, Graves, Jouvel, Koekemoer, Lemze,
  Medezinski, Molino, Moustakas, Ogaz, Riess, Rodney, Rosati, Umetsu, Zheng,
  Zitrin, Bartelmann, Bouwens, Czakon, Golwala, Host, Infante, Jha,
  Jimenez-Teja, Kelson, Lahav, Lazkoz, Maoz, McCully, Melchior, Meneghetti,
  Merten, Moustakas, Nonino, Patel, Reg{\"o}s, Sayers, Seitz, \&
  Wel}]{Postman:2012p27556}
Postman, M., Coe, D., Ben{\'\i}tez, N., {et~al.} 2012, The Astrophysical
  Journal Supplement, 199, 25

\bibitem[{{Queyrel} {et~al.}(2012){Queyrel}, {Contini}, {Kissler-Patig},
  {Epinat}, {Amram}, {Garilli}, {Le F{\`e}vre}, {Moultaka}, {Paioro}, {Tasca},
  {Tresse}, {Vergani}, {L{\'o}pez-Sanjuan}, \& {Perez-Montero}}]{Que++12}
{Queyrel}, J., {Contini}, T., {Kissler-Patig}, M., {et~al.} 2012, \aap, 539,
  A93

\bibitem[{{Refsdal}(1964)}]{Ref64}
{Refsdal}, S. 1964, \mnras, 128, 307

\bibitem[{{Rhoads} {et~al.}(2013){Rhoads}, {Malhotra}, {Stern}, {Dickinson},
  {Pirzkal}, {Spinrad}, {Reddy}, {Hathi}, {Grogin}, {Koekemoer}, {Peth},
  {Cohen}, {Zheng}, {Budavari}, {Ferreras}, {Gardner}, {Gronwall}, {Haiman},
  {K{\"u}mmel}, {Meurer}, {Moustakas}, {Panagia}, {Pasquali}, {Sahu}, {di
  Serego Alighieri}, {Somerville}, {Straughn}, {Walsh}, {Windhorst}, {Xu}, \&
  {Yan}}]{Rho++13}
{Rhoads}, J.~E., {Malhotra}, S., {Stern}, D., {et~al.} 2013, \apj, 773, 32

\bibitem[{{Ricotti} \& {Ostriker}(2004)}]{R+O04}
{Ricotti}, M., \& {Ostriker}, J.~P. 2004, \mnras, 352, 547

\bibitem[Robertson et al.(2015)]{Rob++15} Robertson, B.~E., 
Ellis, R.~S., Furlanetto, S.~R., \& Dunlop, J.~S.\ 2015, \apjl, 802, L19 

\bibitem[Rodney et al.(2015)]{2015arXiv150506211R} Rodney, S.~A., Patel, 
B., Scolnic, D., et al.\ 2015, ApJ, in press, arXiv:1505.06211 

\bibitem[{{Rupke} {et~al.}(2010){Rupke}, {Kewley}, \& {Barnes}}]{RKB10}
{Rupke}, D.~S.~N., {Kewley}, L.~J., \& {Barnes}, J.~E. 2010, \apjl, 710, L156

\bibitem[Shapley(2011)]{Sha11} Shapley, A.~E.\ 2011, \araa, 49, 525

\bibitem[{Schenker {et~al.}(2012)Schenker, Stark, Ellis, Robertson, Dunlop,
  McLure, Kneib, \& Richard}]{Schenker:2012p34406}
Schenker, M.~A., Stark, D.~P., Ellis, R.~S., {et~al.} 2012, The Astrophysical
  Journal, 744, 179

\bibitem[{Schmidt {et~al.}(2014{\natexlab{a}})Schmidt, Treu, Trenti, Bradley,
  Kelly, Oesch, Holwerda, Shull, \& Stiavelli}]{Schmidt:2014p34189}
Schmidt, K.~B., Treu, T., Trenti, M., {et~al.} 2014{\natexlab{a}}, eprint
  arXiv, 1402, 4129

\bibitem[{Schmidt {et~al.}(2014{\natexlab{b}})Schmidt, Treu, Brammer, Brada{\v
  c}, Wang, Dijkstra, Dressler, Fontana, Gavazzi, Henry, Hoag, Jones, Kelly,
  Malkan, Mason, Pentericci, Poggianti, Stiavelli, Trenti, von~der Linden, \&
  Vulcani}]{Schmidt:2014p33661}
Schmidt, K.~B., Treu, T., Brammer, G.~B., {et~al.} 2014{\natexlab{b}}, The
  Astrophysical Journal Letters, 782, L36

\bibitem[{Schmidt {et~al.}(2015)}]{Schmidt+2015}
Schmidt, K.~B., Treu, T., Brada\v{c}~M., {et~al.} 2015, The
Astrophysical Journal, submitted

\bibitem[{Scoville {et~al.}(9)Scoville, Aussel, Brusa, Capak, Carollo, Elvis,
  Giavalisco, Guzzo, Hasinger, Impey, Kneib, Lefevre, Lilly, Mobasher, Renzini,
  Rich, Sanders, Schinnerer, Schminovich, Shopbell, Taniguchi, \&
  Tyson}]{Scoville:9p22538}
Scoville, N., Aussel, H., Brusa, M., {et~al.} 9, The Astrophysical Journal
  Supplement Series, 172, 1

\bibitem[{{Shapley} {et~al.}(2014){Shapley}, {Reddy}, {Kriek}, {Freeman},
  {Sanders}, {Siana}, {Coil}, {Mobasher}, {Shivaei}, {Price}, \& {de
  Groot}}]{Sha++14}
{Shapley}, A.~E., {Reddy}, N.~A., {Kriek}, M., {et~al.} 2014, ArXiv e-prints,
  arXiv:1409.7071

\bibitem[{{Shapley} {et~al.}(2015){Shapley}, {Reddy}, {Kriek}, {Freeman},
  {Sanders}, {Siana}, {Coil}, {Mobasher}, {Shivaei}, {Price}, \& {de
  Groot}}]{Sha++15}
---. 2015, \apj, 801, 88

\bibitem[{{Smith} {et~al.}(2001){Smith}, {Kneib}, {Ebeling}, {Czoske}, \&
  {Smail}}]{Smi++01b}
{Smith}, G.~P., {Kneib}, J.-P., {Ebeling}, H., {Czoske}, O., \& {Smail}, I.
  2001, \apj, 552, 493

\bibitem[{{Sobral} {et~al.}(2013){Sobral}, {Smail}, {Best}, {Geach}, {Matsuda},
  {Stott}, {Cirasuolo}, \& {Kurk}}]{Sob++13}
{Sobral}, D., {Smail}, I., {Best}, P.~N., {et~al.} 2013, \mnras, 428, 1128

\bibitem[{Stark {et~al.}(2011)Stark, Ellis, \& Ouchi}]{Stark:2011p27664}
Stark, D.~P., Ellis, R.~S., \& Ouchi, M. 2011, The Astrophysical Journal
  Letters, 728, L2

\bibitem[{{Steidel} {et~al.}(2014){Steidel}, {Rudie}, {Strom}, {Pettini},
  {Reddy}, {Shapley}, {Trainor}, {Erb}, {Turner}, {Konidaris}, {Kulas}, {Mace},
  {Matthews}, \& {McLean}}]{Ste++14}
{Steidel}, C.~C., {Rudie}, G.~C., {Strom}, A.~L., {et~al.} 2014, \apj, 795, 165

\bibitem[{Straughn {et~al.}(2011)Straughn, Kuntschner, K{\"u}mmel, Walsh,
  Cohen, Gardner, Windhorst, O'Connell, Pirzkal, Meurer, McCarthy, Hathi,
  Malhotra, Rhoads, Balick, Bond, Calzetti, Disney, Dopita, Frogel, Hall,
  Holtzman, Kimble, Mutchler, Paresce, Saha, Silk, Trauger, Walker, Whitmore,
  Young, \& Xu}]{Straughn:2011p8119}
Straughn, A.~N., Kuntschner, H., K{\"u}mmel, M., {et~al.} 2011, The
  Astronomical Journal, 141, 14

\bibitem[{{Sullivan} {et~al.}(2000){Sullivan}, {Ellis}, {Nugent}, {Smail}, \&
  {Madau}}]{Sul+00}
{Sullivan}, M., {Ellis}, R., {Nugent}, P., {Smail}, I., \& {Madau}, P. 2000,
  \mnras, 319, 549

\bibitem[{{Suyu} {et~al.}(2014){Suyu}, {Treu}, {Hilbert}, {Sonnenfeld},
  {Auger}, {Blandford}, {Collett}, {Courbin}, {Fassnacht}, {Koopmans},
  {Marshall}, {Meylan}, {Spiniello}, \& {Tewes}}]{Suy++14}
{Suyu}, S.~H., {Treu}, T., {Hilbert}, S., {et~al.} 2014, \apjl, 788, L35

\bibitem[{Tacchella {et~al.}(2013)Tacchella, Trenti, \&
  Carollo}]{Tacchella:2013p32241}
Tacchella, S., Trenti, M., \& Carollo, C.~M. 2013, The Astrophysical Journal
  Letters, 768, L37

\bibitem[{Tremonti {et~al.}(2004)Tremonti, Heckman, Kauffmann, Brinchmann,
  Charlot, White, Seibert, Peng, Schlegel, Uomoto, Fukugita, \&
  Brinkmann}]{Tremonti:2004p24667}
Tremonti, C.~A., Heckman, T.~M., Kauffmann, G., {et~al.} 2004, The
  Astrophysical Journal, 613, 898

\bibitem[{Trenti \& Stiavelli(2008)}]{Trenti:2008p32309}
Trenti, M., \& Stiavelli, M. 2008, The Astrophysical Journal, 676, 767

\bibitem[{Trenti {et~al.}(2010)Trenti, Stiavelli, Bouwens, Oesch, Shull,
  Illingworth, Bradley, \& Carollo}]{Trenti:2010p29335}
Trenti, M., Stiavelli, M., Bouwens, R.~J., {et~al.} 2010, The Astrophysical
  Journal Letters, 714, L202

\bibitem[{Trenti {et~al.}(2011)Trenti, Bradley, Stiavelli, Oesch, Treu,
  Bouwens, Shull, MacKenty, Carollo, \& Illingworth}]{Trenti:2011p12656}
Trenti, M., Bradley, L.~D., Stiavelli, M., {et~al.} 2011, The Astrophysical
  Journal Letters, 727, L39

\bibitem[{{Treu}(2010)}]{Tre10}
{Treu}, T. 2010, \araa, 48, 87

\bibitem[{{Treu} {et~al.}(2003){Treu}, {Ellis}, {Kneib}, {Dressler}, {Smail},
  {Czoske}, {Oemler}, \& {Natarajan}}]{Tre++03}
{Treu}, T., {Ellis}, R.~S., {Kneib}, J., {et~al.} 2003, \apj, 591, 53

\bibitem[{{Treu} \& {Koopmans}(2002)}]{T+K02b}
{Treu}, T., \& {Koopmans}, L.~V.~E. 2002, \mnras, 337, L6

\bibitem[{Treu {et~al.}(2013)Treu, Schmidt, Trenti, Bradley, \&
  Stiavelli}]{Treu:2013p32132}
Treu, T., Schmidt, K.~B., Trenti, M., Bradley, L.~D., \& Stiavelli, M. 2013,
  The Astrophysical Journal Letters, 775, L29

\bibitem[{Treu {et~al.}(2012)Treu, Trenti, Stiavelli, Auger, \&
  Bradley}]{Treu:2012p12658}
Treu, T., Trenti, M., Stiavelli, M., Auger, M.~W., \& Bradley, L.~D. 2012, The
  Astrophysical Journal, 747, 27

\bibitem[{{Troncoso} {et~al.}(2014){Troncoso}, {Maiolino}, {Sommariva},
  {Cresci}, {Mannucci}, {Marconi}, {Meneghetti}, {Grazian}, {Cimatti},
  {Fontana}, {Nagao}, \& {Pentericci}}]{Tro++14}
{Troncoso}, P., {Maiolino}, R., {Sommariva}, V., {et~al.} 2014, \aap, 563, A58

\bibitem[{{Trump} {et~al.}(2011){Trump}, {Weiner}, {Scarlata}, {Kocevski},
  {Bell}, {McGrath}, {Koo}, {Faber}, {Laird}, {Mozena}, {Rangel}, {Yan},
  {Yesuf}, {Atek}, {Dickinson}, {Donley}, {Dunlop}, {Ferguson}, {Finkelstein},
  {Grogin}, {Hathi}, {Juneau}, {Kartaltepe}, {Koekemoer}, {Nandra}, {Newman},
  {Rodney}, {Straughn}, \& {Teplitz}}]{Tru++11}
{Trump}, J.~R., {Weiner}, B.~J., {Scarlata}, C., {et~al.} 2011, \apj, 743, 144

\bibitem[{Trump {et~al.}(2011)Trump, Weiner, Scarlata, Kocevski, Bell, McGrath,
  Koo, Faber, Laird, Mozena, Rangel, Yan, Yesuf, Atek, Dickinson, Donley,
  Dunlop, Ferguson, Finkelstein, Grogin, Hathi, Juneau, Kartaltepe, Koekemoer,
  Nandra, Newman, Rodney, Straughn, \& Teplitz}]{Trump:2011p10256}
Trump, J.~R., Weiner, B.~J., Scarlata, C., {et~al.} 2011, The Astrophysical
  Journal, 743, 144, ApJ

\bibitem[{{Trump} {et~al.}(2013){Trump}, {Konidaris}, {Barro}, {Koo},
  {Kocevski}, {Juneau}, {Weiner}, {Faber}, {McLean}, {Yan},
  {P{\'e}rez-Gonz{\'a}lez}, \& {Villar}}]{Tru++13}
{Trump}, J.~R., {Konidaris}, N.~P., {Barro}, G., {et~al.} 2013, \apjl, 763, L6

\bibitem[{{Trump} {et~al.}(2014){Trump}, {Barro}, {Juneau}, {Weiner},
  {Luo}, {Brammer}, {Bell}, {Brandt}, {Dekel}, {Guo},
{Hopkins}, {Koo}, {Kocevski}, {MbInstosh}, {Momcheva}, {Faber},
{Ferguson}, {Grogin}, {Kartaltepe}, {Koekemoer}, {Lotz}, {Maseda},
  {Mozena}, {Nandra}, {Rosario},  \& {Zeimann}}]{Trump:2014p41000}
Trump, J.~R., Barro, G., Juneau, S., {et~al.} 2014, The Astrophysical
  Journal, 793, 101
  
\bibitem[{van Dokkum {et~al.}(2011)van Dokkum, Brammer, Fumagalli, Nelson,
  Franx, Rix, Kriek, Skelton, Patel, Schmidt, Bezanson, Bian, da~Cunha, Erb,
  Fan, Schreiber, Illingworth, Labb{\'e}, Lundgren, Magee, Marchesini,
  McCarthy, Muzzin, Quadri, Steidel, Tal, Wake, Whitaker, \&
  Williams}]{vanDokkum:2011p10254}
van Dokkum, P.~G., Brammer, G., Fumagalli, M., {et~al.} 2011, The Astrophysical
  Journal Letters, 743, L15

\bibitem[{Vanzella {et~al.}(2013)Vanzella, Fontana, Zitrin, Coe, Bradley,
  Postman, Grazian, Castellano, Pentericci, Giavalisco, Rosati, Nonino, Smit,
  Balestra, Bouwens, Cristiani, Giallongo, Zheng, Infante, Cusano, \&
  Speziali}]{Vanzella:2013p33637}
Vanzella, E., Fontana, A., Zitrin, A., {et~al.} 2013, eprint arXiv, 1312, 6299

\bibitem[{{Vila-Costas} \& {Edmunds}(1992)}]{VCE92}
{Vila-Costas}, M.~B., \& {Edmunds}, M.~G. 1992, \mnras, 259, 121

\bibitem[{{Vulcani} {et~al.}(2010){Vulcani}, {Poggianti}, {Finn}, {Rudnick},
  {Desai}, \& {Bamford}}]{Vul+10}
{Vulcani}, B., {Poggianti}, B.~M., {Finn}, R.~A., {et~al.} 2010, \apjl, 710, L1

\bibitem[{{Wang} {et~al.}(2015){Wang}, {Hoag}, {Huang}, {Treu}, {Bradac},
  {Schmidt}, {Brammer}, {Vulcani}, {Jones}, {Ryan}, {Amorin}, {Castellano},
  {Fontana}, {Merlin}, \& {Trenti}}]{Wan+15}
{Wang}, X., {Hoag}, A., {Huang}, K., {et~al.} 2015, ApJ, in press,
  arXiv:1504.02405

\bibitem[{{Wright} {et~al.}(2010){Wright}, {Larkin}, {Graham}, \&
  {Ma}}]{Wri++10}
{Wright}, S.~A., {Larkin}, J.~E., {Graham}, J.~R., \& {Ma}, C.-P. 2010, \apj,
  711, 1291

\bibitem[{{Yagi} {et~al.}(2015){Yagi}, {Gu}, {Koyama}, {Nakata}, {Kodama},
  {Hattori}, \& {Yoshida}}]{Yag++15}
{Yagi}, M., {Gu}, L., {Koyama}, Y., {et~al.} 2015, \aj, 149, 36

\bibitem[{{Yuan} {et~al.}(2013){Yuan}, {Kewley}, \& {Rich}}]{YKR11}
{Yuan}, T.-T., {Kewley}, L.~J., \& {Rich}, J. 2013, \apj, 767, 106

\bibitem[{{Zheng} {et~al.}(2011){Zheng}, {Cen}, {Weinberg}, {Trac}, \&
  {Miralda-Escud{\'e}}}]{Zhe+11}
{Zheng}, Z., {Cen}, R., {Weinberg}, D., {Trac}, H., \& {Miralda-Escud{\'e}}, J.
  2011, \apj, 739, 62

\bibitem[{{Zitrin} {et~al.}(2014){Zitrin}, {Redlich}, \& {Broadhurst}}]{ZRT14}
{Zitrin}, A., {Redlich}, M., \& {Broadhurst}, T. 2014, \apj, 789, 51

\end{thebibliography}

\end{document}